\documentclass[prx, twocolumn, superscriptaddress]{revtex4}
\usepackage{graphicx,color}
\usepackage{amssymb}
\usepackage{amsmath}
\usepackage{bm}
\usepackage{hyperref}
\usepackage{tabularx}
\usepackage{multirow}
\usepackage{braket}
\usepackage{xcolor}
\usepackage{gensymb}

\begin{document}

\title{Electronic Density of States of a $U\left(1\right)$ Quantum Spin Liquid with Spinon Fermi Surface. I.  Orbital  Magnetic Field Effects}
\author{Wen-Yu He}\thanks{hewy@shanghaitech.edu.cn}
\affiliation{School of Physical Science and Technology, ShanghaiTech University, Shanghai 201210, China}
\author{Patrick A. Lee}\thanks{palee@mit.edu}
\affiliation{Department of Physics, Massachusetts Institute of Technology, Cambridge, Massachusetts 02139, USA}

\date{\today}

\begin{abstract}
Quantum spin liquid with spinon Fermi surface is an exotic insulator that hosts neutral Fermi surfaces inside the insulating gap. In an external magnetic field, it has been pointed out that the neutral Fermi surfaces are Landau quantized to form Landau levels due to the coupling to the induced emergent gauge magnetic field. In this work, we calculate the electronic density of states (as observed in tunneling experiments) of the quantum spin liquid in an orbital magnetic field. We find that the Landau levels from the neutral Fermi surfaces give rise to a set of steps emerging at the upper and lower Hubbard band edges. Each of the  Hubbard band edge steps further develop into a band edge resonance peak when a weak gauge binding arising from the gauge field fluctuations is taken into account. Importantly, each Hubbard band edge step and its resulting resonance peak in the weak gauge binding are found to have a correspondence Landau level from the neutral Fermi surfaces, so the Hubbard band edge steps and the band edge resonance peaks provide signatures to the unique feature that characterizes the Landau quantization of the in-gap neutral Fermi surfaces in the spin liquid. We further consider the strong gauge binding regime where the band edge resonance peaks  move into the Mott gap and develop into true in-gap bound states. In the strong gauge binding regime, we solve the Landau level spectrum of the in-gap bound states in an orbital magnetic field. For the in-gap bound state with a Mexican hat like band dispersion, we find that the envelop energy to have a state excited from the bound state Landau levels decreases quadratically with the magnetic field. The quadratic decrease behavior of the envelop energy is consistent with the intuition that applying magnetic field localizes the states and energetically promotes the in-gap bound states formation. Finally, we discuss the connection of our results to the electronic density of states spectra measured in the layered 1T-TaS$_2$. We point out that a quantum spin liquid with a quasi-bound state in the upper Hubbard band can give the density of states spectra similar to the one measured in the experiment.
\end{abstract}

\maketitle

\section{Introduction}\label{I}
Quantum spin liquid (QSL), due to its close connections to the high temperature superconductivity phenomenon~\cite{Anderson1,Patrick1} and potential applications in topological quantum computations~\cite{Kitaev,Nayak}, has been a long sought state of matter since Anderson's first proposal in 1973~\cite{Anderson2}. As a QSL is an exotic insulator with no traditional Landau order parameter down to zero temperature~\cite{Patrick1,Balents1,YiZhou}, the search for QSL states in real materials becomes extremely difficult. Among various types of QSLs, the gapless $U\left(1\right)$ QSL with spinon Fermi surface (SFS) is featured by charge neutral spin excitations living on the neutral Fermi surfaces~\cite{Patrick1, Patrick2, Senthil1}, so its experimental identification focuses on detecting the neutral Fermi surfaces inside the insulating gap. 

One seminal idea to detect the neutral Fermi surfaces in the gapless $U\left(1\right)$ QSL is built on the effect of Landau quantization~\cite{Motrunich, Senthil2}. In a $U\left(1\right)$ QSL with SFS, an electron is fractionalized into a spinon and a chargon. The spinon is a charge-neutral fermion that carries spin-1/2, while the chargon is a charged boson that carries the electric charge. The spinon and the chargon are coupled through an emergent $U\left(1\right)$ gauge field~\cite{Patrick2}. In the presence of an external orbital magnetic field $B$, an emergent gauge magnetic field (EGMF) $b$ is induced on the spinons and the remaining magnetic field on the chargons is $B-b$~\cite{Motrunich, Senthil2, Patrick2}. Since the induced EGMF on the spinons Landau quantizes the neutral Fermi surfaces in the $U\left(1\right)$ QSL, the $U\left(1\right)$ QSL was predicted to have quantum oscillations (QOs) of resisitivity and magnetization in spite of its insulating nature~\cite{Motrunich, Senthil2}. However, up to now this effect has not been seen experimentally.

QOs in the insulating states of matter are highly unusual as the canonical understanding of QOs is based on the existence of electronic Fermi surfaces~\cite{Shoenberg}. Recently, a few experimental observations of QOs in insulators were reported~\cite{Sanfeng, Sebastian, LuLi, Ong, LuLi2}, and Landau quantization of the neutral Fermi surfaces inside the insulating gap was suggested to be one possible origin of the observed QOs. However, Landau quantization of the neutral Fermi surfaces is sufficient for QOs in an insulator but not necessary. It has been known that a band insulator with a hybridization gap can also have QOs that arise from the gap size modulation by magnetic field~\cite{Cooper1, FaWang, Patrick3, Wenyu1}. Therefore extrinsic effects need to be ruled out before one can conclude that the observations necessarily indicate the Landau quantization of the in-gap neutral Fermi surfaces. In order to detect the neutral Fermi surfaces in a $U\left(1\right)$ QSL by the effect of Landau quantization, it will be desirable to find  evidence of the Landau quantization of the neutral Fermi surfaces other than QOs.

One possibility is to directly observe the effect of Landau levels by tunneling spectroscopy. For a band insulator in an orbital magnetic field, the electronic density of states (DOS) is known to be a set of discrete Delta function like peaks that originate from the electronic Landau levels (LLs) in each band. For a $U\left(1\right)$ QSL with SFS, since the QSL electron is a composite particle composed of a chargon and a spinon, its electronic DOS in an orbital magnetic field requires a comprehensive consideration of the chargon DOS and the spinon DOS along with the magnetic field partition between them. So far, the nature of the electronic DOS of a $U\left(1\right)$ QSL with SFS in an orbital magnetic field and how it compares to that of a band insulator has not been studied in any detail. This is the main goal of this paper.

\begin{figure}
\centering
\includegraphics[width=3.5in]{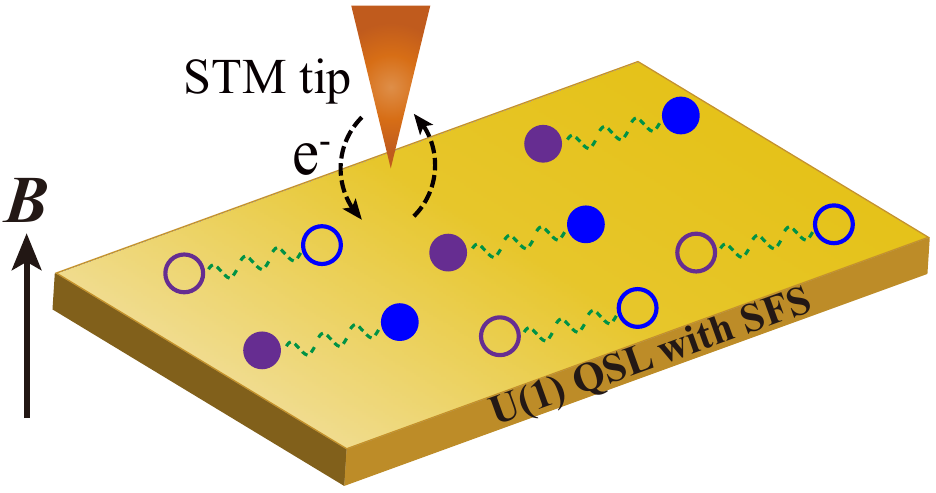}
\caption{The STM setup to measure the local electronic DOS in the QSL. An electron injected into the QSL is fractionalized into a spinon and a chargon. The filled blue and purple circles denote the spinons and doublons respectively. The empty blue and purple ones represent the spinon holes and holons respectively. Wavy lines indicate the spinon chargon attraction due to the $U\left(1\right)$ gauge field fluctuations.}\label{figure1}
\end{figure}

The electron spectal function and the local density of states (DOS) for a Mott insulator with a SFS has been studied in Ref. \cite{Tang} in zero magnetic field. At the mean field level, the electronic spectrum for a given momentum $\bm{k}$ (as measured by angle-resolved photo-emission spectroscopy (ARPES) ) is given by the convolution of the spinon and chargon spectra in frequency and momentum. For the upper and lower Hubbard band (UHB, LHB), the minimum excitation involves a gapped chargon with energy $\Delta$ and a gapless spinon at the spinon Fermi momentum $\bm{k}_{\textrm{F}}$. Therefore the  excitations have an energy threshold at $\Delta$ which occurs in a ring in momentum space with radius given the the Fermi momentum $\bm{k}_{\textrm{F}}$. A continuum of excitation appears above the threshold and the spectral function goes as $\sqrt{E-\Delta}$ above the gap. The local DOS is obtained by integrating over momentum space. A common way to measure the local DOS is by scanning tunneling microscopy (STM) as shown in Fig. \ref{figure1}. The STM tip injects an electron or hole into the system, and breaks up into spinon and chargon. The local DOS was found to increase linearly with energy above the threshold~\cite{Tang}.

Beyond mean field one has to consider the effect of gauge field fluctuations. In Ref.~\cite{Patrick5, XGWen, Tang}, the dominant effect is considered via the screened longitudinal gauge field fluctuations which they model with a short range attraction $U_{\textrm{b}}$. This is illustrated in Fig. \ref{figure1}. Beyond a certain interaction strength, a bound state was found to split off from edge of the Hubbard band. In this paper we explore in greater details the intermediate coupling strength regime and find that a resonance is formed near the band edge.

Recently, several STM measurements have been carried out to detect the possible QSL phases on the surface of bulk 1T-TaS$_2$~\cite{Yayu, Butler2, Butler3, Shichao}, monolayer 1T-TaSe$_2$~\cite{YiChen1, WeiRuan, YiChen2} and 1T/1H-TaS$_2$ heterostructure~\cite {Vano}. 
These data show clear upper and lower Hubbard bands as rather broad peaks in the DOS spectra with rather sharp onset. Interestingly the data on monolayer 1T-TaS$_2$ grown on oriented graphite~\cite{Vano} shows rather linear onsets in both the UHB and LHB, in agreement with the prediction of \cite{Tang}. Furthermore, on the surface of the layered 1T-TaS$_2$, an extra resonance peak with sidebands was found near the UHB edge~\cite{Butler3}. In a subsequent measurement an external magnetic field is applied~\cite{Butler1}, and the UHB edge resonance peak was found to move towards the Mott gap center as the magnetic field increase as $B^2$. These data motivate us to examine the local DOS in greater details, first without a magnetic field and then with an orbital magnetic field. Even in zero $B$ field, the spectrum is far from being free electron like. Therefore we expect more complicated behavior than the naive expectation that the continuum of states will be replaced by a set of discrete LLs.

For a $U\left(1\right)$ QSL in an orbital magnetic field $B$, a physical electron is composed of a spinon in the spinon LLs and a chargon in the chargon LLs, where the spinon LLs and chargon LLs are induced by the EGMF $b$ and the remaining magnetic field $B-b$ respectively. The magnetic field partition between the spinons and the chargons is determined by the Ioffle-Larkin rule~\cite{Ioffe, Patrick4}. For a gapless $U\left(1\right)$ QSL that occurs in the weak Mott regime~\cite{Motrunich2, Senthil3, Senthil4, Wenyu2}, the Ioffle-Larkin rule indicates that the EGMF $b$ dominates over the remaining magnetic field $B-b$. Since the spinon LL spacing is much larger than that of the chargons, each time  a spinon LL is filled will cause a sudden increase in the electronic DOS. Near the energies of Hubbard band edges, such sudden changes of electronic DOS are manifested as a few steps, and those steps have the one to one correspondence to the spinon LLs. Since the orbital magnetic field induced band edge steps in the QSL electronic DOS are in sharp contrast to the discrete  Delta function like peaks in that of a band insulator, those steps represent a unique feature of a gapless $U\left(1\right)$ QSL in an orbital magnetic field.

Next we consider the effect of spinon-chargon binding. We find that when the gauge binding is weak,  with no external magnetic field, the electronic DOS  develop a pair of resonance peaks at the bottom of the UHB and the top of the LHB. As the binding interaction increases, the pair of band edge resonance peaks gradually move inside the Mott gap and eventually develop into a pair of in-gap bound states. 

In the presence of an orbital magnetic field, it is found that for  weak gauge binding  each band edge step in the QSL electronic DOS at zero $U_{\textrm{b}}$ evolves into a resonance peak, which is the precursor of binding between a spinon in the spinon LLs and a chargon. Inherited from the band edge steps at zero $U_{\textrm{b}}$, each band edge resonance peak is intrinsically connected to a LL from the neutral Fermi surfaces as well. For the QSL with a weak gauge binding, the external magnetic field induced Landau quantization of the neutral Fermi surfaces is thus characterized by the set of band edge resonance peaks emerging in the QSL electronic DOS. 

Next we consider the case of strong gauge binding.  With increasing gauge binding the resonance peaks begin to move into the Mott gap. In a magnetic field, these discrete in-gap peaks in a sufficiently large gauge binding correspond to the LLs of the in-gap bound states. By solving the binding equation of spinon LL states and chargons, we obtain the bound state LL spectrum. For the in-gap bound state with a Mexican hat like band dispersion, we find that the envelop energy to have a state excited from the bound state LL spectrum decreases quadratically with $B$. The quadratic decrease of the envelop energy with $B$ matches the intuition that the energy saved in the binding increases as the spinons get more localized in the magnetic field.

The goal of this study is to calculate the electronic DOS of a $U\left(1\right)$ with SFS in an orbital magnetic field. In the first stage the spinon and chargon are treated as non-interacting. In the second stage, the gauge binding from the $U\left(1\right)$ gauge field fluctuations is taken into account to see how the electronic DOS gets affected. In the study, the regimes of zero gauge binding, weak gauge binding and strong gauge binding are all covered so the orbital magnetic field effects on the electronic DOS of a $U\left(1\right)$ QSL with SFS are comprehensively understood. The rest of the paper is organized as follows. In Sec. \ref{II}, the fractionalization of an electron into a spinon and a chargon is introduced for a $U\left(1\right)$ QSL with SFS. In a two-dimensional system, given the spinon DOS and the chargon DOS, the electronic DOS of the gapless $U\left(1\right)$ QSL at $B=0$T is obtained. In Sec. \ref{III}, the gapless $U\left(1\right)$ QSL in an orbital magnetic field is shown to have the Landau quantization that gives rise to steps at the Hubbard band edges. Those band edge steps in the QSL electronic DOS are found to have intrinsic connections to the spinon LLs. In Sec. \ref{IV}, by numerically calculating the electronic DOS of a gapless $U\left(1\right)$ QSL in a triangular lattice, we confirm that the orbital magnetic field induced spinon Landau quantization gives rise to the band edge steps that emerge in the QSL elecronic DOS. In Sec. \ref{V}, the gauge binding $U_{\textrm{b}}$ from the $U\left(1\right)$ gauge field fluctuations is added to the QSL. The QSL electronic DOS at $B=0$T is obtained in both the weak and strong gauge binding regime. The evolution of the QSL electronic DOS with the increase of $U_{\textrm{b}}$ is given. In Sec. \ref{VI}, the orbital magnetic field effects on the QSL electronic DOS are considered in both the weak and strong gauge binding regime. In Sec. \ref{VII}, we deal with the QSL in the strong gauge binding regime. The bound state band dispersions are analyzed in the continuum model at $B=0$T. In a finite magnetic field, the binding equations for the in-gap bound states are derived and the bound state LL spectrum is numerically solved. In Sec. \ref{VIII}, we connect our results of the QSL electronic DOS in an orbital magnetic field to the electronic DOS spectra measured by STM in the layered 1T-TaS$_2$~\cite{Butler1}. In Sec. \ref{IX}, we give a brief conclusion to our results.

\begin{figure*}
\centering
\includegraphics[width=6.8in]{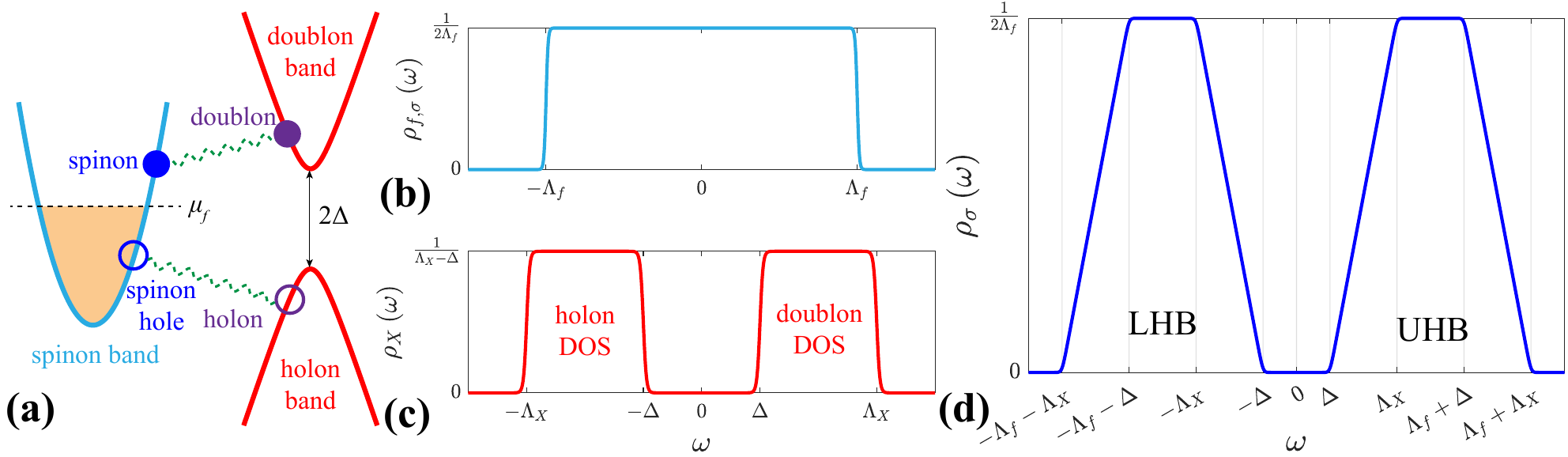}
\caption{Electron fractionalization in a $U\left(1\right)$ QSL with SFS and the resulting electronic DOS, calculated in mean field theory without accounting for gauge field fluctuaions. In the QSL, an electronic excitation is composed of a spinon excitation and a doublon, while a hole state is composed of a spinon hole and a holon. Holons and doublons are the chargons that carry the charge $\pm e$ respectively. In the case of $B=0$T in (a), the spinons, holons and doublons all live in the quadratic bands. The spinons are coupled to the chargons through an emergent $U\left(1\right)$ gauge field that is denoted by the green wavy lines. The quadratic band dispersions in (a) give rise to the constant DOS of the spinons in (b) and that of the chargons in (c). After convolution, the resulting QSL electronic DOS $\rho_\sigma\left(\omega\right)$ in $B=0$T is given in (d).The orange shaded region in (a) represents the spinon Fermi sea filled by the spinons.}\label{figure2}
\end{figure*}

\section{Electron Fractionalization in a QSL and the local Electronic DOS}\label{II}
In a $U\left(1\right)$ QSL with SFS, electrons go through the spin-charge separation and are fractionalized into spinons and chargons. In the slave rotor formalism, the mean field Hamiltonian for the $U\left(1\right)$ QSL with SFS takes the form~\cite{Supplemental}:
\begin{align}\label{mean_H}
H_0=&\sum_{\bm{k}}\epsilon_{\bm{k}}\left(a_{-\bm{k}}a^\dagger_{-\bm{k}}+b^\dagger_{\bm{k}}b_{\bm{k}}\right)+\sum_{\bm{k},\sigma}\xi_{\bm{k}}f^\dagger_{\sigma,\bm{k}}f_{\sigma,\bm{k}},
\end{align}
with $a^{\left(\dagger\right)}_{-\bm{k}}$, $b^{\left(\dagger\right)}_{\bm{k}}$ and $f^{\left(\dagger\right)}_{\sigma,\bm{k}}$ being the annihilation (creation) operators for a holon, doublon and spinon respectively. Here $\sigma=\uparrow/\downarrow$ denotes the spin index. The holons and doublons are the nonrelativistic approximation to the relativistic chargons near the Hubbard band edges~\cite{Patrick2}. A holon carries the charge $+e$ as that in a hole excitation, while a doublon carries the charge $-e$ as that in an electron. For the spinons, the spinon band $\xi_{\bm{k}}$ has the spinon chemical potential $\mu_f$ lying inside the band, so there exist neutral Fermi surfaces that bring about the gapless spin excitations, as is indicated in Fig. \ref{figure2} (a). For the holons and doublons, the energy spectrum $\epsilon_{\bm{k}}$ are gapped so there exists a gap for charge excitations as shown in Fig. \ref{figure2} (a). In the assumption of deconfinement~\cite{XGWen2, SSLee}, the $U\left(1\right)$ QSL with SFS is a charge insulator but exhibits metallic behavior in the spin channel.

In the $U\left(1\right)$ QSL, due to the fractionalization, a physical electron is composed of a spinon and a chargon. In the QSL, to create an electronic state requires to create a spinon and a doublon together, or to create a spinon and simutaneously annihilate a holon. The opeartor to create an electron is written as $c^\dagger_{\bm{k},\bm{k}',\sigma}=f^\dagger_{\sigma,\bm{k}}\left(a_{-\bm{k}'}+b^\dagger_{\bm{k}'}\right)$, so the Matsubara Green's function for an electronic state in the QSL is constructed from the convolution  ~\cite{Tang, Supplemental}
\begin{widetext}
\begin{align}\label{Cov1}
G_{\sigma}\left(i\omega_n,\bm{k},\bm{k}'\right)=&-\frac{1}{\beta}\sum_{\nu_n}G_{f,\sigma}\left(i\omega_n-i\nu_n,\bm{k}\right)\left[G_a\left(-i\nu_n,-\bm{k}'\right)+G_b\left(i\nu_n,\bm{k}'\right)\right],
\end{align}
\end{widetext}
with $\beta=\left(k_{\textrm{b}}T\right)^{-1}$ being the thermodynamic beta, $\omega_n=\left(2n+1\right)\pi\beta^{-1}$ and $\nu_n=2n\pi\beta^{-1}$ being the fermionic and bosonic Matsubara frequencies respectively. Here $G_{f,\sigma}\left(i\omega_n,\bm{k}\right)=\left(i\omega_n-\xi_{\bm{k}}\right)^{-1}$ is the spinon Matsubara Green's function, $G_{a}\left(-i\nu_n,-\bm{k}\right)=\left(-i\nu_n-\epsilon_{\bm{k}}\right)^{-1}$ is the holon Matsubara Green's function and $G_{b}\left(i\nu_n,\bm{k}\right)=\left(i\nu_n-\epsilon_{\bm{k}}\right)^{-1}$ is the doublon Matsubara Green's function. After summing over the Matsubara frequencies, one can perform the analytic continuation $i\omega_n\rightarrow\omega+i0^+$ to get the retarded electronic Green's function:
\begin{align}\label{GR1}
G_\sigma^{\textrm{R}}\left(\omega,\bm{k},\bm{k}'\right)=&\frac{n_{\textrm{F}}\left(\xi_{\bm{k}}\right)+n_{\textrm{B}}\left(\epsilon_{\bm{k}'}\right)}{\omega+i0^+-\xi_{\bm{k}}+\epsilon_{\bm{k}'}}+\frac{n_{\textrm{F}}\left(-\xi_{\bm{k}}\right)+n_{\textrm{B}}\left(\epsilon_{\bm{k}'}\right)}{\omega+i0^+-\xi_{\bm{k}}-\epsilon_{\bm{k}'}},
\end{align}
with $n_{\textrm{F}}\left(\xi\right)=\frac{1}{2}\left(1-\tanh\frac{\beta\xi}{2}\right)$ and $n_{\textrm{B}}\left(\epsilon\right)=\frac{1}{2}\left(\coth\frac{\beta\epsilon}{2}-1\right)$ being the Fermi distribution function and Bose distribution function respectively. In this section we discuss the local DOS as measured by an STM experiment. 
For a QSL with translational symmetry, the local electronic DOS counts all the allowed $\bm{k}$ and $\bm{k}'$ modes, so the number of electronic states per energy per unit cell is~\cite{Supplemental}
\begin{align}\nonumber\label{Local_DOS1}
\rho_\sigma\left(\omega\right)=&\frac{1}{N^2}\sum_{\bm{k},\bm{k}'}n_{\textrm{F}}\left(\xi_{\bm{k}}\right)\delta\left(\omega-\xi_{\bm{k}}+\epsilon_{\bm{k}'}\right)\\
&+\frac{1}{N^2}\sum_{\bm{k},\bm{k}'}n_{\textrm{F}}\left(-\xi_{\bm{k}}\right)\delta\left(\omega-\xi_{\bm{k}}-\epsilon_{\bm{k}'}\right),
\end{align}
with $N$ being the number of lattice sites. From here we refer the number of states per energy per unit cell to the DOS. In Eq. \ref{Local_DOS1}, the Bose factor has been dropped in the low temperature regime $k_{\textrm{b}}T\ll\textrm{min}\left[\epsilon_{\bm{k}}\right]=\Delta$. Here $\Delta$ denotes the Mott gap.

The QSL electronic DOS obtained in Eq. \ref{Local_DOS1} provides a clear picture about how the compositions of the spinons and the chargons contribute to the total electronic DOS. In the $U\left(1\right)$ QSL with SFS, a spinon hole inside the spinon Fermi sea together with a holon forms a hole state, while a spinon excitation above the spinon Fermi level combined with a doublon gives rise to an electronic excitation, as is illustrated in Fig. \ref{figure2} (a). Since the combination of a spinon state with a quasi-momentum $\bm{k}$ and a chargon state with a quasi-momentum $\bm{k}'$ is arbitrary, the total number of the combinations gives the number of states in the QSL. The first term in Eq. \ref{Local_DOS1} counts the number of hole states, so it gives the DOS in the lower Hubbard band (LHB). Similarly, the second term in Eq. \ref{Local_DOS1} counts the number of electronic excitations, so it gives the DOS in the upper Hubbard band (UHB).

By applying Eq. \ref{Local_DOS1}, one can obtian the electronic DOS of a two-dimensional $U\left(1\right)$ QSL with SFS. The spinon band and the chargon band in the QSL can be approximated by the quadratic dispersions: $\xi_{\bm{k}}=\frac{\hbar^2\bm{k}^2}{2m_f}-\mu_f$ and $\epsilon_{\bm{k}}=\frac{\hbar^2\bm{k}^2}{2m_X}+\Delta$ respectively. In the quadratic band approximation, as long as the energy lies inside the bands, the DOS always takes the constant value. The spinon DOS takes $\rho_{f,\sigma}\left(\omega\right)=\frac{m_fA_{\textrm{c}}}{2\pi\hbar^2}$ for $\omega\in\left[-\Lambda_f,\Lambda_f\right]$, as plotted in Fig. \ref{figure2} (b). For the chargons, the holons in the range $\omega\in\left[-\Lambda_X-\Delta,-\Delta\right]$ and the doublons in the range $\omega\in\left[\Delta,\Lambda_X+\Delta\right]$ both contribute to the constant DOS $\rho_X\left(\omega\right)=\frac{m_XA_{\textrm{c}}}{2\pi\hbar^2}$ as plotted in Fig. \ref{figure2} (c). Here $A_{\textrm{c}}$ is the unit cell area, $\Lambda_f=\frac{\pi\hbar^2}{m_fA_{\textrm{c}}}$ and $\Lambda_X=\frac{2\pi\hbar^2}{m_XA_{\textrm{c}}}+\Delta$ are the energy cut-offs introduced for the bands of the spinons and the chargons respectively. Both the spinon DOS and the chargon DOS show an abrupt increase from zero at the band edges. Given the spinon DOS and the chargon DOS, the QSL electronic DOS is then evaluated and plotted in Fig. \ref{figure2} (d). One can observe that the electronic DOS shows finite slopes at the Hubbard band edges ~\cite{Tang}, which is consistent with Eq. \ref{Local_DOS1} that the number of states increases linearly from zero as the energy crosses from $|\omega|=\Delta$ into the bulk Hubbard bands. 
Here the simple quadratic approximation for the spinon and chargon band dispersions has been adopted, but one can see in Sec. \ref{IV} below that the quadratic band approximation captures all the key features present in the more realistic QSL electronic DOS calculated in a lattice model.

\begin{figure*}
\centering
\includegraphics[width=6.8in]{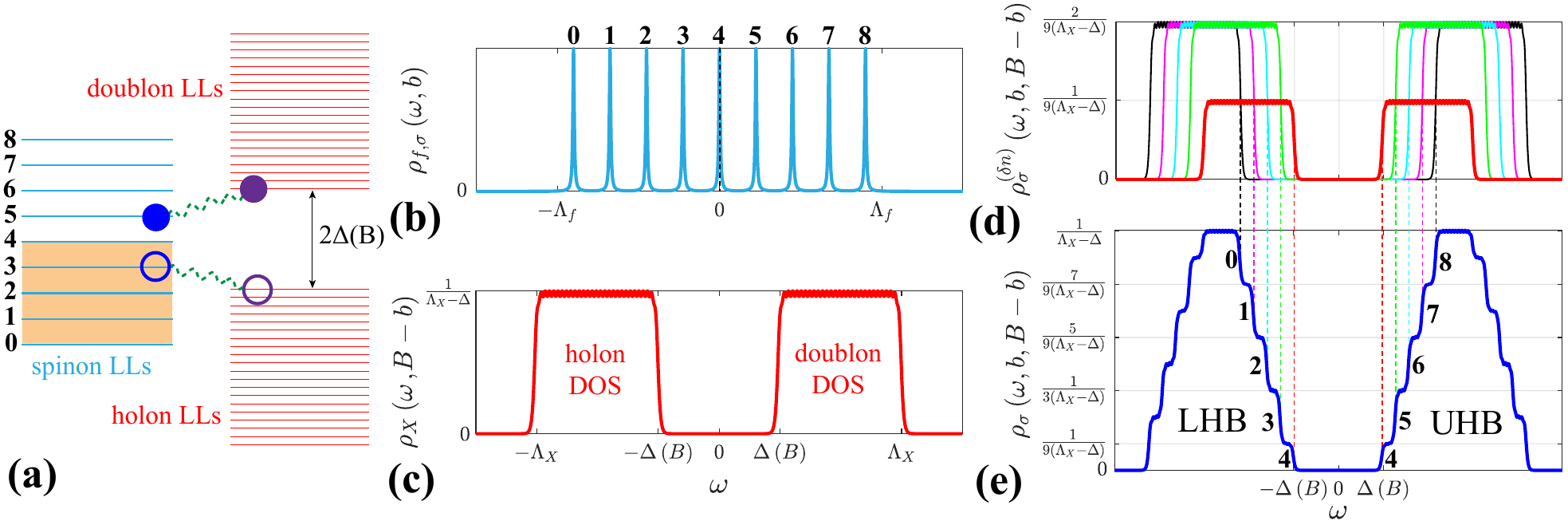}
\caption{Landau quantizations in the QSL and the DOS. This figure is designed to be compared with Fig. \ref{figure2} (a)-(d) which describe the case of zero magnetic field. In a finite orbital magnetic field $B$, the bands of the spinons and chargons are Landau quantized to form the LLs in (a). For the QSL that occurs in the weak Mott regime, the spinon LL spacing $\hbar\omega_f$ is larger than the chargon LL spacing $\hbar\omega_X$. In the temperature region $\hbar\omega_X<k_{\textrm{b}}T\ll\hbar\omega_f$, the spinon LLs give rise to a set of Dirac Delta function like peaks in the spinon DOS $\rho_{f,\sigma}\left(\omega,b\right)$ in (b), while the chargon LLs are thermally broadened, generating the chargon DOS $\rho_X\left(\omega,B-b\right)$ in (c) with a few ripples representing the chargon LLs. In (b), the black dashed line denotes the spinon chemical potential $\mu_f\left(b\right)$, and the $n=4$ spinon LL is assumed to be half filled. In this case, a hole state in the LHB is composed of a spinon hole in the $n=4,3,2,1,0$ LLs and a holon. An electronic excitation in the UHB is composed of a spinon excitation in the $n=4,5,6,7,8$ LLs and a doublon. The resulting electronic DOS is $\rho_\sigma\left(\omega,b,B-b\right)=\sum_{\delta n=0}^4\rho^{\left(\delta n\right)}_\sigma\left(\omega,b,B-b\right)$, which includes 5 terms. The DOS $\rho^{\left(\delta n\right)}_\sigma\left(\omega,b,B-b\right)$ with $\delta n=0,1,2,3,4$ correspond to the red, green, cyan, magenta, and black lines respectively in (d). Summing over all the spinon LL index as indicated in Eq. \ref{rho_sigma_simple} gives the total electronic DOS $\rho_\sigma\left(\omega,b,B-b\right)$ in (e), which shows a few steps near the Hubbard band edges. The Hubbard band edge steps in (e) originate from the abrupt change of $\rho^{\left(\delta n\right)}_\sigma\left(\omega,b,B-b\right)$ in (d) as demonstrated by the dashed lines in (d) and (e). Each Hubbard band edge step is labeled by one spinon LL index. The orange shaded region in (a) means that the spinon LL states there are occupied.}\label{figure3}
\end{figure*}

\section{Landau Quantization and the Electronic DOS of a QSL}\label{III}
An orbital magnetic field $B$ applied to a $U\left(1\right)$ QSL is divided into two parts: an EGMF $b$ on the spinons and a remaining magnetic field $B-b$ on the chargons~\cite{Patrick2, Senthil1, Motrunich}. The spinons and the chargons are both Landau quantized as schematically shown in Fig. \ref{figure3} (a), so the mean field Hamiltonian in Eq. \ref{mean_H} now becomes
\begin{widetext}
\begin{align}
H_0\left(b,B-b\right)=&\sum_{n,m}\epsilon_n\left(a_{n,m}a^\dagger_{n,m}+b^\dagger_{n,m}b_{n,m}\right)+\sum_{n,m,\sigma}\xi_{n}f^\dagger_{n,m,\sigma}f_{n,m,\sigma}.
\end{align}
\end{widetext}
Here $\epsilon_n$ are the chargon LLs induced by $B-b$ and $\xi_n$ represent the spinon LLs arising from the EGMF $b$. The operators $a^{\left(\dagger\right)}_{n,m}$, $b^{\left(\dagger\right)}_{n,m}$ and $f^{\left(\dagger\right)}_{n,m}$ are the annihilation (creation) operator for a holon, doublon and spinon in the $n$th LL respectively. The index $m$ counts the LL degeneracy.

As the orbital magnetic field $B$ introduces Landau quantization, an electronic state in the $U\left(1\right)$ QSL is now composed of a spinon LL state and a chargon LL state. The creation operator of an electronic state now takes the form $c^\dagger_{n,m,n',m',\sigma}=f^\dagger_{n,m,\sigma}\left(a_{n',m'}+b^\dagger_{n',m'}\right)$. Similar to the case of $B=0$T, the electronic Matsubara Green's function in an orbital magnetic field $B$ is constructed from the convolution~\cite{Supplemental}
\begin{widetext}
\begin{align}
G_{n,m,n',m',\sigma}\left(i\omega_n\right)=-\frac{1}{\beta}\sum_{\nu_n}G_{f,n,m,\sigma}\left(i\omega_n-i\nu_n\right)\left[G_{a,n',m'}\left(-i\nu_n\right)+G_{b,n',m'}\left(i\nu_n\right)\right],
\end{align}
\end{widetext}
with $G_{f,n,m,\sigma}\left(i\omega_n\right)=\left(i\omega_n-\xi_n\right)^{-1}$, $G_{a,n,m}\left(-i\nu_n\right)=\left(-i\nu_n-\epsilon_n\right)^{-1}$ and $G_{b,n,m}\left(i\nu_n\right)=\left(i\nu_n-\epsilon_n\right)^{-1}$ being the Matsubara Green's function of the spinon, holon and doublon LL states respectively. After performing the Matsubara frequency summation, one gets the retarded Green's function for the QSL electronic state:
\begin{align}
G^{\textrm{R}}_{n,m,n',m',\sigma}\left(\omega\right)=&\frac{n_{\textrm{F}}\left(\xi_n\right)+n_{\textrm{B}}\left(\epsilon_{n'}\right)}{\omega+i0^+-\xi_n+\epsilon_{n'}}+\frac{n_{\textrm{F}}\left(-\xi_n\right)+n_{\textrm{B}}\left(\epsilon_{n'}\right)}{\omega+i0^+-\xi_n-\epsilon_{n'}}.
\end{align}
For the QSL in an orbital magnetic field $B$, as the number of electronic states counts all the states in the LLs, the electronic DOS takes the form
\begin{align}\nonumber\label{Local_DOS2}
\rho_\sigma\left(\omega,b,B-b\right)=&\frac{1}{N^2}\sum_{n,m,n',m'}n_{\textrm{F}}\left(\xi_n\right)\delta\left(\omega-\xi_n+\epsilon_{n'}\right)\\
&+\frac{1}{N^2}\sum_{n,m,n',m'}n_{\textrm{F}}\left(-\xi_n\right)\delta\left(\omega-\xi_n-\epsilon_{n'}\right).
\end{align}
Here the Bose factor has been dropped in the low temperature regime as is done in the case of $B=0$T.

Comparing Eq. \ref{Local_DOS2} and Eq. \ref{Local_DOS1}, one finds that the electronic DOS in a finite $B$ has a similar form to the electronic DOS at $B=0$T, except that the quasi-momentum in Eq. \ref{Local_DOS1} are replaced by the LL index in Eq. \ref{Local_DOS2}. The two terms in Eq. \ref{Local_DOS2} have the similar physical meaning as those in Eq. \ref{Local_DOS1}. In an orbital magnetic field $B$, both the spinon bands and the chargon bands are Landau quantized to form LLs as schematically shown in Fig. \ref{figure3} (a). For the $U\left(1\right)$ QSL in a magnetic field, a hole state is composed of a spinon hole from a filled (or partial filled) spinon LL and a holon LL state. The number of hole states equals to the total combinations of the occupied spinon LL states and the hole LL states, which is captured by the first term in Eq. \ref{Local_DOS2}. Similarly, a spinon excitation in an empty (or partial filled) spinon LL combined with a doublon LL state gives rise to an electronic excitation in the QSL. The number of electronic excitations is the total combinations of the unoccupied spinon LL states and the doublon LL states, which corresponds to the second term in Eq. \ref{Local_DOS2}. Therefore the LHB DOS and the UHB DOS in a finite orbital magnetic field are given by the first and second terms in Eq. \ref{Local_DOS2} respectively.

In order to obtain the QSL electronic DOS from Eq. \ref{Local_DOS2}, one needs to know the magnetic field partition between the spinons and the chargons. It is known from the Ioffe-Larkin rule~\cite{Ioffe, Patrick4} that the EGMF $b$ takes $b=\alpha B=\chi_X B/\left(\chi_X+\chi_f\right)$, with $\chi_f$ and $\chi_X$ being the diamagnetic susceptibility of the spinons and the chargons respectively. In the temperature region where $k_{\textrm{b}}T$ is larger than the LL spacing, the diamagnetic susceptibilty of the spinon is $\chi_f=\frac{e^2}{12\pi m_f}$, and that of the chargon has been calculated to be $\chi_X=\frac{e^2v^2}{24\pi \Delta}$~\cite{ZhehaoDai} with $v$ being the relativistic chargon velocity. Now the ratio $\alpha$ takes $\alpha=m_fv^2/\left(m_fv^2+2\Delta\right)$. It is clear that the ratio $\alpha$ approaches to $1$ in the small gap limit. As the gapless $U\left(1\right)$ QSL phase tends to occur in the weak Mott regime~\cite{Motrunich2, Senthil3, Senthil4, Wenyu2}, it is reasonable to assume that the EGMF $b$ on the spinons dominates over the remaining $B-b$ on the chargons. It indicates that the spinon LL spacing is much larger than the chargon LL spacing in reality.

In the quadratic band approximation, the LL spectrum of the spinon and the chargon are $\xi_n=\left(n+\frac{1}{2}\right)\hbar\omega_f-\mu_f\left(b\right)$ and $\epsilon_n=\left(n+\frac{1}{2}\right)\hbar\omega_X+\Delta$ respectively. Here $\omega_f=\frac{eb}{m_f}$ and $\omega_X=\frac{e\left(B-b\right)}{m_X}$ are the cyclotron frequencies of the spinon and chargon respectively. The spinon chemical potential $\mu_f\left(b\right)$ is determined by the equation $\sum_{n=0}^{\infty}1/n_{\textrm{F}}\left(\xi_n\right)=\nu$ with $\nu=\frac{\mu_f}{\hbar\omega_f}$ being the spinon LL filling factor~\cite{Supplemental}. Due to the spinon Landau quantization, the chemical potential $\mu_f\left(b\right)$ oscillates with $b$ and approaches to the spinon Fermi energy $\mu_f$ when $b\rightarrow0$. The evolution of $\mu_f\left(b\right)$ with $b$ can be found in Fig. S1 in the Supplemental Material~\cite{Supplemental}. In the temperature region $\hbar\omega_X<k_{\textrm{b}}T\ll\hbar\omega_f$, the spinon DOS $\rho_{f,\sigma}\left(\omega,b\right)$ is composed of a set of Dirac Delta function like peaks as is schematically plotted in Fig. \ref{figure3} (b), while the chargon LL peaks are smoothed by the thermal fluctuations. As a result, the chargon DOS $\rho_X\left(\omega, B-b\right)$ in a finite magnetic field in Fig. \ref{figure2} (c) is almost the same as that in Fig. \ref{figure3} (c) with $B=0$T . The DOS $\rho_X\left(\omega, B-b\right)$ in Fig. \ref{figure3} (c) differs from that in Fig. \ref{figure2} (c) only in two apsects: 1) in Fig. \ref{figure3} (c), the thermally smoothed LL peaks appear like ripples; 2) in Fig. \ref{figure3} (c), the Mott gap $\Delta\left(B\right)=\Delta+\frac{1}{2}\hbar\omega_X$ increaes with $B$. The linear increase of Mott gap with $B$ reflects the internal distribution of $B$ on the chargons.

In an orbital magnetic field, suppose that the QSL has its spinons filled up to the $n=n_0$ spinon LL, the electronic DOS calculated from Eq. \ref{Local_DOS2} reads~\cite{Supplemental}
\begin{widetext}
\begin{align}\label{rho_sigma_simple}
\rho_\sigma\left(\omega,b,B-b\right)=&\sum_{n=0}^{n_0}\frac{\lambda_n}{\nu}\rho_{h}\left[\omega+\left(n_0-n\right)\hbar\omega_f,B-b\right]+\sum_{n=n_0}^{n_{\textrm{c}}}\frac{1-\lambda_n}{\nu}\rho_{d}\left[\omega-\left(n-n_0\right)\hbar\omega_f,B-b\right],
\end{align}
\end{widetext}
with $\lambda_n$ being the filling of the $n$th spinon LL and $\nu=\sum_n\lambda_n$. Here $\rho_{h}\left(\omega,B-b\right)$ and $\rho_{d}\left(\omega,B-b\right)$ denote the DOS of the holons and doublons respectively. Please note that the summation of the holon DOS and doublon DOS gives the chargon DOS: $\rho_X\left(\omega,B-b\right)=\rho_{h}\left(\omega,B-b\right)+\rho_{d}\left(\omega,B-b\right)$, which can be inferred from Fig. \ref{figure2} (c) and Fig,. \ref{figure3} (c). The $n_{\textrm{c}}$ is introduced as a cut-off in the spinon LL because the spinon band has a finite band width.

To illustrate the QSL electronic DOS in an orbital magnetic field, we consider a spinon system that has its $n=4$ LL half filled as indicated in Fig. \ref{figure3} (b). The total QSL electronic DOS calculated from Eq. \ref{rho_sigma_simple} is written as $\rho_\sigma\left(\omega,b,B-b\right)=\sum_{\delta n=0}^4\rho^{\left(\delta n\right)}_\sigma\left(\omega,b,B-b\right)$, which includes 5 terms:
\begin{align}\label{rho_delta_n}
\rho_\sigma^{\left(\delta n\right)}\left(\omega,b,B-b\right)=\left\{\begin{matrix}
\frac{1}{9}\rho_X\left(\omega,B-b\right),& \delta n=0,\\ 
\frac{2}{9}\rho_{h}\left(\omega+\delta n\hbar\omega_f,B-b\right)\\
+\frac{2}{9}\rho_{d}\left(\omega-\delta n\hbar\omega_f,B-b\right), & \delta n\neq0.
\end{matrix}\right.
\end{align}
Here we have set $\delta n=\left|n-n_0\right|$. The DOS $\rho_\sigma^{\left(\delta n\right)}\left(\omega,b,B-b\right)$ with $\delta n=0,1,2,3,4$ are plotted in red, green, cyan, magenta and black respectively in Fig. \ref{figure3} (d). By performing the summation over all the spinon LL index $n$, we eventually arrive at the QSL electronic DOS plotted in Fig. \ref{figure3} (e). Importantly, since the chargon DOS $\rho_X\left(\omega,B-b\right)$ shows an abrupt increase at the threshold energy $\omega=\pm\Delta\left(B\right)$, the QSL electronic DOS obtained by summing all the energy shifted chargon DOS in Eq. \ref{rho_delta_n} exhibits a set of steps emerging near the Hubbard band edges, as can be seen in Fig. \ref{figure3} (e). Here we have assumed the half filling of the $n=4$ spinon LL, so equal number of spinon holes and spinon excitations in the $n=4$ LL are involved in the formation of physical electronic states. Therefore the resulting two steps at $\omega=\pm\Delta\left(B\right)$ in Fig. \ref{figure3} (e) are of the same height. More generally, the height of the two steps at $\omega=\pm\Delta\left(B\right)$ differs as the filling $\lambda_{n_0}$ deviates from 1/2. Specifically, the step at $-\Delta\left(B\right)$ increases from 0 to $\frac{1}{\nu\left(\Lambda_X-\Delta\right)}$ and the step at $\Delta\left(B\right)$ accordingly decreases from $\frac{1}{\nu\left(\Lambda_X-\Delta\right)}$ to 0 as the filling $\lambda_{n_0}$ increases from 0 to 1. When the applied magnetic field is so small that $\hbar\omega_f<k_{\textrm{b}}T$, those steps would be thermally smoothed and the electronic DOS consistently approaches to the case of $B=0$T shown in Fig. \ref{figure2} (d). Importantly, Eq. \ref{rho_sigma_simple} for the electronic DOS of the QSL with SFS applies to arbitrary magnetic field partition between the spinons and chargons. In an orbital magnetic field, as long as the EGMF $b$ on the spinons dominates over the remaining field $B-b$ on the chargons, the Hubbard band edge steps arising from the spinon LLs can always be identified in the electronic DOS spectra, given the temperature respecting $k_{\textrm{b}}T\ll\hbar\omega_f$.

\section{The QSL Electronic DOS in a Triangular Lattice Model}\label{IV}
\begin{figure}
\centering
\includegraphics[width=3.5in]{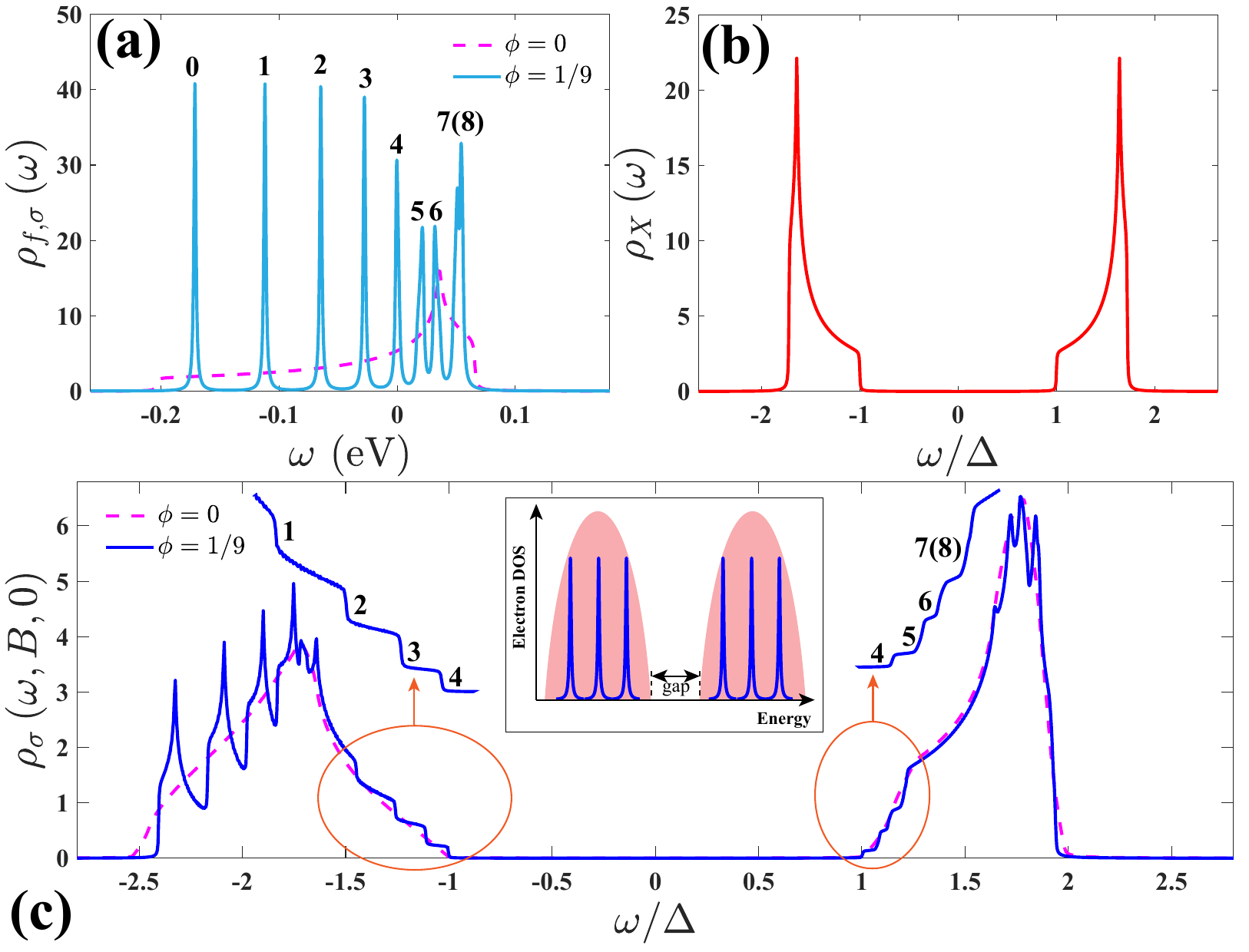}
\caption{(a) The DOS of the spinons in a triangular lattice. Given a magnetic flux ratio $\phi=1/9$, the continuous spinon DOS is Landau discretized into 9 peaks. (b) The chargon DOS in the triangular lattice. (c) The QSL electronic DOS calculated in the triangular lattice model. The spinon Landau quantization induces a few steps near the Hubbard band edges, giving the unique feature of the spinon Landau quantization in the QSL. The middle inset is the schematic showing of the electronic DOS of a band insulator. In an orbital magnetic field $B$, the continuous spectrum (pink shaded regions) below and above the gap are Landau quantized into Dirac Delta function like peaks colored in blue. In the magnetic field, the QSL electronic DOS exhibits the unique feature that is completely different from the electronic DOS of a band insulator.}\label{figure4}
\end{figure}

To verify the two-dimensional QSL electronic DOS obtained in Sec. \ref{II} and \ref{III}, we consider a gapless $U\left(1\right)$ QSL in a triangular lattice and perform a more realistic calculation for the electronic DOS in the lattice model. The band dispersions for the spinon and the chargon in the trangular lattice are taken to be
\begin{align}\label{spinon_band}
\xi_{\bm{k}}=&-2t_f\left(2\cos\frac{1}{2}k_xa\cos\frac{\sqrt{3}}{2}k_ya+\cos k_xa\right)-\mu_f,\\
\epsilon_{\bm{k}}=&-2t_X\left(2\cos\frac{1}{2}k_xa\cos\frac{\sqrt{3}}{2}k_ya+\cos k_xa-3\right)+\Delta,
\end{align}
where $a$ is the lattice constant. The band structure parameters are set to be $t_f=0.03$ eV, $t_X=0.02$ eV and $\Delta=0.25$ eV. Here the spinon chemical potential takes $\mu_f=0.025$ eV to make the spinon band half-filled. At zero magnetic field, the spinon DOS reads
\begin{align}
\rho_{f,\sigma}\left(\omega\right)=-\frac{1}{N\pi}\sum_{\bm{k}}\textrm{Im}\frac{1}{\omega+i0^+-\xi_{\bm{k}}}
\end{align}
and is plotted as the dashed magenta line in Fig. \ref{figure4} (a). The chargon DOS takes 
\begin{align}\label{rho_X}
\rho_X\left(\omega\right)=-\frac{1}{N\pi}\sum_{\bm{k}}\textrm{Im}\left(\frac{1}{\omega+i0^++\epsilon_{\bm{k}}}+\frac{1}{\omega+i0^+-\epsilon_{\bm{k}}}\right)
\end{align}
and is plotted in Fig. \ref{figure4} (b). The spinon DOS at the band bottom in Fig. \ref{figure4} (a) and that of the chargon at $\omega=\pm\Delta$ in Fig. \ref{figure4} (b) both show an abrupt increase from zero, which agree well with the quadratic band approximation results. The QSL electronic DOS $\rho_\sigma\left(\omega\right)$ in the triangular lattice can then be obtained through Eq. \ref{Cov1}, Eq. \ref{GR1} and Eq. \ref{Local_DOS1}. In Fig. \ref{figure4} (c), $\rho_\sigma\left(\omega\right)$ is plotted as the magenta dashed line. At the threshold energy $\omega=\pm\Delta$, the QSL electronic DOS $\rho_\sigma\left(\omega\right)$ from the lattice model also shows the finte slope, consistent with that from the quadratic band approximation in Sec. \ref{II}.

For the QSL in an orbital magnetic field $B$, since the magnetic field breaks the lattice translational symmetry, its mean field Hamiltonian in Eq. \ref{mean_H} needs to be reconstructed as~\cite{Supplemental}
\begin{align}\nonumber\label{hat_H}
\hat{H}_0=&\sum_{\bm{k}}\left[\hat{a}_{-\bm{k}}\hat{h}_X\left(\bm{k}\right)\hat{a}^\dagger_{-\bm{k}}+\hat{b}_{\bm{k}}^\dagger\hat{h}_X\left(\bm{k}\right)\hat{b}_{\bm{k}}\right]\\
&+\sum_{\sigma,\bm{k}}\hat{f}^\dagger_{\sigma,\bm{k}}\hat{h}_f\left(\bm{k}\right)\hat{f}_{\sigma,\bm{k}}.
\end{align}
Here the annihilation (creation) opeartors $\hat{a}^{\left(\dagger\right)}_{-\bm{k}}$, $\hat{b}^{\left(\dagger\right)}_{\bm{k}}$ and $\hat{f}^{\left(\dagger\right)}_{\sigma,\bm{k}}$ are column vectors with each element representing a state at one site in the magnetic unit cell. The matrices $\hat{h}_f\left(\bm{k}\right)$ and $\hat{h}_X\left(\bm{k}\right)$ represent the mean field tight binding Hamiltonian for the spinons in the EGMF $b$ and the chargons in the remaining field $B-b$ respectively. In principle, both $\hat{h}_f\left(\bm{k}\right)$ and $\hat{h}_X\left(\bm{k}\right)$ should be the Hofstadter Hamiltonian matrices in the triangular lattice, but it is extremely difficult to deal with the two gauge fields $b$ and $B-b$ simultaneously in the same lattice. Since the EGMF $b$ is supposed to be much larger than the remaining magnetic field $B-b$, we push it to the limit of $b\rightarrow B$ so that in the lattice model the orbital magnetic field is only acted on the spinons.

The detail form of the spinon Hofstadter Hamiltonian matrix $\hat{h}_f\left(\bm{k}\right)$ in a rational magnetic flux ratio $\phi=\frac{eB}{h}\frac{\sqrt{3}a^2}{2}$ is given in the Supplemental Materials~\cite{Supplemental}. The chargon mean field Hamiltonian matrix $\hat{h}_X\left(\bm{k}\right)$ in zero magnetic flux can be also found in the Supplemental Mateirals~\cite{Supplemental}. Here $\hat{h}_X\left(\bm{k}\right)$ is constructed in the magnetic unit cell so that a chargon can get combined with a spinon in the  same site to form a physical electron. Now for the multi-band system, the spinon DOS is
\begin{align}\label{rho_f_b}
\rho_{f,\sigma}\left(\omega, b=B\right)=&-\frac{1}{N\pi}\sum_{\bm{k}}\textrm{tr}\textrm{Im}\frac{1}{\omega+i0^+-\hat{h}_f\left(\bm{k}\right)}
\end{align}
and the chargon DOS $\rho_X\left(\omega, B-b=0\right)$ takes the same value as that calculated in Eq. \ref{rho_X}. The electronic retarded Green's function matrix elements for the multi-band system are found to be~\cite{Supplemental}
\begin{widetext}
\begin{align}\label{G_kl}
\hat{G}^{\textrm{R}}_{\sigma,k,l}\left(\omega,\bm{k},\bm{k}'\right)=&\sum_{i,i'}\hat{U}_{f,k,i}\left(\bm{k}\right)\hat{U}_{X,k,i'}\left(\bm{k}'\right)\left[\frac{n_{\textrm{F}}\left(\xi_{\bm{k},i}\right)+n_{\textrm{B}}\left(\epsilon_{\bm{k}',i'}\right)}{\omega+i0^+-\xi_{\bm{k},i}+\epsilon_{\bm{k}',i'}}+\frac{n_{\textrm{F}}\left(-\xi_{\bm{k},i}\right)+n_{\textrm{B}}\left(\epsilon_{\bm{k}',i'}\right)}{\omega+i0^+-\xi_{\bm{k},i}-\epsilon_{\bm{k}',i'}}\right]\hat{U}^\ast_{f,i,l}\left(\bm{k}\right)\hat{U}^\ast_{X,i',l}\left(\bm{k}'\right)
\end{align}
\end{widetext}
with $\hat{U}_{f}\left(\bm{k}\right)$ and $\hat{U}_X\left(\bm{k}\right)$ being the unitary matrices that diagonalize the mean field Hamiltonian $\hat{h}_f\left(\bm{k}\right)$ and $\hat{h}_X\left(\bm{k}\right)$ respectively. Here $k$, $l$ represent the matrix element index in $\hat{G}^{\textrm{R}}_\sigma\left(\omega,\bm{k},\bm{k}'\right)$. The subscripts $i$, $i'$ denote the $i$th and $i'$th eigenvalues of $\hat{h}_f\left(\bm{k}\right)$ and $\hat{h}_X\left(\bm{k}\right)$ respectively. The QSL electronic DOS in the lattice model is then derived to be
\begin{align}\label{rho_sigma_B}
\rho_\sigma\left(\omega, B,0\right)=&-\frac{1}{N^2\pi}\sum_{\bm{k},\bm{k}'}\textrm{tr}\textrm{Im}\hat{G}^{\textrm{R}}_\sigma\left(\omega,\bm{k},\bm{k}'\right).
\end{align}

The computation of Eq. \ref{G_kl} is extremely heavy when the magnetic unit cell has a large size in a small magnetic flux. In the simulation, we take the magnetic flux ratio to be $\phi=1/9$, which is the limit of our computating power. Given $\phi=1/9$, the spinon DOS calculated from Eq. \ref{rho_f_b} is plotted as the blue line in Fig. \ref{figure4} (a). It can be seen that the continuous spinon DOS at $\phi=0$ is discretized into 9 peaks. The spinon band in Eq. \ref{spinon_band} deviates from the quadratic dispersion in high energy, so the Landau discretized peaks in Fig. \ref{figure4} (a) are not uniformly distributed. The corresponding QSL electronic DOS calculated from Eq. \ref{G_kl} and \ref{rho_sigma_B} at $\phi=1/9$ is plotted as the blue line in Fig. \ref{figure4} (c), where one can see clearly that 8 steps emerge near the Hubbard band edges. As labeled in Fig. \ref{figure4} (c), each step corresponds to one discrete peak in the spinon DOS. In Fig. \ref{figure4} (c), the step from the $0$th peak in Fig. \ref{figure4} (a) is merged into the bulk LHB so it is difficult to identify. The step from the $7$th spinon LL gets mixed with that from the $8$th spinon LL because of the small energy spacing between the two spinon LLs. Physically, each edge step emerging in the QSL electronic DOS indicates a sudden change in the number of electronic states as the integer part of the spinon LL filling factor $\nu$ changes by 1. Those steps reflect the electron fractionalization in the QSL. In sharp contraste to the edge steps emerging in the QSL electronic DOS, the electronic DOS of a band insulator in an orbital magnetic field always shows discrete Dirac Delta function like peaks inside the bands, as is schematically shown in the inset of Fig. \ref{figure4} (c). The discrete peaks in Fig. \ref{figure4} (c) inset originate from the electronic LLs inside the bands. In the presence of an orbital magnetic field, such completely different features between the electronic DOS of a QSL and a band insulator therefore provide a further diagonosis to the ground state of an insulator that exhibits QOs. The edge steps in the QSL electronic DOS are regarded as the unique feature that characterizes Landau quantization of the in-gap neutral Fermi surfaces.

\section{Spinon chargon attraction: case of zero magnetic field}\label{V}
In a $U\left(1\right)$ QSL, the spinon and the chargon that are fractionalized from an electron both couple to an emerging $U\left(1\right)$ gauge field~\cite{Patrick2}, so fluctuations of the $U\left(1\right)$ gauge field in turn affect the composite electronic state. For a gapless $U\left(1\right)$ QSL in an orbital magnetic field, it has been predicted in Sec. \ref{III} and \ref{IV} that the QSL electronic DOS is characterized by a few steps emerging near the Hubbard band edges, so one question to ask is how those steps evolve as the gauge field fluctuations are turned on. In this section, we first proceed with the case of zero magnetic field.

Near the band edge energies $\omega=\pm\Delta$, the longitudinal component of the gauge field fluctuations is supposed to have the dominant effect~\cite{Patrick5, XGWen, Tang}, because the transverse components of the gauge field fluctuations are negligible due to the small current-current correlations there. The longitudinal gauge field fluctuations generate a gauge binding interaction $U_{\textrm{b}}$~\cite{Tang, Wenyu3} that couples the spinon and the chargon:
\begin{align}\label{H_int}
H_{\textrm{int}}=&\frac{U_{\textrm{b}}}{N}\sum_{\sigma,\bm{k},\bm{q},\bm{q}'}f^\dagger_{\sigma,\bm{k}-\bm{q}}f_{\sigma,\bm{k}-\bm{q}'}\left(a_{-\bm{q}}a^\dagger_{-\bm{q}'}-b^\dagger_{\bm{q}}b_{\bm{q}'}\right),
\end{align}
so the mean field Hamiltonian for the QSL becomes $H=H_0+H_{\textrm{int}}$. The original gauge binding interaction is like a Coulomb interaction, but the itinerant spinons can screen the gauge binding and make it a short range onsite interaction. The gauge binding interaction strength depends on the screening of the SFS~\cite{Supplemental}. The gauge binding interaction $U_{\textrm{b}}$ changes the QSL electronic DOS to be~\cite{Supplemental}
\begin{align}\nonumber
\tilde{\rho}_\sigma\left(\omega\right)=&-\frac{1}{N\pi}\sum_{\bm{k}}\textrm{Im}\frac{\frac{1}{N}\sum_{\bm{q}}G^{\textrm{R}}_{h,\sigma}\left(\omega,\bm{k}-\bm{q},\bm{q}\right)}{1-\frac{U_{\textrm{b}}}{N}\sum_{\bm{q}}G^{\textrm{R}}_{h,\sigma}\left(\omega,\bm{k}-\bm{q},\bm{q}\right)}\\
&-\frac{1}{N\pi}\sum_{\bm{k}}\textrm{Im}\frac{\frac{1}{N}\sum_{\bm{q}}G^{\textrm{R}}_{d,\sigma}\left(\omega,\bm{k}-\bm{q},\bm{q}\right)}{1+\frac{U_{\textrm{b}}}{N}\sum_{\bm{q}}G^{\textrm{R}}_{d,\sigma}\left(\omega,\bm{k}-\bm{q},\bm{q}\right)},
\end{align}
where $G^{\textrm{R}}_{h,\sigma}\left(\omega,\bm{k},\bm{k}'\right)$ and $G^{\textrm{R}}_{d,\sigma}\left(\omega,\bm{k},\bm{k}'\right)$ are the retarded Green's function for the states in the LHB and UHB respectively. The expressions of $G^{\textrm{R}}_{h,\sigma}\left(\omega,\bm{k},\bm{k}'\right)$ and $G^{\textrm{R}}_{d,\sigma}\left(\omega,\bm{k},\bm{k}'\right)$ are given in the first and second term in Eq. \ref{GR1} respectively.

\begin{figure*}
\centering
\includegraphics[width=6.8in]{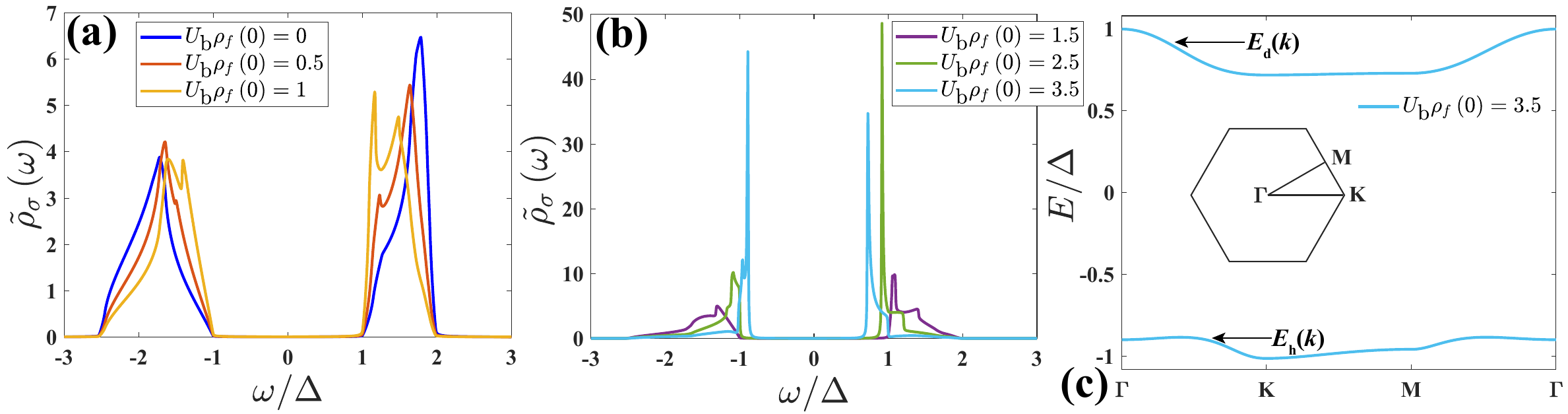}
\caption{(a) The QSL electronic DOS in different gauge binding interactions at $B=0$T. A pair of band edge resonance peaks develop as the gauge binding $U_{\textrm{b}}\rho_f\left(0\right)$ increases from 0 to 1. (b) The evolution of the band edge resonance peaks as the gauge binding $U_{\textrm{b}}\rho_f\left(0\right)$ further increases. Given a sufficiently large gauge binding, the band edge resonance peaks move inside the Mott gap and become in-gap peaks. The pair of in-gap peaks at $U_{\textrm{b}}\rho_f\left(0\right)=3.5$ indicate a pair of in-gap bound states. (c) The band dispersions of the in-gap bound states formed at $U_{\textrm{b}}\rho_f\left(0\right)=3.5$. The inset denotes the Brillouin zone of the triangular lattice. Here the binding equations in Eq. \ref{hole_band_dispersion} and \ref{electron_band_dispersion} are solved in zero temperature.}\label{figure5}
\end{figure*}

For the gapless $U\left(1\right)$ QSL in the triangular lattice, the QSL electronic DOS at the gauge binding $U_{\textrm{b}}\rho_f\left(0\right)=1$ is plotted in Fig. \ref{figure5} (a). Here $\rho_f\left(0\right)=\rho_{f,\uparrow}\left(0\right)+\rho_{f,\downarrow}\left(0\right)$ denotes the total spinon DOS at the spinon Fermi level. The value $U_{\textrm{b}}\rho_f\left(0\right)=1$ is an ideal case where the SFS brings about the Thomas-Fermi type screening~\cite{Supplemental}. In Fig. \ref{figure5} (a), it is observed that as the gauge binding interaction increases from zero, a pile-up of spectral weight is transferred from the bulk Hubbard bands to the band edges, which eventually gives rise to a pair of band edge resonance peaks. The DOS spectra is similar to that of a magnetic impurity embeded in a QSL, but a larger gauge binding is required to have a pair of band edge resonance peaks induced in the pristine QSL~\cite{Wenyu3, YiChen2}. Physically, the gauge binding $U_{\textrm{b}}\rho_f\left(0\right)=1$ promotes the binding of a spinon hole and a holon to form a hole state and also the binding of a spinon and a doublon to form an electronic state, but $U_{\textrm{b}}\rho_f\left(0\right)=1$ is not sufficiently large to generate real bound states. Therefore, the pair of band edge resonance peaks at $U_{\textrm{b}}\rho_f\left(0\right)=1$ in Fig. \ref{figure5} (a) represent the precursors of the bound states. When the gauge binding interaction increases, the band edge resonance peaks are found to further move towards the Mott gap as can be seen in Fig. \ref{figure5} (b). Given a sufficiently large gauge binding $U_{\textrm{b}}\rho_f\left(0\right)=3.5$, the peaks in the QSL electronic DOS are mainly localized inside the Mott gap. The in-gap peaks in Fig. \ref{figure5} (b) indicate the formation of real in-gap bound states at $U_{\textrm{b}}\rho_f\left(0\right)=3.5$. Importantly, the bound states are in-gap itinerant electronic states that have band dispersions. The binding equations that determine the band dispersions of the bound states are derived to be~\cite{Supplemental}
\begin{align}\label{hole_band_dispersion}
\frac{1}{U_{\textrm{b}}}-\frac{A_{\textrm{c}}}{\left(2\pi\right)^2}\int\frac{n_{\textrm{F}}\left(\xi_{\bm{k}-\bm{q}}\right)+n_{\textrm{B}}\left(\epsilon_{\bm{q}}\right)}{E_{h}\left(\bm{k}\right)+i0^+-\xi_{\bm{k}-\bm{q}}+\epsilon_{\bm{q}}}d^2\bm{q}=&0,\\\label{electron_band_dispersion}
\frac{1}{U_{\textrm{b}}}+\frac{A_{\textrm{c}}}{\left(2\pi\right)^2}\int\frac{n_{\textrm{F}}\left(-\xi_{\bm{k}-\bm{q}}\right)+n_{\textrm{B}}\left(\epsilon_{\bm{q}}\right)}{E_{d}\left(\bm{k}\right)+i0^+-\xi_{\bm{k}-\bm{q}}-\epsilon_{\bm{q}}}d^2\bm{q}=&0,
\end{align}
where $E_{h}\left(\bm{k}\right)$ and $E_{d}\left(\bm{k}\right)$ are the band dispersions of the hole bound state above the LHB and the electronic bound state below the UHB respectively. The bound state bands at $U_{\textrm{b}}\rho_f\left(0\right)=3.5$ solved from Eq. \ref{hole_band_dispersion} and \ref{electron_band_dispersion} are plotted in Fig. \ref{figure5} (c). There are two Van Hove singularities in $E_{h}\left(\bm{k}\right)$ and one in $E_{d}\left(\bm{k}\right)$, so the electronic DOS shows two peaks in $\omega<0$ and one in $\omega>0$. Near $\Gamma$, the band $E_{h}\left(\bm{k}\right)$ shows a flipped Mexican hat like shape, while the band $E_{d}\left(\bm{k}\right)$ is quadratic with a negative effective mass. The two shapes of bound state bands shown in Fig. \ref{figure5} (c) are quite representative for the spinon chargon bound states. In fact, the specific bound state band shape is affected by many factors such as the gauge binding strength, the dispersions of the spinon and chargon bands, the spinon chemical potential, and temperture, etc. In Sec. \ref{VII} below, the bound state band shape is further analyzed in the continuum model near $\Gamma$.

\section{ Weak spinon chargon binding  in a Magnetic Field : Band Edge Resonance Peaks}\label{VI}
Given a weak gauge binding interaction that arises from the $U\left(1\right)$ gauge field fluctuations, it has been found in Sec. \ref{V} that the QSL electronic DOS at $B=0$T has a pair of resonance peaks develop at the Hubbard band edges. In the presence of a finite orbital magnetic field, when the magnetic flux ratio $\phi=\frac{eB}{h}\frac{\sqrt{3}a^2}{2}$ is rational, we can proceed to deal with the gauge binding effect in the lattice model. In the lattice model of a QSL that couples with an orbital magnetic field, the gauge binding term in Eq. \ref{H_int} is changed to be
\begin{align}
\hat{H}_{\textrm{int}}=\frac{U_{\textrm{b}}}{N'}\sum_{\sigma, \bm{k},\bm{q},\bm{q}',i}\hat{f}^\dagger_{\sigma,\bm{k}-\bm{q},i}\hat{f}_{\sigma,\bm{k}-\bm{q}',i}\left(\hat{a}_{-\bm{q},i}\hat{a}^\dagger_{-\bm{q}',i}-\hat{b}^\dagger_{\bm{q},i}\hat{b}_{\bm{q}',i}\right),
\end{align}
where the subscript $i$ labels the $i$th elements in the column vectors $\hat{a}_{-\bm{k}}$, $\hat{b}_{\bm{k}}$ and $\hat{f}_{\sigma,\bm{k}}$. Here $N'=N/I$ is the number of magnetic unit cells and $I$ is the number of lattice sites in one magnetic unit cell. Now the QSL mean field Hamiltonian becomes $\hat{H}=\hat{H}_0+\hat{H}_{\textrm{int}}$. For the QSL with multiple spinon and chargon bands, the gauge binding interaction $U_{\textrm{b}}$ changes the QSL electronic DOS in Eq. \ref{rho_sigma_B} to be~\cite{Supplemental}
\begin{widetext}
\begin{align}\nonumber\label{hat_rho_B}
\tilde{\rho}_\sigma\left(\omega,B,0\right)=&-\frac{1}{N\pi}\sum_{\bm{k}}\textrm{tr}\textrm{Im}\left\{\left[\frac{1}{N}\sum_{\bm{q}}\hat{G}^{\textrm{R}}_{h,\sigma}\left(\omega,\bm{k}-\bm{q},\bm{q}\right)\right]\left[1-\frac{U_{\textrm{b}}}{N'}\sum_{\bm{q}}\hat{G}^{\textrm{R}}_{h,\sigma}\left(\omega,\bm{k}-\bm{q},\bm{q}\right)\right]^{-1}\right\}\\
&-\frac{1}{N\pi}\sum_{\bm{k}}\textrm{tr}\textrm{Im}\left\{\left[\frac{1}{N}\sum_{\bm{q}}\hat{G}^{\textrm{R}}_{d,\sigma}\left(\omega,\bm{k}-\bm{q},\bm{q}\right)\right]\left[1+\frac{U_{\textrm{b}}}{N'}\sum_{\bm{q}}\hat{G}^{\textrm{R}}_{d,\sigma}\left(\omega,\bm{k}-\bm{q},\bm{q}\right)\right]^{-1}\right\},
\end{align}
\end{widetext}
where the elements of the retarded Green's function matrix $\hat{G}^{\textrm{R}}_{h,\sigma}\left(\omega,\bm{k},\bm{k}'\right)$ and $\hat{G}^{\textrm{R}}_{d,\sigma}\left(\omega,\bm{k},\bm{k}'\right)$ are given by the first and second term in Eq. \ref{G_kl} respectively.

In the weak gauge binding $U_{\textrm{b}}\rho_f\left(0\right)=1$, the QSL electronic DOS at $\phi=0$ and $\phi=1/9$ is plotted in Fig. \ref{figure6} (a) and (b) respectively. The electronic DOS spectra of $\phi=1/9$ has a similar shape as that of $\phi=0$, but more band edge resonance peaks apppear in the spectra of $\phi=1/9$ in Fig. \ref{figure6} (b). The band edge resonance peaks near the LHB top and the UHB bottom are zoomed in Fig. \ref{figure6} (c) and (d) respectively, with the electronic DOS spectra $\rho_\sigma\left(\omega,B,0\right)$ at $\phi=1/9$ in Fig. \ref{figure2} (c) plotted for comparison. From Fig. \ref{figure6} (c) and (d), one can find that the weak gauge binding $U_{\textrm{b}}\rho_f\left(0\right)=1$ evolves each Hubbard band edge step into a resonance peak. Physically, since the gauge binding $U_{\textrm{b}}$ tends to bind  spinon LL states with chargons, each resonance peak near the LHB band edge in Fig. \ref{figure6} (c) corresponds to the quasi-binding of a spinon hole LL state and a holon. Similarly, each resonance peak near the UHB band edge in Fig. \ref{figure6} (d) indicates a quasi-bound state of a spinon LL state and a doublon. As a result, in an orbital magnetic field, both the weak gauge binding induced multiple band edge resonance peaks and the emerging band edge steps at zero gauge binding have the intrinsic connection to the spinon LLs, as is labeled in Fig. \ref{figure6} (c) and (d).

In the weak gauge binding $U_b\rho_f\left(\omega\right)=1$, each resulting band edge resonance peak appears almost at the same energy as that of the band edge step at $U_b\rho_f\left(\omega\right)=0$. When the applied orbital magnetic field $B$ changes, the energy spacing between the adjacent band edge resonance peak in Fig. \ref{figure6} (b) and (c) changes linearly with $B$. In the weak gauge binding regime, since the loosely quasi-bound spinon chargon pairs have negligible energy change, the energy dependence of the band edge resonance peaks on $B$ follows that of the spinon LLs. In the strong gauge binding regime, all the band edge resonance peaks move inside the Mott gap and evolve into in-gap peaks as shown in Fig. \ref{figure7} (a) and (b). Those in-gap peaks originate from the in-gap bound state LLs. Since real in-gap bound states are formed, the energy saved in the binding, namely the binding energy, plays an important role in determining the bound state energy. In contrast to the linear $B$ energy dependence of the band edge resonance peaks in the weak gauge binding regime, the interplay between the binding energy and the orbital magnetic field $B$ in the strong gauge binding regime complicates the $B$ dependence of the in-gap bound state LL spectrum.

\section{Strong spinon chargon binding in a magnetic field: Bound state band dispersions and the Landau levels}\label{VII}
In order to get the in-gap bound state LL spectrum, the bound state band dispersions in $B=0$T need to be analyzed first. In the strong gauge binding regime, a spinon hole gets bound with a holon to form a hole bound state above the LHB, while an electronic bound state below the UHB arises from the binding of a spinon and a doublon. The binding process of a spinon hole and a holon is similar to that of a spinon-doublon bound state, so in the below we mainly focus on the electronic bound state formed by a spinon and a doublon. The binding of a spinon hole and a holon can be found in the Supplemental Materials~\cite{Supplemental}. 

\subsection{The Mexican hat like band dispersion}

To proceed, we consider the continuum model near $\Gamma$, where the spinon band and the chargon band are approximated by the quadratic dispersions: $\xi_{\bm{k}}=\frac{\hbar^2\bm{k}^2}{2m_f}-\mu_f$ and $\epsilon_{\bm{k}}=\frac{\hbar^2\bm{k}^2}{2m_X}+\Delta$ respectively. In the continuum description of the gapless $U\left(1\right)$ QSL, the threshold energy to excite an electron with a quasi-momenta $\bm{k}$ is $E_{\textrm{th}}\left(\bm{k}\right)=\frac{\hbar^2\left(|\bm{k}|-|\bm{k}_{\textrm{F}}|\right)^2}{2m_X}+\Delta$~\cite{Supplemental}, where $\bm{k}_{\textrm{F}}$ is the spinon Fermi wave vector. In Fig. \ref{figure8} (a), the black dashed line denotes the threshold energy $E_{\textrm{th}}\left(\bm{k}\right)$, and the green shaded region above $E_{\textrm{th}}\left(\bm{k}\right)$ corresponds to the continuous electronic excitation spectrum. In the presence of a sufficiently large gauge binding, an electronic state gets dragged down and becomes a bound state with the band dispersion just below $E_{\textrm{th}}\left(\bm{k}\right)$. In the continuum description, the electronic bound state binding equation in Eq. \ref{electron_band_dispersion} is changed to be
\begin{align}\label{Electron_bound_quadratic}
\frac{1}{U_{\textrm{b}}}+\frac{A_{\textrm{c}}}{\left(2\pi\right)^2}\int_0^{|\bm{q}|=k_{\textrm{c}}}\frac{n_{\textrm{F}}\left(-\xi_{\bm{k}-\bm{q}}\right)+n_{\textrm{B}}\left(\epsilon_{\bm{q}}\right)}{E_{d}\left(\bm{k}\right)+i0^+-\xi_{\bm{k}-\bm{q}}-\epsilon_{\bm{q}}}d^2\bm{q}=&0,
\end{align}
with $k_{\textrm{c}}$ being the momenta cut-off. Given the band parameters $m_f/m_X=3$, $\Delta/\mu_f=1/2$, $U_{\textrm{b}}\rho_f\left(0\right)=2.9$ and the cut-off $k_{\textrm{c}}=2.6|\bm{k}_{\textrm{F}}|$, one gets the electronic bound state band with a Mexican hat like dispersion shown in Fig. \ref{figure8} (a). The evolution of the electronic bound state band as the gauge binding interaction increases can be found in Fig. S2 in the Supplemental Materials~\cite{Supplemental}. The Mexican hat like shape of the electronic bound state band is inherited from the threshold energy $E_{\textrm{th}}\left(\bm{k}\right)$, and it is one reprsentative bound state band dispersion.

In order to get the LL spectrum of the electronic bound state with the band dispersion $E_{d}\left(\bm{k}\right)$, it is instructive to first consider Landau quantization of a free charged particle with a general energy dispersion $E\left(\bm{k}\right)$ that depends on $\bm{k}^2$ only. It is straightforward to show that a simple substitution $\bm{k}^2=2l_B^{-2}\left(n+\frac{1}{2}\right)$~\cite{Supplemental} gives the $n$th LL $\tilde{E}_n\left(B\right)=E\left[\sqrt{2l_B^{-2}\left(n+\frac{1}{2}\right)}\right]$ of the free charged particle. Here $l_B=\sqrt{\frac{\hbar}{eB}}$ is the magnetic length. For the Mexican hat like band dispersion shown in Fig. \ref{figure8} (a), the corresponding spectrum $\tilde{E}_n\left(B\right)$ is plotted in Fig. \ref{figure8} (b). Importantly, for a Mexican hat like band $E\left(\bm{k}\right)=E_{d}\left(\bm{k}\right)$ that takes the band minimum at $|\bm{k}|=k_m$, the resulting spectrum $\tilde{E}_n\left(B\right)$ takes the same minimum value at $1/B=\frac{2\pi e}{\hbar \pi k^2_m}\left(n+\frac{1}{2}\right)$ with $n=0, 1, 2, \dots$. Therefore, the edge of the spectrum $\tilde{E}_n\left(B\right)$ oscillates in $1/B$ as can be seen in Fig. \ref{figure8} (b). The oscillation frequency is $\tilde{F}=\frac{\pi\hbar k^2_m}{2\pi e}$, where the wave vector at the band minimum plays the role of the Fermi wave vector in a metal. The constant band minimum $E_{\textrm{min}}$ defines the envelop of the band edge oscillation.

Interestingly, the LL spectrum $\tilde{E}_n\left(B\right)$ of a free charged particle indicates that for insulating systems with a Mexican hat like conduction or valence band, the resulting LL spectrum would induce  insulating gap modulations so that the thermally activated resistivity oscillates with $B$. The oscillation of the resistivity ratio to the back ground resistivity is detectable even at low temperatures~\cite{Patrick3, Wenyu1}. An example of this kind of Mexican hat like band may be found in the biased Bernal bilayer graphene~\cite{Neto}. It will be interesting to search for this effect experimentally. This  provides a new mechanism for observing  QOs in band insulators because unlike previous proposals, the gap is not generated by hybridization~\cite{Cooper1, FaWang, Patrick3, Wenyu1}.

\begin{figure}
\centering
\includegraphics[width=3.5in]{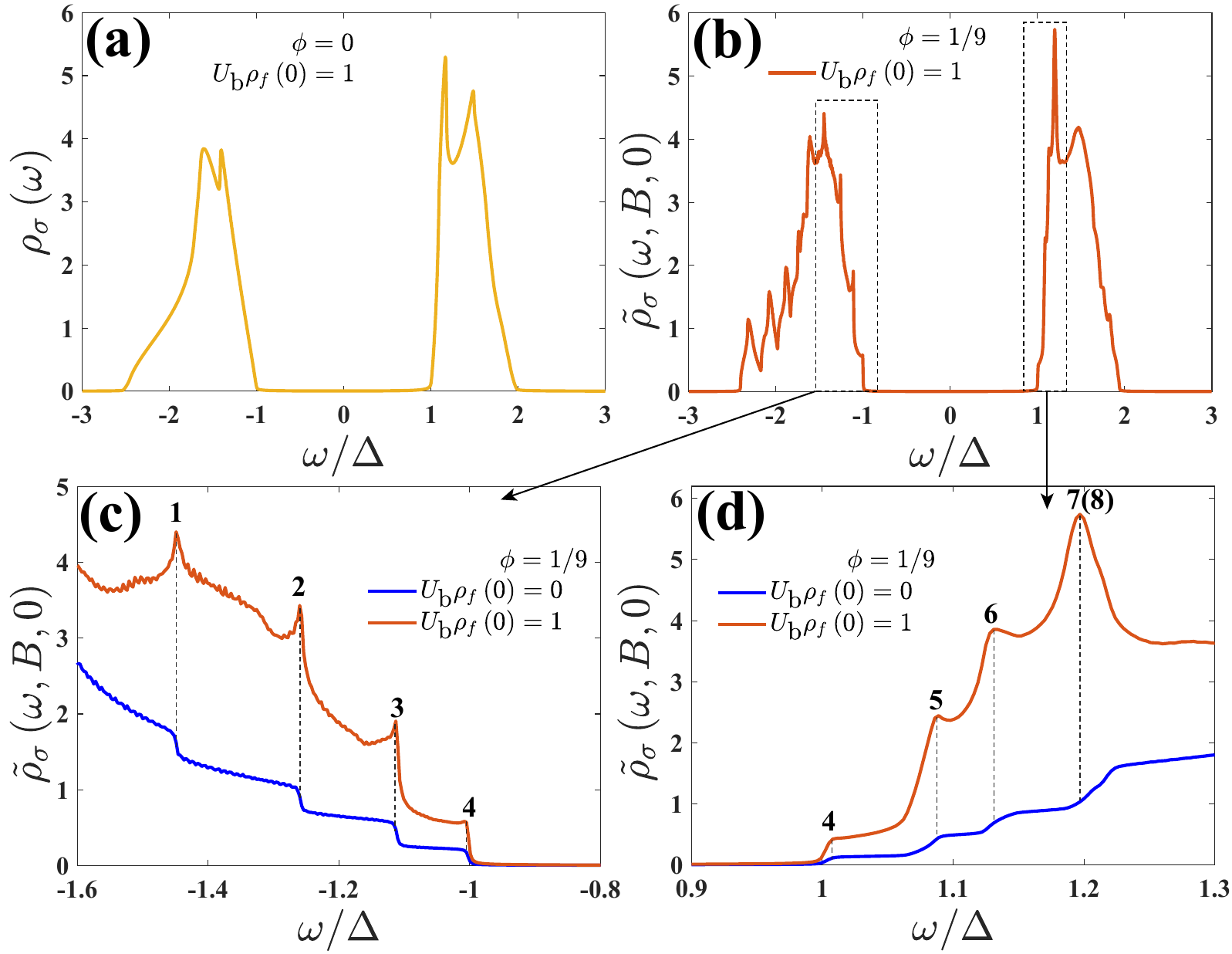}
\caption{(a) The QSL electronic DOS at $U_{\textrm{b}}\rho_f\left(0\right)=1$ in zero magnetic flux. (b) The QSL electronic DOS at $U_{\textrm{b}}\rho_f\left(0\right)=1$ with the magnetic flux ratio being $\phi=1/9$. At both the LHB top and the UHB bottom, the electric DOS spectra in (b) shows a few resonance peaks. (c) and (d) zoom in the DOS spectra in the black dashed rectangles in (b). The electronic DOS $\rho_\sigma\left(\omega,B,0\right)$ at $\phi=1/9$ with zero gauge binding is plotted in blue for comparison. The resonance peaks at the Hubbard band edges are found to appear at the same energies as those of the edge steps in the electronic DOS $\rho_\sigma\left(\omega,B,0\right)$. It indicates that in an orbital magnetic field,  each edge step at zero $U_{\textrm{b}}$ in the QSL electronic DOS evolves into a resonance peak in the presence of  weak gauge interaction $U_{\textrm{b}}$}\label{figure6}
\end{figure}

\subsection{Landau quantization of the in-gap bound state}
The spectrum $\tilde{E}_n\left(B\right)$ obtained through the simple subsitution cannot give the correct LLs of the electronic bound state that is composed of a spinon and a doublon in the QSL. For the electronic bound state formed in an orbital magnetic field, the magnetic field is divided into two parts acting on the spinons and the doublons respectively. The effect of the orbital magnetic field is two fold. First, the constituent particles, namely the spinons and the doublons, all have the LLs formed in the spectrum which increase their energy. Thus the energy of the composite particle is expected to increase. Second, the wave functions of the spinons and the doublons become more localized, which tend to increase the binding energy, leading to a decrease of the energy of the composite particle. Therefore these two effects tend to compete. As a result, the bound state Landau quantization is a much more complicated problem that we will deal with in the below.

For the gapless $U\left(1\right)$ QSL in an orbital magnetic field $B$, the gauge binding occurs between the spinon LL states and the chargon LL states. In the LL basis, the gauge binding term takes the form
\begin{widetext}
\begin{align}
H_{\textrm{int}}\left(b,B-b\right)=\sum_{\sigma,n_1,n_2,n_1',n_2',m_1,m_2,m_1',m_2'}\tilde{U}_{\textrm{b},n_1,n_2,n_1',n_2',m_1,m_2,m_1',m_2'}f^\dagger_{n_1,m_1,\sigma}f_{n_2,m_2,\sigma}\left(a_{n_1',m_1'}a^\dagger_{n_2',m_2'}-b^\dagger_{n_1',m_1'}b_{n_2',m_2'}\right),
\end{align}
\end{widetext}
with the interaction matrix elements given in the Supplemental Materials~\cite{Supplemental}. In principle, the LL spectrum of the bound states can be obtained through diagonalizing the mean field Hamiltonian $H\left(b,B-b\right)=H_0\left(b,B-b\right)+H_{\textrm{int}}\left(b,B-b\right)$. However, it is extremely challenging to fully diagonalize $H\left(b,B-b\right)$ given an arbitrary magnetic field partition. In the weak Mott regime of the QSL, since the EGMF $b$ dominates over the remaining $B-b$, we proceed with the limiting case of $b\rightarrow B$. Calculations about the bound state LL spectrum in the opposite limiting case of $b\rightarrow0$T can be found in the Supplemental Materials~\cite{Supplemental}.

The limiting case offers great simplification because the kernal of the integral equation factorizes, leading to the much simpler binding equation below for the electronic bound state energies $E_{d,n}\left(B\right)$~\cite{Supplemental}:
\begin{widetext}
\begin{align}\label{LL_electronic_bound}
\frac{1}{U_{\textrm{b}}}+\frac{A_{\textrm{c}}}{\left(2\pi\right)^2}\sum_{n'}\int_0^{|\bm{q}|=k_{\textrm{c}}}\frac{n_{\textrm{F}}\left(-\xi_{n'}\right)+n_{\textrm{B}}\left(\epsilon_{\bm{q}}\right)}{E_{d,n}\left(B\right)+i0^+-\xi_{n'}-\epsilon_{\bm{q}}}D_{n,n'}\left(q\right)D_{n',n}\left(-q\right)\exp\left(-l_B^2\bm{q}^2/2\right)d^2\bm{q}=&0.
\end{align}
\end{widetext}
Here the function $D_{n,n'}\left(q\right)$ respects the relation $D_{n,n'}\left(q\right)=D_{n',n}\left(-q^\ast\right)$ and its expression for $n\geqslant n'$ is $D_{n,n'}\left(q\right)=\sqrt{n'!/n!}\left(-l_Bq/2\right)^{n-n'}L_{n'}^{n-n'}\left(l_B^2qq^\ast/2\right)$ with $q=q_x+iq_y$. The function $L_n^m\left(x\right)$ is the associated Laguerre function. The binding equation for the hole bound state can be found in the Supplemental Materials~\cite{Supplemental}.

By numerically solving Eq. \ref{LL_electronic_bound}, the resulting LL spectrum $E_{d,n}\left(B\right)$ that correspond to the bound state dispersion in Fig. \ref{figure8} (a) is plotted in Fig. \ref{figure8} (c) and (d). It can be seen in Fig. \ref{figure8} (c) that the band edge of $E_{d,n}\left(B\right)$ oscillates in $1/B$ as well. In the small magnetic field region, the electronic bound state LL spectrum $E_{d,n}\left(B\right)$ approaches to the spectrum $\tilde{E}_n\left(B\right)$ as $B\rightarrow0$T. As the magnetic field increases, the band edge oscillation of $E_{d,n}\left(B\right)$ intensifies and the frequency approaches to $F=\frac{\pi\hbar\bm{k}_{\textrm{F}}^2}{2\pi e}$, which equals to the oscillation frequency of the spinon chemical potential $\mu_f\left(b=B\right)$~\cite{Supplemental}. It is consistent with the fact that the memory of the spinon chemical potential oscillation is retained after forming the bound state, so the band edge oscillation of $E_{d,n}\left(B\right)$ originates intrinsically from Landau quantization of the spinon Fermi surface.

\begin{figure}
\centering
\includegraphics[width=3.5in]{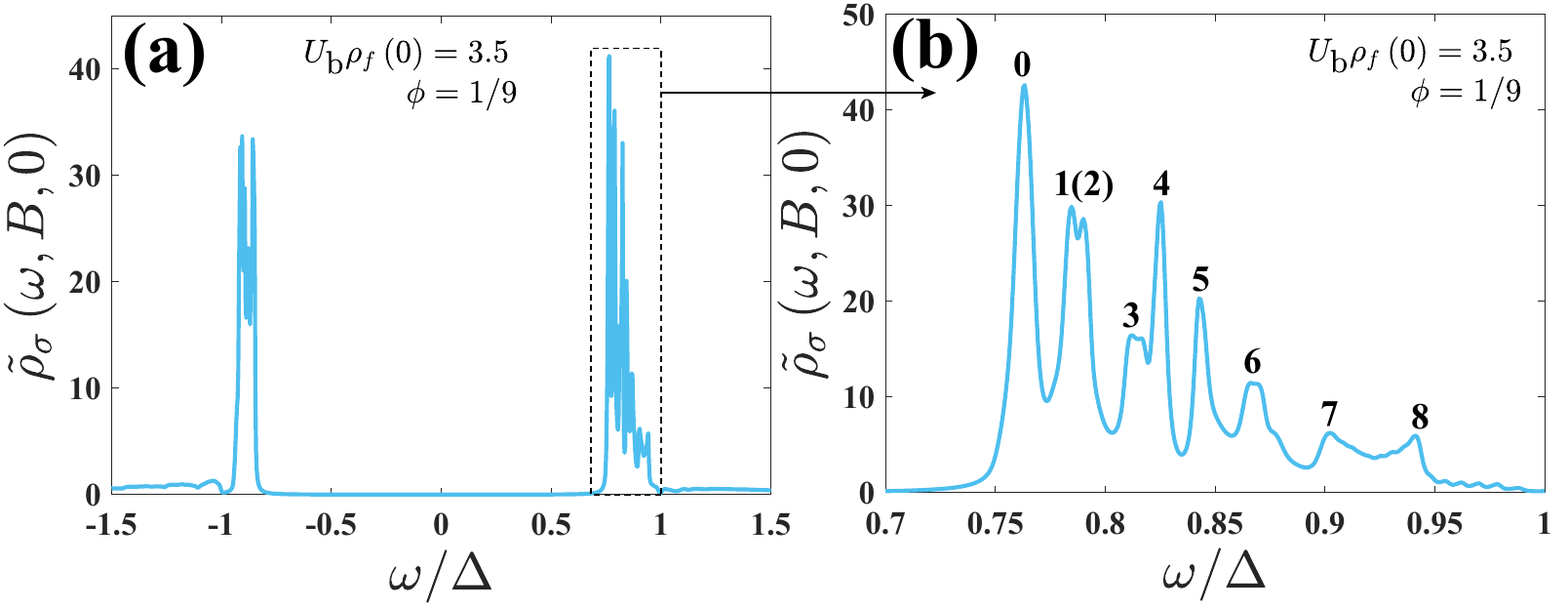}
\caption{(a) The QSL electronic DOS at $U_{\textrm{b}}\rho_f\left(0\right)=3.5$ with the magnetic flux being $\phi=1/9$. As the gauge binding is sufficiently large to induce real in-gap bound states, the electronic DOS is composed of in-gap peaks that originate from the bound state LLs. (b) Zoom-in of  black dashed rectangle in (a). The discrete peaks indicate Landau quantization of the electronic bound state below the UHB. Given the magnetic flux $\phi=1/9$, there are 9 bound state LLs as labeled.}\label{figure7}
\end{figure}

Importantly, the envelop energy of $E_{d,n}\left(B\right)$ is found to decrease quadratically with $B$ as seen in Fig. \ref{figure8} (d), indicating an increase in the binding energy. We believe that the reason behind is the orbital magnetic field effect on the binding energy. In the limit of $b\rightarrow B$ we considered, the orbital magnetic field $B$ only acts on the spinons. The magnetic field on the spinons induces the spinon chemical potential to oscillate and makes the spinon wave function more and more localized as $B$ increases. Intuitively, binding a doublon with a more spatially localized spinon is energetically more favorable, so the binding energy increases as $B$ increases. The increase of the binding energy brings down the electronic bound state energy. In a finite temperature, the binding energy at finite $B$ must be smoothly connected to that at $B=0$T, so a quadratic $B$ term, which is the lowest allowed order, must be involved in the binding energy. For the electronic bound state with a Mexican hat like band dispersion, when the effect of the orbital magnetic field on the binding energy is not considered, the envelop energy of the spectrum $\tilde{E}_n\left(B\right)$ is the constant $E_{\textrm{min}}$. After the quadratic $B$ dependent binding energy is taken into account, it introduces a term that decreases quadratically in $B$ to the electronic bound state energy, so the resulting envelop energy of the electronic bound state LL spectrum solved from Eq. \ref{LL_electronic_bound} exhibits the quadratic decrease in $B$. 

\begin{figure}
\centering
\includegraphics[width=3.5in]{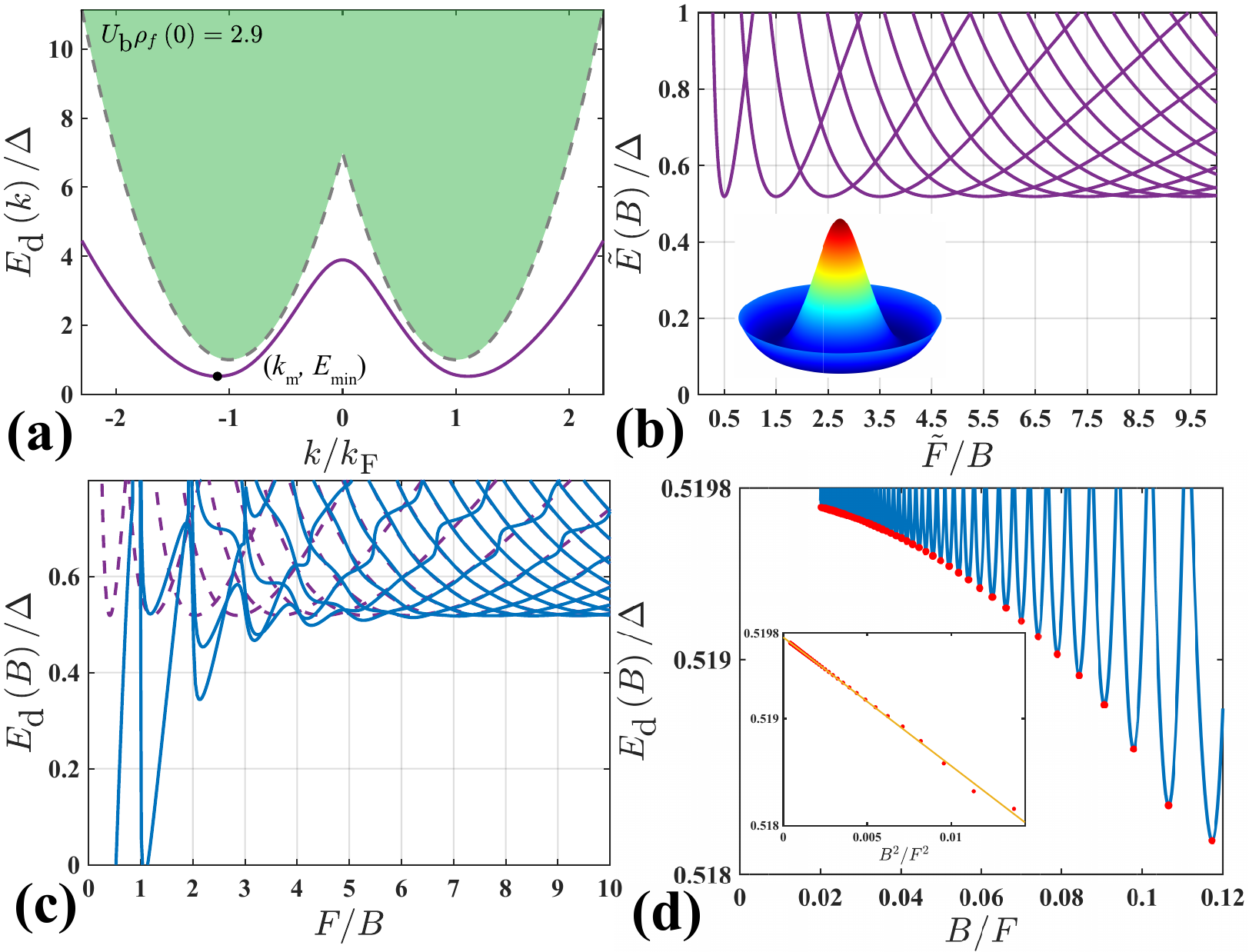}
\caption{(a) The electronic bound state band dispersion numerically solved from Eq. \ref{LL_electronic_bound}. A Mexican hat like bound state band dispersion emerges from the threshold energy $E_{\textrm{th}}\left(\bm{k}\right)$ that is denoted by the black dashed line. The black dot on the bound state band denotes the bound state band minimum $E_{\textrm{min}}$ at $|\bm{k}|=k_{\textrm{m}}$. The green shaded region corresponds to the continuous spectrum of the electronic excitations in the UHB. (b) The LL spectrum of a charged particle that has the same band dispersion $E\left(\bm{k}\right)=E_{d}\left(\bm{k}\right)$ as shown in (a). The inset in the lower left corner shows the two dimensional plot of the bound state band in (a), which has the shape of a Mexican hat. Note that the gap is periodically modulated in $1/B$. (c) The LL spectrum of the electronic bound state with the band dispersion shown in (a). The LL spectrum is numerically solved from Eq. \ref{LL_electronic_bound}. The LL spectrum in (b) is plotted as purple dashed lines for comparison. At $B\rightarrow0$, the electronic LLs $E_{d,n}\left(B\right)$  approaches to $\tilde{E}_n\left(B\right)$. (d) The electronic bound state LL spectrum in (c) plotted as a function of $B/F$. The minimum of each LL is labeled by a red dot, and the red dots are plotted in $B^2/F^2$ in the inset. In the inset, the minimum values of LLs exhibit the linearly decrease in $B^2/F^2$, which indicates that the envelop energy of the electronic bound state LL spectrum decrease quadratically in $B$. Here the temperature has been fixed to be $k_{\textrm{b}}T/\mu_f=0.05$ in all the calculations.}\label{figure8}
\end{figure}

\begin{figure}
\centering
\includegraphics[width=3.5in]{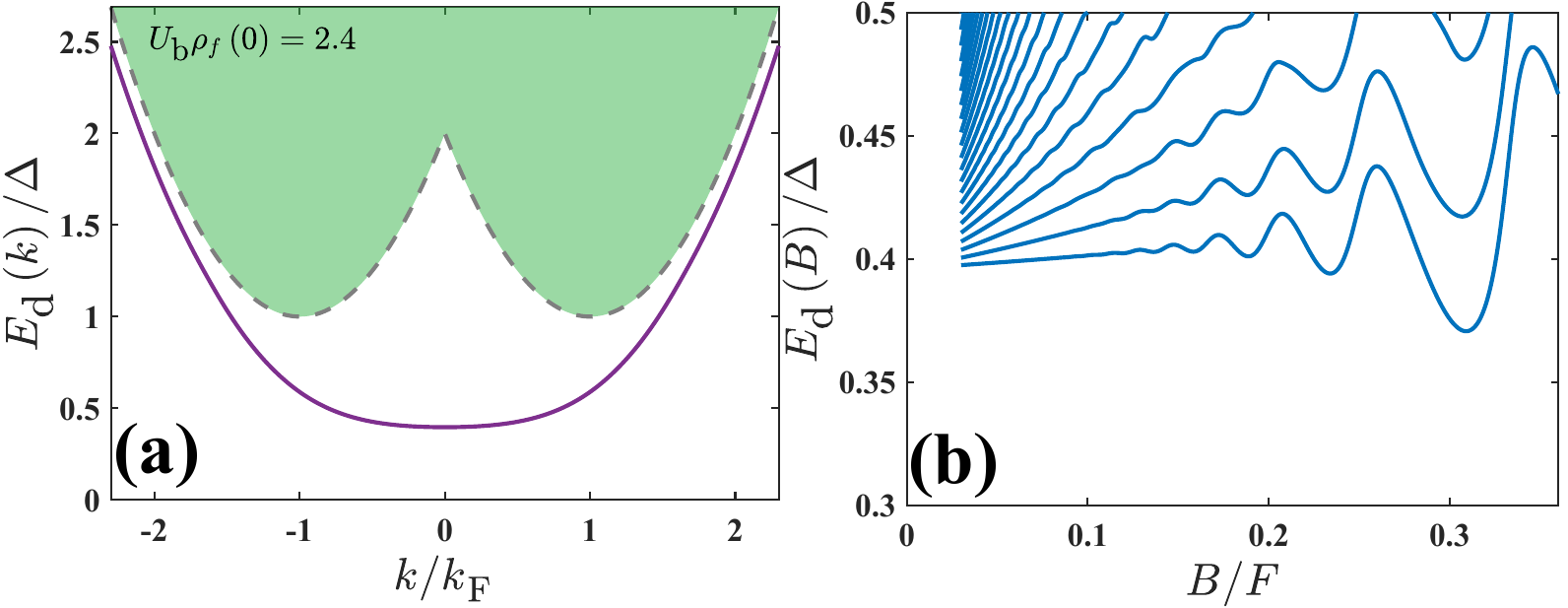}
\caption{(a) The electronic bound state band with a quadratic dispersion. The band parameters used in solving Eq. \ref{Electron_bound_quadratic} are $m_f/m_X=1/2$, $\Delta/\mu_f=1/2$, $U_{\textrm{b}}\rho_f\left(0\right)=2.4$, $k_{\textrm{b}}T/\mu_f=0.05$. The momenta cut-off is set to be $k_{\textrm{c}}=2.6|\bm{k}_{\textrm{F}}|$. (b) The LL spectrum corresponding to the electronic bound state band shown in (a). In the region $B\rightarrow0$T, the electronic bound state energy levels exhibit the linear increase in $B$. As the magnetic field $B$ increases, the bound state energy levels start to show the oscillation that originates from the spinon chemical potential oscillation. When the magnetic field $B$ is sufficiently large, the envelop energy of the oscillation turns to decrease in $B$ because the orbital magnetic field induced energy saving plays the dominant role in the binding.}\label{figure9}
\end{figure}

It is important to note that the quadratic decrease of the electronic bound state envelop energy in $B$ is unique to the bound state that has a Mexican hat like band dispersion at $B=0$T. In fact, the quadratic decrease of the  envelop energy of the bound statewith $B$ stems from two indispensable factors: one is the the quadratic increase of the binding energy in $B$; the other is the constant envelop energy $E_{\textrm{min}}$ of the spectrum $\tilde{E}_n\left(B\right)$. To demonstrate this point, we consider the case where the electronic bound state has a quadratic band dispersion shown in Fig. \ref{figure9} (a). This occurs by tuning $U_{\textrm{b}}$. As shown in Fig. \ref{figure9} (b), the bound state LL spectrum $E_{d, n}\left(B\right)$ approaches to $\tilde{E}_n\left(B\right)$ as $B\rightarrow0$T. For the bound state in Fig. \ref{figure9} (a), since all the energy levels in $\tilde{E}_n\left(B\right)$ increase linearly in $B$, in the $B\rightarrow0$T regime the linear $B$ increase dominates over the quadratic B dependent term in the binding energy. As a result, the LL spectrum $E_{d, n}\left(B\right)$ of an electronic bound state with a quadratic band dispersion retains the linear $B$ increase in the $B\rightarrow0$T regime, as can be seen in Fig. \ref{figure9} (b). As $B$ further increases, the band edge of the LL spectrum $E_{d, n}\left(B\right)$ starts to show the oscillation that comes from the spinon chemical potential oscillation. Interestingly, in Fig. \ref{figure9} (b) the envelop energy of the oscillation is seen to change from increasing in $B$ to decreasing in $B$ as $B$ continues increasing, indicating that the orbital magnetic field promoted energy saving in the binding finally becomes dominant. It again matches the intuition that the binding saves more energy as $B$ increases regardless of the resulting bound state band dispersion. The above analysis on the electronic bound state LL spectrums applies to the hole bound state LL spectrum as well~\cite{Supplemental}.

\section{Connection to experiments.}\label{VIII}
\begin{figure}
\centering
\includegraphics[width=3.5in]{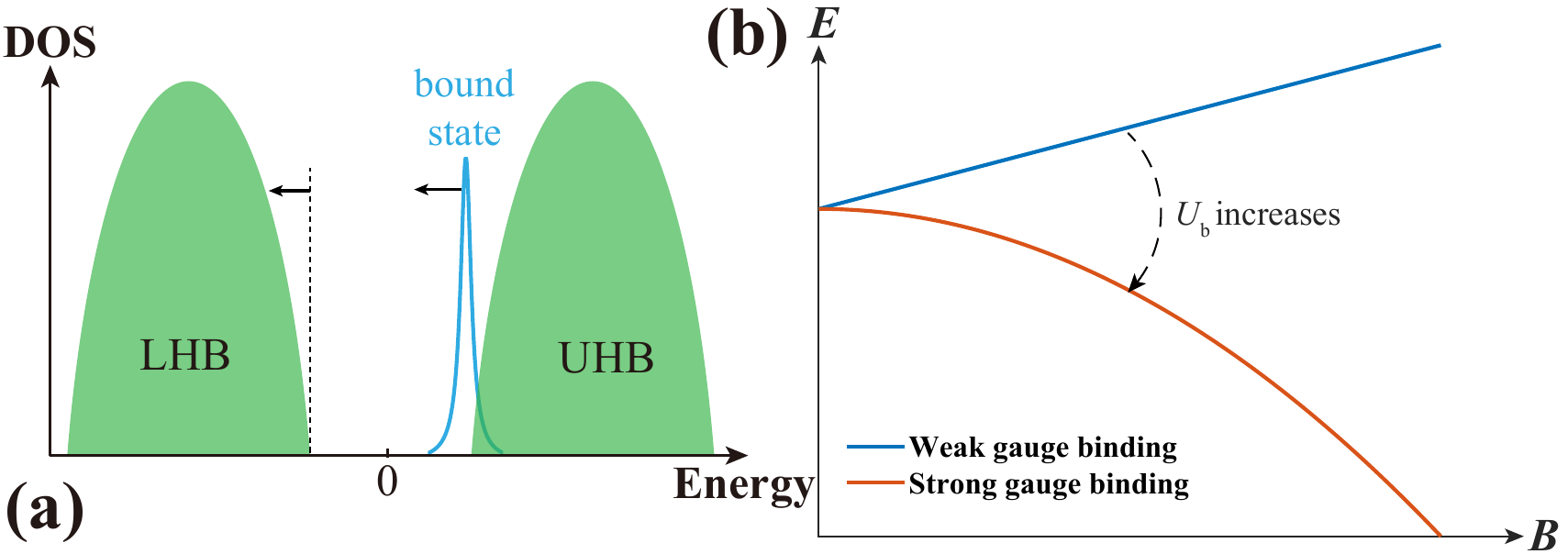}
\caption{(a) The schematic showing of the layered 1T-TaS$_2$ electronic DOS spectra measured in the experiment~\cite{Butler3,Butler1}. A resonance peak emerges at the UHB edge but no such resonance peak is observed at the LHB edge. By applying a magnetic field, the energy of the resonance peak is observed to decrease and so does that of the LHB edge, even though the effect on the LHB edge is much smaller. The direction of the shift of the UHB edge resonance peak and the LHB edge with the applied magnetic field is indicated by the black arrows. Importantly, the energy of the UHB edge peak exhibits a quadratic decrease with $B$ in the experiment. (b) Schematic drawing of the theoretical prediction for the energy dependence of the UHB edge peak on the applied magnetic field. In the weak gauge binding regime, the energy of the UHB edge peak increases linearly as that of the spinon LL. In the strong gauge binding regime, the energy of the UHB edge peak is determined by the envelop energy of the bound state LL spectrum, so it decreases quadratically with $B$. As the gauge binding interaction increases, the UHB edge peak energy is expected to change from the linear increase with $B$ to the quadratic decrease with $B$.}\label{figure10}
\end{figure}

We have studied the effects of orbital magnetic field and gauge field fluctuations on the electronic DOS of the $U\left(1\right)$ QSL with SFS. For the electronic DOS spectra, one widely used technique to detect the local electronic DOS in experiment is the STM. Given a simple metal as the STM tip, the differential conductance in the setup is~\cite{Flensberg}
\begin{align}
\frac{dI}{dV}\propto\int_{-\infty}^\infty-\frac{\partial n_{\textrm{F}}\left(\omega+eV\right)}{\partial\omega}\sum_\sigma\tilde{\rho}_{\sigma}\left(\omega,b,B-b\right)d\omega,
\end{align}
where $\tilde{\rho}_{\sigma}\left(\omega,b,B-b\right)$ is the QSL electronic DOS that covers both the effects of orbital magnetic field and gauge binding. As analyzed in Sec. \ref{II} and \ref{VI}, the DOS spectra of a  $U\left(1\right)$ QSL with SFS at $B=0$T and $U_{\textrm{b}}\rho_f\left(0\right)=0$ is composed of two dome like regions separated by a Mott gap as schematically illustrated in Fig. \ref{figure10} (a).

As mentioned in Sec. \ref{I}, recent STM measurements on the bulk 1T-TaS$_2$, monolayer 1T-TaSe$_2$~\cite{YiChen1, WeiRuan, YiChen2} and 1T/1H-TaS$_2$ heterostructure~\cite {Vano} all show clear Hubbard band edges in the electronic DOS spectra. Specifically, on the surface of the layered 1T-TaS$_2$, an extra resonance peak with sidebands was found near the UHB edge~\cite{Butler3}. In a subsequent measurement in an external magnetic field~\cite{Butler1}, the UHB edge resonance peak was found to move towards the Mott gap center as the magnetic field increases, and its energy exhibits a quadratic decrease with $B$. For the LHB edge, it was observed to move away from the Mott gap center, albeit at a much smaller rate. The evolution of the DOS spectra in an external magnetic field observed in the experiment~\cite{Butler1} is schematically indicated in Fig. \ref{figure10} (a).

The experimental observation of the LHB edge and the UHB edge resonance peak evolving in the same direction in an external magnetic field is a surprising result. It indicates that the energy cost to excite a hole increases with the external magnetic field but that to excite an electron decreases. Normally we expect the electron energy to increase due to orbital effects of a magnetic field. Hence the observation for the UHB is highly surprising.  We would like to interprete the  DOS spectra observed in the experiment assuming the material is a gapless $U\left(1\right)$ QSL. The UHB edge resonance peak is then interpreted as a quasi-bound state of a spinon and a doublon. In the experiment the peak is accompanied by a series of sidebands which has been interpreted in analogy with phonon sidebands but using the amplitude mode of the charge density wave instead of phonons~\cite{Butler3}. Note that in this scenario the electronic mode is usually required to be  almost localized  with a linewidth less than the phonon energy. This indicates that the electron is not in a propagating band, as the rather broad Hubbard band suggests. Instead, this fits our scenario that the electron is in a resonant or near bound state. Experimentally the spectral weight of the resonance and the sideband in the UHB is rather small. The weight may be comparable to the case $U_{\textrm{b}}\rho_f\left(0\right)=1$ as shown in Fig. \ref{figure5} (a) and Fig. \ref{figure6} (a). This puts us in the intermediate binding regime where the state is not fully bound, but appears as a near edge resonance. Unfortunately we do not have a quantitative theory for the orbital magnetic field induced evolution of the UHB edge resonance in this intermediate regime. Recall that in the strong  binding regime, the LL spectrum of the real bound state with a Mexican hat-like band dispersion has the envelop energy decrease quadratically with $B$, so the thermally smoothed peak of the bound state LLs is expected to have its energy decrease quadratically with $B$ as well. For a general quasi-bound state near the UHB edge, the energy of the resonance peak therefore changes from the linear increase with $B$ to the quadratic decrease with $B$ as the gauge binding increases, which is schematically plotted in Fig. \ref{figure10} (b). By interpolating between the strong and weak binding limits, we may argue that the behavior observed in the experiment is closer to the strong binding case, and a $B^2$ decrease of the energy is expected.  For the LHB edge, there is no resonance peak observed in the experiment at zero $B$, so the gauge binding of the spinon holes and the holons is presumably smaller. In a magnetic field we predict an increase in the excitation energy, hence the threshold should move away from the gap center, in agreement with experiment. The $B$ dependence is small and whether it is $B^2$ or not is less certain experimentally. Thus qualitatively a model based on spinon Ferm surface and spinon chargon binding may provide an explanation for the unusual features of the experiment:  the existence of the side-bands and the magnetic field dependence.

In the layered 1T-TaS$_2$ DOS spectra measured in the experiment~\cite{Butler1}, no Zeeman spin splitting was observed. The temperature in the measurement is $T=1.5$K and the largest applied magnetic field is $B=12$T. It is possible that the thermal fluctuations smear the spin split levels, so a spin polarized STM measurement is neededed to resolve the Zeeman effect. The thermal fluctuations may also be the reason why the band edge steps and multiple resonance peaks cannot be identified in the electronic DOS spectra in the magnetic field. In the current work, we mainly focus on the orbital effect of the applied magnetic field. The Zeeman field effect on the electronic DOS spectra a in the QSL with SFS is studied in our companion paper~\cite{Wenyu4}.

In this work, the QSL electronic DOS features of the Hubbard band edge steps and resonance peaks in a magnetic field are both found to have the intrinsic connection to the spinon LLs induced by the EGMF $b$ on the spinons. In reality, the remaining $B-b$ on the chargons induces LLs as well, but the DOS features from the spinon LLs will always be maintained as long as the LL spacing respects $\hbar\omega_f\gg\hbar\omega_X$. To have the DOS features of band edge steps and resonance peaks well identified in the measurements, the temperture should lie in the range $k_{\textrm{b}}T\ll\hbar\omega_f$. The spinon bandwidth is set by the exchange scale which is much smaller than the usual Fermi energy. However, in cases such as 1T-TaS$_2$, 1T-TaSe$_2$ and the organics, the unit cell size is large, making the spinon effective mass $m_f$ a bit larger than the free electron mass $m_e$. Assuming the spinon effective mass to be $m_f\approx2m_e$, one can find that the spinon Landau level spacing in $b\rightarrow B=1$T is 0.06 meV. In an experimental accessible magnetic field $B=10$T, the suitable temperature range to carry out the STM measurement is then estimated to be $k_{\textrm{b}}T/\hbar\omega_f<1/10$, which gives $T<0.7$K. In order to identify the DOS features at the Hubbard band edges in the spectroscopy, a resolution better than 0.6 meV is required.

In the DOS spectra, each step or resonance peak near the Hubbard band edges is intrinsically connected to one spinon LL. However, it is not always true that each spinon LL can have a corresponding step or resonance peak emerging near the Hubbard band edges. The QSL simulated in a triangular lattice in Sec. \ref{IV} already shows in Fig. \ref{figure4} (c) that the $0$th spinon LL, which is the farthest one from the spinon chemical potential, has the corresponding step merged inside the bulk LHB and cannot be identified in the spectrum. By comparing Fig. \ref{figure2} (d) and Fig. \ref{figure3} (e), one can find that the prerequisite to have all the steps or resonance peaks from the spinon LLs identified at the Hubbard band edges is that the chargon band width is larger than the energy between the farthest spinon LL and the spinon chemical potential, namely $\Lambda_X-\Delta<\textrm{Min}\left[\Lambda_f+\mu_f,\Lambda_f-\mu_f\right]$.

\section{Conclusions}\label{IX}
In this paper, we have studied the electronic DOS of a $U\left(1\right)$ QSL in an orbital magnetic field. The QSL electronic DOS spectra is found to have the characteristic Hubbard band edge steps induced by the spinon Landau quantization in the magnetic field. The band edge steps are further found to evolve into resonance peaks when the $U\left(1\right)$ interaction between spinons and chargons due to the weakly fluctuating gauge field is included. In an orbital magnetic field, the QSL electronic DOS features of the band edge steps and the resonance peaks behave completely differently from the discrete Delta like peaks in the DOS of a band insulator, so the finding provides a way to distinguish the QSL with neutral Fermi surfaces from a band insulator. In the case of strong $U\left(1\right)$ gauge field fluctuations, the large gauge binding can induce in-gap bound states and the LL spectrum of the in-gap bound states in a magnetic field are solved. For an in-gap bound state with a Mexican hat like band dispersion, the local DOS exhibits a peak which moves as $B^2$ towards the gap center  with the magnetic field.

Recently, apart from the Mott physics in the 1T-TaS$_2$ and 1T-TaSe$_2$ family, new two-dimensional Mott insulating states have been reported in both the two-dimensional Moir\'e systems~\cite{Pasupathy, Fai} and the monolayer 1T-NbSe$_2$~\cite{Takahashi, Mengke, Yeliang, YingShuang}. For those newly emerging Mott insulators, our study of the QSL electronic DOS spectra in an orbital magnetic field suggests that a tunneling measurement on the electronic DOS would serve as a great diagonosis to whether there exist neutral Fermi surfaces inside the insulating gap.

\section*{ACKNOWLEDGEMENTS}
The authors thank C. J. Butler and T. Hanaguri for sharing their unpublished data. W.-Y. He thanks Yang Qi and Yuanbo Zhang for helpful discussions. W.-Y. He acknowledges the start-up grant of ShanghaiTech University. P. A. Lee acknowledges support by DOE office of Basic Sciences grant number DE-FG02-03ER46076. Part of the computing for this work was performed on the HPC platform of ShanghaiTech University.

\onecolumngrid
\clearpage
\begin{center}

{\bf Supplemental Material for ``Electronic Density of States of a $U\left(1\right)$ Quantum Spin Liquid with Spinon Fermi Surface. I. Orbital  Magnetic Field Effects''}

\end{center}

\maketitle
\setcounter{equation}{0}
\setcounter{figure}{0}
\setcounter{table}{0}
\setcounter{page}{1}
\setcounter{section}{0}
\makeatletter

\renewcommand{\theequation}{S\arabic{equation}}
\renewcommand{\thefigure}{S\arabic{figure}}
\renewcommand{\thetable}{S\arabic{table}}
\renewcommand{\bibnumfmt}[1]{[S#1]}
\renewcommand{\citenumfont}[1]{S#1}

\section{Slave Rotor Mean Field Description for the $U\left(1\right)$ Quantum Spin Liquid}
In the slave rotor mean field description, the partition function for the $U\left(1\right)$ quantum spin liquid (QSL) with spinon Fermi surface (SFS) takes the form:
\begin{align}
Z=&\int\mathcal{D}\left[f^\dagger_{i,\sigma}\left(\tau\right),f_{i,\sigma}\left(\tau\right),X^\ast\left(\tau\right),X\left(\tau\right),a_{ij}\left(\tau\right),a_{0,i}\left(\tau\right)\right]\int d\lambda_i\exp\left(-S\right),
\end{align}
with the action $S_0$ being~\cite{Lee01_Supp,Florens_Supp}
\begin{align}\nonumber\label{S_act0}
S_0=&\int_0^\beta\left[\sum_{i,\sigma}f^\dagger_{\sigma,i}\left(\partial_\tau-h_i+iea_{0,i}-\mu\right)f_{\sigma,i}-\sum_{\left\langle i, j \right\rangle,\sigma}\left(t_{f,ij}e^{i\frac{e}{\hbar}\int_{\bm{r}_i}^{\bm{r}_j}\bm{a}\left(\bm{R},\tau\right)\cdot d\bm{R}}f^\dagger_{\sigma,i}f_{\sigma,j}+t^\ast_{f,ij}e^{-i\frac{e}{\hbar}\int_{\bm{r}_i}^{\bm{r}_j}\bm{a}\left(\bm{R},\tau\right)\cdot d\bm{R}}f^\dagger_{j,\sigma}f_{i,\sigma}\right)\right]d\tau\\\nonumber
&+\int_0^\beta\left\{\frac{1}{U}\sum_i\left[\left(\partial_\tau+h_i-iea_{0,i}\right)X_i\right]\left[\left(\partial_\tau-h_i+iea_{0,i}\right)X_i^\ast\right]-\sum_{\left\langle ij \right\rangle}\left[t_{X,ij}e^{-i\frac{e}{\hbar}\int_{\bm{r}_i}^{\bm{r}_j}\left[\bm{A}\left(\bm{R},\tau\right)-\bm{a}\left(\bm{R},\tau\right)\right]\cdot d\bm{R}}X_iX^\ast_j\right.\right.\\
&\left.\left.+t^\ast_{X,ij}e^{i\frac{e}{\hbar}\int_{\bm{r}_i}^{\bm{r}_j}\left[\bm{A}\left(\bm{R},\tau\right)-\bm{a}\left(\bm{R},\tau\right)\right]\cdot d\bm{R}}X_jX^\ast_i\right]+\sum_i\lambda_i\left(X^\ast_iX_i-1\right)\right\}.
\end{align}
Here $f^{\left(\dagger\right)}_{\sigma, i}$ and $X^{\left(\ast\right)}_i$ are to annihilate (create) a spinon and a chargon at the site $\bm{r}_i$ respectively, $\sigma=\uparrow/\downarrow$ denotes the spin index, and $t_{f, ij}$, $t_{X,ij}$ are the mean field hopping parameters. In the slave rotor formalism, the physical electronic operators have the composite form: $c^\dagger_{\sigma,i}=f^\dagger_{\sigma,i}X^\ast_i$, $c_{\sigma,i}=f_{\sigma,i}X_i$. The variables $h_i$ and $\lambda_i$ in Eq. \ref{S_act0} are the Lagrantian multipliers that give the constraint $\sum_\sigma c^\dagger_{\sigma,i}c_{\sigma,i}=\sum_{\sigma}f^\dagger_{\sigma,i}f_{\sigma,i}=1$ and $X^\ast_iX_i=1$. The vector gauge field $\bm{A}$ is from the external orbital magnetic field. The gauge field $\left(a_0, \bm{a}\right)$ is the emergent $U\left(1\right)$ gauge field in the quantum spin liquid. One can check that the action is invariant under the local $U\left(1\right)$ gauge transformation:
\begin{align}
f_{\sigma,i}\rightarrow e^{i\frac{e}{\hbar}\chi\left(\bm{r}_i,\tau\right)}f_{\sigma,i},\quad X_i\rightarrow e^{-i\frac{e}{\hbar}\chi\left(\bm{r}_i,\tau\right)}X_i,\quad \bm{a}\rightarrow\bm{a}-\nabla\chi\left(\bm{r},\tau\right),\quad a_0\rightarrow a_0-\frac{1}{\hbar}\partial_\tau\chi\left(\bm{r},\tau\right).
\end{align}
The Lagrangian multipliers $h_i$, $\lambda_i$ can be determined through the saddle point approximation in the mean field theory~\cite{Lee01_Supp,Florens_Supp}. For simplicity, we assume that the Lagrangian multipliers are spatially uniform: $h_i=h$, $\lambda_i=\lambda$. Then the QSL action $S_0$ is divided into two terms: $S_0=S_f+S_X$. One is for the spinons:
\begin{align}
S_f=&\int_0^\beta\left[\sum_{\sigma,i}f^\dagger_{\sigma,i}\left(\partial_\tau+iea_{0,i}-\mu_f\right)f_{\sigma,i}-\sum_{\left\langle i, j \right\rangle,\sigma}\left(t_{f,ij}e^{i\frac{e}{\hbar}\int_{\bm{r}_i}^{\bm{r}_j}\bm{a}\left(\bm{R},\tau\right)\cdot d\bm{R}}f^\dagger_{\sigma,i}f_{\sigma,j}+t^\ast_{f,ij}e^{-i\frac{e}{\hbar}\int_{\bm{r}_i}^{\bm{r}_j}\bm{a}\left(\bm{R},\tau\right)\cdot d\bm{R}}f^\dagger_{\sigma,j}f_{\sigma,i}\right)\right]d\tau
\end{align}
with $\mu_f=\mu+h$, and the one for the chargons reads
\begin{align}\nonumber
S_X=&\int_0^\beta\left\{\frac{1}{U}\sum_i\left[\left(\partial_\tau+h-iea_{0,i}\right)X_i\right]\left[\left(\partial_\tau-h+iea_{0,i}\right)X^\ast_i\right]+\lambda\sum_iX^\ast_iX_i\right.\\
&\left.-\sum_{\left\langle ij \right\rangle}\left[t_{X,ij}e^{-i\frac{e}{\hbar}\int_{\bm{r}_i}^{\bm{r}_j}\left[\bm{A}\left(\bm{R},\tau\right)-\bm{a}\left(\bm{R},\tau\right)\right]\cdot d\bm{R}}X_iX^\ast_j+t^\ast_{X,ij}e^{i\frac{e}{\hbar}\int_{\bm{r}_i}^{\bm{r}_j}\left[\bm{A}\left(\bm{R},\tau\right)-\bm{a}\left(\bm{R},\tau\right)\right]\cdot d\bm{R}}X_jX^\ast_i\right]\right\}d\tau.
\end{align}
Importantly, the chargons described by the action $S_X$ are relativistic. The relativistic chargons includes the bolon branch and the doublon branch. We would like to reconstruct the bosonic action $S_X$ and make it expressed directly in terms of the holons and doublons.

Given the relativistic action $S_X$ of the chargons, one can change the imaginary time back to the real time $\tau=it/\hbar$ and write down the Lagrangian as
\begin{align}\nonumber
L_X=&\frac{\hbar^2}{U}\sum_i\left[\left(\partial_t+\frac{i}{\hbar}h-i\frac{e}{\hbar}\tilde{a}_{0,i}\right)X_i\right]\left[\left(\partial_t-\frac{i}{\hbar}h+i\frac{e}{\hbar}\tilde{a}_{0,i}\right)X_i^\ast\right]-\lambda\sum_iX_i^\ast X_i\\
&+\sum_{\left\langle i,j \right\rangle}\left[t_{X,ij}e^{-i\frac{e}{\hbar}\int_{\bm{r}_i}^{\bm{r}_j}\left[\bm{A}\left(\bm{R},t\right)-\bm{a}\left(\bm{R},t\right)\right]\cdot d\bm{R}}X_iX^\ast_j+t^\ast_{X,ij}e^{i\frac{e}{\hbar}\int_{\bm{r}_i}^{\bm{r}_j}\left[\bm{A}\left(\bm{R},t\right)-\bm{a}\left(\bm{R},t\right)\right]\cdot d\bm{R}}X_jX^\ast_i\right],
\end{align}
which is similar to the bosonic Klein Gordon field. Here the Wick rotation has changed the scalar gauge field to be $\tilde{a}_{0,i}=ia_{0,i}$. We can define the canonical momentum from the Lagrangian as
\begin{align}
\Pi_i=\frac{\partial L_X}{\partial\dot{X}_i}=\frac{\hbar^2}{U}\left(\partial_t-\frac{i}{\hbar}h+i\frac{e}{\hbar}\tilde{a}_{0,i}\right)X_i^\ast,\quad\quad \Pi^\dagger_i=\frac{\partial L_X}{\partial\dot{X}_i^\ast}=\frac{\hbar^2}{U}\left(\partial_t+\frac{i}{\hbar}h-i\frac{e}{\hbar}\tilde{a}_{0,i}\right)X_i.
\end{align}
The canonical momentum and the canonical coordinate respect the commutation relation: $\left[X_i, X_j^\ast\right]=\left[\Pi,\Pi^\dagger_j\right]=0$, $\left[X_i,\Pi_j\right]=\left[X^\ast_i,\Pi^\dagger_j\right]=i\hbar\delta_{ij}$. The Hamiltonian can then be obtained from the Legendre transformation
\begin{align}\nonumber
H_X=&\sum_i\left(\Pi^\dagger_{i}\dot{X}_i^\ast+\Pi_i\dot{X}_i\right)-L_X\\\nonumber
=&\frac{U}{\hbar^2}\sum_i\Pi^\dagger_i\Pi_i+\lambda\sum_iX^\ast_iX_i+\sum_i\left(\frac{ih}{\hbar}-\frac{ie}{\hbar}\tilde{a}_{0,i}\right)\left(\Pi^\dagger_iX_i^\ast-\Pi_iX_i\right)\\
&-\sum_{\left\langle i,j \right\rangle}\left[t_{X,ij}e^{-i\frac{e}{\hbar}\int_{\bm{r}_i}^{\bm{r}_j}\left[\bm{A}\left(\bm{R},t\right)-\bm{a}\left(\bm{R},t\right)\right]\cdot d\bm{R}}+t^\ast_{X,ij}e^{i\frac{e}{\hbar}\int_{\bm{r}_i}^{\bm{r}_j}\left[\bm{A}\left(\bm{R},t\right)-\bm{a}\left(\bm{R},t\right)\right]\cdot d\bm{R}}X_jX^\ast_i\right].
\end{align}
In the canonical quantization Language, bosonic operators can be defined in terms of the canonical momentum and coordinate as
\begin{align}
a_i^{\left(\dagger\right)}=\frac{1}{\sqrt{2\hbar}}\left[\hbar^{\frac{1}{2}}\left(\frac{\lambda}{U}\right)X_i^\ast+i\hbar^{-\frac{1}{2}}\left(\frac{\lambda}{U}\right)^{-\frac{1}{4}}\Pi_i\right]^{\left(\dagger\right)},\quad b_i^{\left(\dagger\right)}=\frac{1}{\sqrt{2\hbar}}\left[\hbar^{\frac{1}{2}}\left(\frac{\lambda}{U}\right)X_i+i\hbar^{-\frac{1}{2}}\left(\frac{\lambda}{U}\right)^{-\frac{1}{4}}\Pi^\dagger_i\right]^{\left(\dagger\right)},
\end{align}
and one can easily check that $a_i$, $a_i^\dagger$, $b_i$, $b_i^\dagger$ respect the bosonic commutation relations: $\left[a_i,a^\dagger_j\right]=\left[b_i,b^\dagger_j\right]=0$, $\left[a_i,a_j\right]=\left[b_i,b_j\right]=0$. With the new bosonic operator defined, the Hamiltonian $H_X$ is rewritten as
\begin{align}\nonumber
H_X=&\sqrt{U\lambda}\sum_i\left(a_ia^\dagger_i+b^\dagger_ib_i\right)-\sum_i\left(h-e\tilde{a}_{0,i}\right)\left(a_ia^\dagger_i-b^\dagger_ib_i-1\right)-\frac{1}{2}\sqrt{\frac{U}{\lambda}}\sum_{\left\langle i, j \right\rangle}\left[t_{X,ij}e^{-i\frac{e}{\hbar}\int_{\bm{r}_i}^{\bm{r}_j}\left[\bm{A}\left(\bm{R},t\right)-\bm{a}\left(\bm{R},t\right)\right]\cdot d\bm{R}}\begin{pmatrix}
a_j & b^\dagger_j
\end{pmatrix}\begin{pmatrix}
1 & 1 \\ 1 & 1
\end{pmatrix}\begin{pmatrix}
a^\dagger_i \\ b_i
\end{pmatrix}\right.\\
&\left.+t^\ast_{X,ij}e^{i\frac{e}{\hbar}\int_{\bm{r}_i}^{\bm{r}_j}\left[\bm{A}\left(\bm{R},t\right)-\bm{a}\left(\bm{R},t\right)\right]\cdot d\bm{R}}\begin{pmatrix}
a_i & b^\dagger_i
\end{pmatrix}\begin{pmatrix}
1 & 1 \\ 1 & 1
\end{pmatrix}\begin{pmatrix}
a^\dagger_j \\ b_j
\end{pmatrix}\right].
\end{align}
Here $a_{i}^{\left(\dagger\right)}$ and $b_i^{\left(\dagger\right)}$ are the annihilation (creation) operators of the holons and doublons respectively. In terms of the holon and doublon operators, now the chargon action $S_X$ is rewritten as
\begin{align}\nonumber\label{S_X_tot}
S_X=&\int_0^\beta\left\{\sum_i\begin{pmatrix}
a_i & b^\dagger_i
\end{pmatrix}\begin{pmatrix}
-\partial_\tau+\sqrt{U\lambda}-\left(h-iea_{0,i}\right) & 0 \\
0 & \partial_\tau+\sqrt{U\lambda}+\left(h-iea_{0,i}\right)
\end{pmatrix}\begin{pmatrix}
a^\dagger_i \\ b_i
\end{pmatrix}\right.\\
&\left.-\frac{1}{2}\sqrt{\frac{U}{\lambda}}\sum_{\left\langle i, j \right\rangle}\left[t_{X,ij}e^{-i\frac{e}{\hbar}\int_{\bm{r}_i}^{\bm{r}_j}\left[\bm{A}\left(\bm{R},t\right)-\bm{a}\left(\bm{R},t\right)\right]\cdot d\bm{R}}\begin{pmatrix}
a_j & b^\dagger_j
\end{pmatrix}\begin{pmatrix}
1 & 1 \\ 1 & 1
\end{pmatrix}\begin{pmatrix}
a^\dagger_i \\ b_i
\end{pmatrix}+t^\ast_{X,ij}e^{i\frac{e}{\hbar}\int_{\bm{r}_i}^{\bm{r}_j}\left[\bm{A}\left(\bm{R},t\right)-\bm{a}\left(\bm{R},t\right)\right]\cdot d\bm{R}}\begin{pmatrix}
a_i & b^\dagger_i
\end{pmatrix}\begin{pmatrix}
1 & 1 \\ 1 & 1
\end{pmatrix}\begin{pmatrix}
a^\dagger_j \\ b_j
\end{pmatrix}\right]\right\}.
\end{align}

The bosonic action $S_X$ in Eq. \ref{S_X_tot} has considered the correlations between the holons and doublons, which come from the relativistic nature of the chargons. In the energies near the Hubbard band edges, the correlations between the holons and doublons are negligible~\cite{Lee01_Supp}, so the off-diagonal terms in Eq. \ref{S_X_tot} can be dropped. In the absence of external magnetic field, we first drop the gauge fields so the QSL action $S_0$ in Eq. \ref{S_act0} take the simple form
\begin{align}\nonumber\label{S_QSL2}
S_0=&\int_0^\beta\left[\sum_{i,\sigma}f^\dagger_{\sigma,i}\left(\partial_\tau-\mu_f\right)f_{\sigma,i}-\sum_{\left\langle i, j \right\rangle, \sigma}\left(t_{f,ij}f^\dagger_{i,\sigma}f_{j,\sigma}+t^\ast_{f,ij}f^\dagger_{j,\sigma}f_{i,\sigma}\right)+\sum_ia_i\left(-\partial_\tau+\sqrt{U\lambda}-h\right)a^\dagger_i+\sum_ib^\dagger_i\left(\partial_\tau+\sqrt{U\lambda}+h\right)b_i\right.\\
&\left.-\frac{1}{2}\sqrt{\frac{U}{\lambda}}\sum_{\left\langle i, j \right\rangle}\left(t_{X,ij}a_ja^\dagger_i+t^\ast_{X,ij}a_ia^\dagger_j\right)-\frac{1}{2}\sqrt{\frac{U}{\lambda}}\sum_{\left\langle i, j \right\rangle}\left(t_{X,ij}b^\dagger_jb_i+t^\ast_{X,ij}b^\dagger_ib_j\right)\right]d\tau.
\end{align}
After Fourier transformation, the QSL action in Eq. \ref{S_QSL2} becomes
\begin{align}\label{Simple_S0}
S_0=&-\sum_{\bm{k},\sigma,\omega_n}f^\dagger_{\sigma,\bm{k},n}\left(i\omega_n-\xi_{\bm{k}}\right)f_{\sigma,\bm{k},n}-\sum_{\bm{k},\nu_n}a_{-\bm{k},-n}\left(-i\nu_n-\epsilon_{\bm{k}}\right)a^\dagger_{-\bm{k},n}-\sum_{\bm{k},\nu_n}b^\dagger_{\bm{k},n}\left(i\nu_n-\epsilon_{\bm{k}}\right)b_{\bm{k},n},
\end{align}
with $\xi_{\bm{k}}$ and $\epsilon_{\bm{k}}$ being the band dispersions of the spinons and chargons respectively.
Therefore, the corresponding QSL mean field Hamiltonian takes the form
\begin{align}\label{QSL_mean0}
H_0=\sum_{\bm{k}}\epsilon_{\bm{k}}\left(a_{-\bm{k}}a^\dagger_{-\bm{k}}+b^\dagger_{\bm{k}}b_{\bm{k}}\right)+\sum_{\bm{k},\sigma}\xi_{\bm{k}}f^\dagger_{\sigma,\bm{k}}f_{\sigma,\bm{k}}.
\end{align}

\section{Electronic Green's Function in the $U\left(1\right)$ Quantum Spin Liquid}
In the $U\left(1\right)$ QSL with SFS, creating an electron is to create a spinon and simultaneously create a doublon or annihilate a holon. The creation operator for an electronic state takes the form $c^\dagger_{\bm{k},\bm{k}',\sigma}=f^\dagger_{\sigma,\bm{k}}\left(a_{-\bm{k}'}+b^\dagger_{\bm{k}'}\right)$. By definition, the electronic Matsubara Green's function in the QSL is
\begin{align}\nonumber\label{Green_mats1}
G_{\sigma}\left(i\omega_n,\bm{k},\bm{k}'\right)=&-\int_0^\beta\left\langle c_{\bm{k},\bm{k}',\sigma}\left(\tau\right)c^\dagger_{\bm{k},\bm{k}',\sigma}\left(0\right) \right\rangle_0e^{i\omega_n\tau}d\tau\\\nonumber
=&-\int_0^\beta\left\langle f_{\sigma,\bm{k}}\left(\tau\right)f^\dagger_{\sigma,\bm{k}}\left(0\right) \right\rangle_0\left[\left\langle a^\dagger_{-\bm{k}'}\left(\tau\right)a_{-\bm{k}'}\left(0\right) \right\rangle_0+\left\langle b_{\bm{k}'}\left(\tau\right)b^\dagger_{\bm{k}'}\left(0\right) \right\rangle_0\right]e^{i\omega_n\tau}d\tau\\
=&-\frac{1}{\beta}\sum_{\nu_n}G_{f,\sigma}\left(i\omega_n-i\nu_n,\bm{k}\right)\left[G_a\left(-i\nu_n,-\bm{k}'\right)+G_b\left(i\nu_n,\bm{k}'\right)\right],
\end{align}
where $G_{f,\sigma}\left(i\omega_n,\bm{k}\right)$ is the spinon Matsubara Green's function
\begin{align}
G_{f,\sigma}\left(i\omega_n,\bm{k}\right)=-\int_0^\beta\left\langle f_{\sigma,\bm{k}}\left(\tau\right)f^\dagger_{\sigma,\bm{k}}\left(0\right) \right\rangle_0 e^{i\omega_n\tau}d\tau=\frac{1}{i\omega_n-\xi_{\bm{k}}},
\end{align}
$G_a\left(-i\nu_n,-\bm{k}\right)$ is the holon Matsubara Green's function
\begin{align}
G_a\left(-i\nu_n,-\bm{k}\right)=&-\int_0^\beta\left\langle a^\dagger_{-\bm{k}}\left(\tau\right)a_{-\bm{k}}\left(0\right)\right\rangle_0e^{-i\nu_n\tau}d\tau=\frac{1}{-i\nu_n-\epsilon_{\bm{k}}},
\end{align}
and $G_b\left(i\nu_n,\bm{k}\right)$ is the doublon Matsubara Green's function
\begin{align}
G_b\left(i\nu_n,\bm{k}\right)=-\int_0^\beta\left\langle b_{\bm{k}}\left(\tau\right)b^\dagger_{\bm{k}}\left(0\right) \right\rangle_0 e^{i\nu_n\tau}d\tau=\frac{1}{i\nu_n-\epsilon_{\bm{k}}}.
\end{align}
Here $\left\langle \dots \right\rangle_0$ denotes the thermal average calcualted from $S_0$ in Eq. \ref{Simple_S0}. After Mastubara frequency summation, one can get the form of the electronic Matsubara Green's function in Eq. \ref{Green_mats1} to be
\begin{align}
G_{\sigma}\left(i\omega_n,\bm{k},\bm{k}'\right)=&\frac{n_{\textrm{F}}\left(\xi_{\bm{k}}\right)+n_{\textrm{B}}\left(\epsilon_{\bm{k}'}\right)}{i\omega_n-\xi_{\bm{k}}+\epsilon_{\bm{k}'}}+\frac{n_{\textrm{F}}\left(\xi_{\bm{k}}\right)+n_{\textrm{B}}\left(\epsilon_{\bm{k}'}\right)}{i\omega_n-\xi_{\bm{k}}-\epsilon_{\bm{k}'}}.
\end{align}
Here $n_{\textrm{F}}\left(\xi\right)=\frac{1}{2}\left(1-\tanh\frac{1}{2}\beta\xi\right)$ and $n_{\textrm{B}}\left(\epsilon\right)=\frac{1}{2}\left(\coth\frac{1}{2}\beta\epsilon-1\right)$ are the Fermi-Dirac and Bose-Einstein distribution functions respectively. The retarded electronic Green's function can then obtained through the analytic continuation $i\omega_n\rightarrow\omega+i0^+$:
\begin{align}\label{GR0}
G^{\textrm{R}}_\sigma\left(\omega,\bm{k},\bm{k}'\right)=&\frac{n_{\textrm{F}}\left(\xi_{\bm{k}}\right)+n_{\textrm{B}}\left(\epsilon_{\bm{k}'}\right)}{\omega+i0^+-\xi_{\bm{k}}+\epsilon_{\bm{k}'}}+\frac{n_{\textrm{F}}\left(\xi_{\bm{k}}\right)+n_{\textrm{B}}\left(\epsilon_{\bm{k}'}\right)}{\omega+i0^+-\xi_{\bm{k}}-\epsilon_{\bm{k}'}}.
\end{align}
The electronic density of states (DOS) per unit cell then takes
\begin{align}
\rho_\sigma\left(\omega\right)=&-\frac{1}{N^2\pi}\sum_{\bm{k},\bm{k}'}\textrm{Im}G^{\textrm{R}}_\sigma\left(\omega,\bm{k},\bm{k}'\right)=-\frac{1}{N^2\pi}\sum_{\bm{k},\bm{k}'}\textrm{Im}\frac{n_{\textrm{F}}\left(\xi_{\bm{k}}\right)+n_{\textrm{B}}\left(\epsilon_{\bm{k}'}\right)}{\omega+i0^+-\xi_{\bm{k}}+\epsilon_{\bm{k}'}}-\frac{1}{N^2\pi}\sum_{\bm{k},\bm{k}'}\textrm{Im}\frac{n_{\textrm{F}}\left(\xi_{\bm{k}}\right)+n_{\textrm{B}}\left(\epsilon_{\bm{k}'}\right)}{\omega+i0^+-\xi_{\bm{k}}-\epsilon_{\bm{k}'}}.
\end{align}
In the temperature region $k_{\textrm{b}}T\ll\textrm{min}\left[\epsilon_{\bm{k}}\right]=\Delta$, the Bose factor is $n_{\textrm{B}}\left(\epsilon_{\bm{k}}\right)\rightarrow0$, so the QSL electronic DOS becomes
\begin{align}
\rho_\sigma\left(\omega\right)=\frac{1}{N^2}\sum_{\bm{k},\bm{k}'}n_{\textrm{F}}\left(\xi_{\bm{k}}\right)\delta\left(\omega-\xi_{\bm{k}}+\epsilon_{\bm{k}'}\right)+\frac{1}{N^2}\sum_{\bm{k},\bm{k}'}n_{\textrm{F}}\left(-\xi_{\bm{k}}\right)\delta\left(\omega-\xi_{\bm{k}}-\epsilon_{\bm{k}}\right).
\end{align}

\section{Electronic DOS of the $U\left(1\right)$ Quantum Spin Liquid in an Orbital Magnetic Field}
In an orbital magnetic field $B$, an emergent gauge magnetic field (EGMF) $b$ is induced on the spinons and the remaining magnetic field on the chargons is $B-b$. The QSL mean field Hamiltonian in Eq. \ref{QSL_mean0} is now Landau quantized to be
\begin{align}\label{H_LLs}
H_0\left(b,B-b\right)=\sum_{\sigma,n,m}\xi_nf^\dagger_{\sigma,n,m}f_{\sigma,n,m}+\sum_{n,m}\epsilon_n\left(a_{n,m}a^\dagger_{n,m}+b^\dagger_{n,m}b_{n,m}\right).
\end{align}
Here $\xi_n$ is the $n$th spinon LL induced by the EGMF $b$, $\epsilon_n$ is the $n$th chargon LL induced by $B-b$ and the index $m$ counts the LL degeneracy. Similar to the case of $B=0$T, the composite electronic creation operator takes the form $c^\dagger_{n,m,n',m',\sigma}=f^\dagger_{n,m,\sigma}\left(a_{n',m'}+b^\dagger_{n',m'}\right)$, so the electronic Matsubara Green's function is constructed as
\begin{align}\nonumber\label{G_nm}
G_{n,m,n',m',\sigma}\left(i\omega_n\right)=&-\int_0^\beta\left\langle c_{n,m,n',m',\sigma}\left(\tau\right)c^\dagger_{n,m,n',m',\sigma}\left(\tau\right) \right\rangle_0e^{i\omega_n\tau}d\tau\\
=&-\frac{1}{\beta}\sum_{\nu_n}G_{f,n,m,\sigma}\left(i\omega_n-i\nu_n\right)\left[G_{a,n',m'}\left(-i\nu_n\right)+G_{b,n',m'}\left(i\nu_n\right)\right].
\end{align}
Here the Matsubara Green's functions are defined as
\begin{align}
G_{f,n,m,\sigma}\left(i\omega_n\right)=&-\int_0^\beta\left\langle f_{n,m,\sigma}\left(\tau\right)f^\dagger_{n,m,\sigma}\left(0\right) \right\rangle=\frac{1}{i\omega_n-\xi_n},\\
G_{a,n,m}\left(-i\nu_n\right)=&-\int_0^\beta\left\langle a^\dagger_{n,m}\left(\tau\right)a_{n,m}\left(0\right) \right\rangle_0e^{-i\nu_n\tau}d\tau=\frac{1}{-i\nu_n-\epsilon_n},\\
G_{b,n,m}\left(i\nu_n\right)=&-\int_0^\beta\left\langle b_{n,m}\left(\tau\right)b^\dagger_{n,m}\left(0\right)\right\rangle_0e^{i\nu_N\tau}d\tau=\frac{1}{i\nu_n-\epsilon_n}.
\end{align}
After summing over the Matsubara frequency, the electronic Matsubara Green's function in Eq. \ref{G_nm} gets the form
\begin{align}
G_{n,m,n',m',\sigma}\left(i\omega_n\right)=\frac{n_{\textrm{F}}\left(\xi_n\right)+n_{\textrm{B}}\left(\epsilon_{n'}\right)}{i\omega_n-\xi_n+\epsilon_{n'}}+\frac{n_{\textrm{F}}\left(-\xi_n\right)+n_{\textrm{B}}\left(\epsilon_{n'}\right)}{i\omega_n-\xi_n-\epsilon_{n'}}.
\end{align}
By analytic continuation $i\omega_n\rightarrow\omega+i0^+$, we obtain the retarded electronic Green's function
\begin{align}
G^{\textrm{R}}_{n,m,n',m',\sigma}\left(\omega\right)=&\frac{n_{\textrm{F}}\left(\xi_n\right)+n_{\textrm{B}}\left(\epsilon_{n'}\right)}{\omega+i0^+-\xi_n+\epsilon_{n'}}+\frac{n_{\textrm{F}}\left(-\xi_n\right)+n_{\textrm{B}}\left(\epsilon_{n'}\right)}{\omega+i0^+-\xi_n-\epsilon_{n'}}.
\end{align}
The corresponding electronic DOS of the QSL in a magnetic field is then calculated as
\begin{align}\nonumber\label{QSL_DOS0}
\rho_\sigma\left(\omega,b,B-b\right)=&-\frac{1}{N^2\pi}\sum_{n,m,n',m'}\textrm{Im}G^{\textrm{R}}_{n,m,n',m',\sigma}\left(\omega\right)\\
=&\frac{1}{N^2}\sum_{n,m,n',m'}n _{\textrm{F}}\left(\xi_n\right)\delta\left(\omega-\xi_n+\epsilon_{n'}\right)+\frac{1}{N^2}\sum_{n,m,n',m'}n_{\textrm{F}}\left(-\xi_n\right)\delta\left(\omega-\xi_n-\epsilon_{n'}\right).
\end{align}
Here the Bose factor is also dropped in the low temperature region as has been done in the case of $B=0$T.

\begin{figure}
\centering
\includegraphics[width=2.8in]{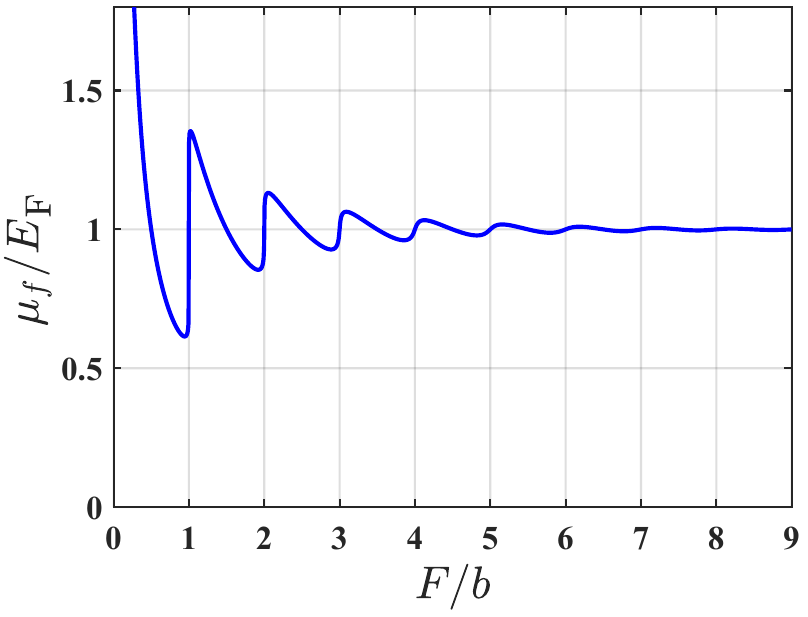}
\caption{The spinon chemical potential oscillation induced by the EGMF $b$. The spinon chemical potential $\mu_f\left(b\right)$ is solved from the self-consistent equation: $\sum_{n=0}^{\infty}1/n_{\textrm{F}}\left(\xi_n\right)=\nu$ with the filling factor $\nu$ being $\nu=\frac{\mu_f}{\hbar\omega_f}$. The temperature is fixed to be $k_{\textrm{b}}T/\mu_f=0.3$ in the calculation. The oscillation frequency $F$ is determined by the Onsager theorem: $F=\frac{\hbar S_{\textrm{F}}}{2\pi e}$ with $S_{\textrm{F}}$ being the area covered by the spinon Fermi surface area.}\label{figureS1}
\end{figure}

Near the band bottom, the spinon band $\xi_{\bm{k}}$ and the chargon band $\epsilon_{\bm{k}}$ can be approximated by the quadratic dispersions: $\xi_{\bm{k}}=\frac{\hbar^2\bm{k}^2}{2m_f}-\mu_f$, $\epsilon_{\bm{k}}=\frac{\hbar^2\bm{k}^2}{2m_X}+\Delta$. In an orbital magnetic field $B$, the spinon band is Landau quantized as $\xi_{n}=\left(n+\frac{1}{2}\right)\hbar\omega_f-\mu_f\left(b\right)$, while the chargon band LLs are $\epsilon_n=\left(n+\frac{1}{2}\right)\hbar\omega_X+\Delta$. Here the cyclotron frequencies are $\omega_f=\frac{eb}{m_f}$, $\omega_X=\frac{e\left(B-b\right)}{m_X}$. The spinon chemical potential $\mu_f\left(b\right)$ is determined by the equation $\sum_{n=0}^\infty1/n_{\textrm{F}}\left(\xi_n\right)=\nu$ with $\nu=\frac{\mu_f}{\hbar\omega_f}$ being the spinon LL filling factor. The numerically solved spinon chemical potential $\mu_f\left(b\right)$ at $k_{\textrm{b}}T/\mu_f=0.3$ is plotted in Fig. \ref{figureS1}. It oscillates with $b$ and approaches to $\mu_f$ as $b\rightarrow0$T. In the orbital magnetic field $B$, it is known that the DOS of the spinons, holons and doublons can be obtained as
\begin{align}
\rho_{f,\sigma}\left(\omega,b\right)=\frac{1}{N}\sum_{n,m}\delta\left(\omega-\xi_n\right),\quad\rho_{h}\left(\omega,B-b\right)=\frac{1}{N}\sum_{n,m}\delta\left(\omega+\epsilon_n\right),\quad\rho_d\left(\omega,B-b\right)=\frac{1}{N}\sum_{n,m}\delta\left(\omega-\epsilon_n\right),
\end{align}
so the QSL electronic DOS in Eq. \ref{QSL_DOS0} can be further simplified to be
\begin{align}\nonumber
\rho_\sigma\left(\omega,b,B-b\right)=&\frac{1}{N}\sum_{n,m}n_{\textrm{F}}\left(\xi_n\right)\frac{1}{N}\sum_{n',m'}\delta\left(\omega-\xi_n+\epsilon_{n'}\right)+\frac{1}{N}\sum_{n,m}n_{\textrm{F}}\left(-\xi_n\right)\frac{1}{N}\sum_{n',m'}\delta\left(\omega-\xi_n-\epsilon_{n'}\right)\\\nonumber
=&\frac{D}{N}\sum_{n=0}^{n_0}\lambda_n\rho_h\left[\omega+\left(n_0-n\right)\hbar\omega_f,B-b\right]+\frac{D}{N}\sum_{n=n_0}^{n_{\textrm{c}}}\left(1-\lambda_n\right)\rho_d\left[\omega-\left(n-n_0\right)\hbar\omega_f,B-b\right]\\
=&\sum_{n=0}^{n_0}\frac{\lambda_n}{\nu}\rho_h\left[\omega+\left(n_0-n\right)\hbar\omega_f,B-b\right]+\sum_{n=n_0}^{n_{\textrm{c}}}\frac{1-\lambda_n}{\nu}\rho_d\left[\omega-\left(n-n_0\right)\hbar\omega_f,B-b\right].
\end{align}
Here $\lambda_n$ is the filling of the $n$th spinon LL and $\nu=\sum_n\lambda_n$. In the derivation we have used $\nu=\frac{N}{D}$ with $D$ being the LL degeneracy. The LL spacing and the temperature respect: $\hbar\omega_X<k_{\textrm{b}}T\ll\hbar\omega_f$. As the spinon band has a finite width, the $n_{\textrm{c}}$ is introduced as a cut-off in the spinon LL.

\section{The $U\left(1\right)$ Quantum Spin Liquid in a Triangular Lattice }
In a triangular lattice, the band dispersions of the spinon band and the chargon band are taken to be
\begin{align}
\xi_{\bm{k}}=-2t_f\left(2\cos\frac{1}{2}k_xa\cos\frac{\sqrt{3}}{2}k_ya+\cos k_xa\right)-\mu_f,\quad \epsilon_{\bm{k}}=-2t_X\left(2\cos\frac{1}{2}k_xa\cos\frac{\sqrt{3}}{2}k_ya+\cos k_x-3\right)+\Delta.
\end{align}
Here only nearest neighbor hopping is considered: $t_{f,ij}=t_f$, $t_{X,ij}=t_X$. The band parameters are set to be $t_f=0.03$ eV, $t_X=0.02$ eV, $\Delta=0.25$ eV and $\mu_f=0.025$ eV. The lattice constant is $a$. In an orbital magnetic field $B$, for the QSL that emerges in the weak Mott regime, the spinon LL spacing $\hbar\omega_f$ is much larger than that of the chargon LL, so we push it to the limit $b\rightarrow B$. In the Landau gauge, given a rational magnetic flux ratio $\phi=\frac{eB}{h}\frac{\sqrt{3}a^2}{2}=\frac{1}{q}$, the spinon magnetic unit cell has the size $qa$ along $x$ and $\sqrt{3}a$ along $y$. There are 2$q$ sites in one magnetic unit cell. In the magnetic field, the mean field Hamiltonian in Eq. \ref{QSL_mean0} is now rewritten as
\begin{align}\label{hat_H0}
\hat{H}_0=\sum_{\sigma,\bm{k}}\hat{f}^\dagger_{\sigma,\bm{k}}\hat{h}_f\left(\bm{k}\right)\hat{f}_{\sigma,\bm{k}}+\sum_{\bm{k}}\left[\hat{a}_{-\bm{k}}\hat{h}_X\left(\bm{k}\right)\hat{a}^\dagger_{-\bm{k}}+\hat{b}^\dagger_{\bm{k}}\hat{h}_X\left(\bm{k}\right)\hat{b}_{\bm{k}}\right].
\end{align}
Here the spinon Hamiltonian matrix $\hat{h}_f\left(\bm{k}\right)$ takes the form
\begin{align}\label{h_f}
\hat{h}_f\left(\bm{k}\right)=\begin{pmatrix}
0 & B_1e^{-i\frac{1}{2}k_xa} & -t_fe^{-ik_xa} & 0 & \dots & 0 & -t_fe^{ik_xa} & B_{2q}e^{i\frac{1}{2}k_xa} \\
B_1e^{i\frac{1}{2}k_xa} & 0 & B_2e^{-i\frac{1}{2}k_xa} & -te^{-i\frac{1}{2}k_xa} & & & 0 & -t_fe^{ik_xa} \\
-t_fe^{ik_xa} & B_2e^{i\frac{1}{2}k_xa} & 0 & \ddots & \ddots & & & 0 \\
0 & -t_fe^{i\frac{1}{2}k_xa} & \ddots  & \ddots & & & & \vdots \\
\vdots & & \ddots & & & & & \\
-t_fe^{-ik_xa} & 0 & \dots & & & B_{2q-2}e^{i\frac{1}{2}k_xa} & 0 & B_{2q-1}e^{-i\frac{1}{2}k_xa} \\
B_{2q}e^{-i\frac{1}{2}k_xa} & -t_fe^{-ik_xa} & & & & & B_{2q-1}e^{i\frac{1}{2}k_xa} & 0
\end{pmatrix}-\mu_f
\end{align}
with $B_m=-2t_f\cos\left[\frac{\sqrt{3}}{2}k_ya+\left(m-\frac{1}{2}\right)\pi\phi\right]$. The chargon Hamiltonian matrix $\hat{h}_X\left(\bm{k}\right)$ is constructed in the same magnetic unit cell as that of $\hat{h}_f\left(\bm{k}\right)$. The matrix $\hat{h}_X\left(\bm{k}\right)$ has the form
\begin{align}\label{h_X}
\hat{h}_X\left(\bm{k}\right)=\begin{pmatrix}
0 & B_Xe^{-i\frac{1}{2}k_xa} & -t_Xe^{-ik_xa} & 0 & \dots & 0 & -t_Xe^{ik_xa} & B_Xe^{i\frac{1}{2}k_xa} \\
B_Xe^{i\frac{1}{2}k_xa} & 0 & B_Xe^{-i\frac{1}{2}k_xa} & -t_Xe^{-i\frac{1}{2}k_xa} & & & 0 & -t_Xe^{ik_xa} \\
-t_Xe^{ik_xa} & B_Xe^{i\frac{1}{2}k_xa} & 0 & \ddots & \ddots & & & 0 \\
0 & -t_Xe^{i\frac{1}{2}k_xa} & \ddots  & \ddots & & & & \vdots \\
\vdots & & \ddots & & & & & \\
-t_Xe^{-ik_xa} & 0 & \dots & & & B_Xe^{i\frac{1}{2}k_xa} & 0 & B_Xe^{-i\frac{1}{2}k_xa} \\
B_Xe^{-i\frac{1}{2}k_xa} & -t_Xe^{-ik_xa} & & & & & B_Xe^{i\frac{1}{2}k_xa} & 0
\end{pmatrix}+6t_X+\Delta
\end{align}
with $B_X=-2t_X\cos\frac{\sqrt{3}}{2}k_ya$. Please note that the spinon, holon and doublon operators in Eq. \ref{hat_H0} are in the vector form:
\begin{align}
\hat{f}^{\left(\dagger\right)}_{\sigma,\bm{k}}=&\left[f_{\sigma,\bm{k},1}, f_{\sigma,\bm{k},2},\dots,f_{\sigma,\bm{k},2q-1},f_{\sigma,\bm{k},2q}\right]^{\left(\dagger\right)},\\
\hat{a}^{\left(\dagger\right)}_{\bm{k}}=&\left[a_{\bm{k},1}, a_{\bm{k},2},\dots,a_{\bm{k},2q-1},a_{\bm{k},2q}\right]^{\left(\dagger\right)},\\
\hat{b}^{\left(\dagger\right)}_{\bm{k}}=&\left[b_{\bm{k},1}, b_{\bm{k},2},\dots,b_{\bm{k},2q-1},b_{\bm{k},2q}\right]^{\left(\dagger\right)},
\end{align}
with the subscript $1,2,\dots,2q-1,2q$ labeling the lattice sites inside a magnetic unit cell.

With the spinon and chargon Hamiltonian matrices given, the Matsubara Green's function matrices are defined as
\begin{align}
\hat{G}_{f,\sigma}\left(i\omega_n,\bm{k}\right)=\frac{1}{i\omega_n-\hat{h}_f\left(\bm{k}\right)},\quad\hat{G}_a\left(-i\nu_n,-\bm{k}\right)=-\frac{1}{i\nu_n+\hat{h}_X\left(\bm{k}\right)},\quad\hat{G}_b\left(i\nu_n,\bm{k}\right)=\frac{1}{i\nu_n-\hat{h}_X\left(\bm{k}\right)}.
\end{align}
We know that the Hamiltonian matrices $\hat{h}_f\left(\bm{k}\right)$ and $\hat{h}_X\left(\bm{k}\right)$ can be diagonalized by the unitary transformation
\begin{align}
\hat{U}_f\left(\bm{k}\right)\hat{h}_f\left(\bm{k}\right)\hat{U}_f\left(\bm{k}\right)=&\textrm{diag}\left[\xi_{\bm{k},1},\xi_{\bm{k},2},\dots,\xi_{\bm{k},2q}\right],\\
\hat{U}_X^\dagger\left(\bm{k}\right)\hat{h}_X\left(\bm{k}\right)\hat{U}_X\left(\bm{k}\right)=&\textrm{diag}\left[\epsilon_{\bm{k},1},\epsilon_{\bm{k},2},\dots,\epsilon_{\bm{k},2q}\right],
\end{align}
with $\hat{U}_f\left(\bm{k}\right)\hat{U}_f\left(\bm{k}\right)=1$ and $\hat{U}_X\left(\bm{k}\right)\hat{U}_X\left(\bm{k}\right)=1$ being the unitary matrices. With the eigen-vectors that diagonalize the spinon and chargon Hamiltonian matrices, one can write down the elements in the Matsubara Green's function matrices:
\begin{align}\label{GG1}
\hat{G}_{f,\sigma,k,l}\left(i\omega_n,\bm{k}\right)=&\sum_j\frac{\hat{U}_{f,k,j}\left(\bm{k}\right)\hat{U}^\ast_{f,j,l}\left(\bm{k}\right)}{i\omega_n-\xi_{\bm{k},j}},\quad\hat{G}_{a,k,l}\left(-i\nu_n,-\bm{k}\right)=\sum_j\frac{\hat{U}_{X,k,j}\left(\bm{k}\right)\hat{U}^\ast_{X,j,l}\left(\bm{k}\right)}{-i\nu_n-\epsilon_{\bm{k},j}},\\\label{GG2}
\hat{G}_{b,k,l}\left(i\nu_n,\bm{k}\right)=&\sum_j\frac{\hat{U}_{X,k,j}\left(\bm{k}\right)\hat{U}^\ast_{X,j,l}\left(\bm{k}\right)}{i\nu_n-\epsilon_{\bm{k},j}}.
\end{align}
Now the electronic creation operator is the vector form: $\hat{c}^\dagger_{\bm{k},\bm{k}',\sigma}=\hat{f}^\dagger_{\sigma,\bm{k}}\left(\hat{a}_{-\bm{k}'}+b^\dagger_{\bm{k}'}\right)$, so the electronic Matsuabara Green's function matrix can also be constructed as
\begin{align}\nonumber
\hat{G}_{\sigma,k,l}\left(i\omega_n,\bm{k},\bm{k}'\right)=&-\int_0^\beta\left\langle \hat{c}_{\bm{k},\bm{k}',\sigma,k,l}\left(\tau\right)\hat{c}^\dagger_{\bm{k},\bm{k}',\sigma,k,l}\left(0\right) \right\rangle_0e^{i\omega_n\tau}d\tau\\\nonumber
=&-\int_0^\beta\left\langle \hat{f}_{\sigma,\bm{k},k}\left(\tau\right)\hat{f}^\dagger_{\sigma,\bm{k},l}\left(0\right) \right\rangle_0\left[\left\langle \hat{a}^\dagger_{-\bm{k}',k}\left(\tau\right)a_{-\bm{k}',l}\left(0\right) \right\rangle_0+\left\langle b_{\bm{k}',k}\left(\tau\right)b^\dagger_{\bm{k}',l}\left(0\right) \right\rangle_0\right]e^{i\omega_n\tau}d\tau\\
=&-\frac{1}{\beta}\sum_{\nu_n}\hat{G}_{f,\sigma,k,l}\left(i\omega_n,\bm{k}\right)\left[\hat{G}_{a,k,l}\left(-i\nu_n,-\bm{k}'\right)+\hat{G}_{b,k,l}\left(i\nu_n,\bm{k}'\right)\right].
\end{align}
Subsituting the Matsbara Green's function matrix elements in Eq. \ref{GG1} and \ref{GG2} into $\hat{G}_{\sigma,k,l}\left(i\omega_n,\bm{k},\bm{k}'\right)$ and performing the Matsubara frequency summation, one can get
\begin{align}\nonumber
\hat{G}_{\sigma,k,l}\left(i\omega_n,\bm{k},\bm{k}'\right)=&\frac{1}{\beta}\sum_{\nu_n,i,i'}\hat{U}_{f,k,i}\left(\bm{k}\right)\hat{U}_{X,k,i'}\left(\bm{k}'\right)\left(\frac{1}{i\omega_n-i\nu_n-\xi_{\bm{k},i}}\frac{1}{i\nu_n+\epsilon_{\bm{k}',i'}}+\frac{1}{i\omega_n-i\nu_n-\xi_{\bm{k},i}}\frac{1}{i\nu_n-\epsilon_{\bm{k}',i'}}\right)\hat{U}^\ast_{f,i,l}\left(\bm{k}\right)\hat{U}^\ast_{X,i',l}\left(\bm{k}'\right)\\
=&\sum_{i,i'}\hat{U}_{f,k,i}\left(\bm{k}\right)\hat{U}_{X,k,i'}\left(\bm{k}'\right)\left[\frac{n_{\textrm{F}}\left(\xi_{\bm{k},i}\right)+n_{\textrm{B}}\left(\epsilon_{\bm{k},i'}\right)}{i\omega_n-\xi_{\bm{k},i}+\epsilon_{\bm{k}',i'}}+\frac{n_{\textrm{F}}\left(-\xi_{\bm{k},i}\right)+n_{\textrm{B}}\left(\epsilon_{\bm{k},i'}\right)}{i\omega_n-\xi_{\bm{k},i}-\epsilon_{\bm{k}',i'}}\right]\hat{U}^\ast_{f,i,l}\left(\bm{k}\right)\hat{U}^\ast_{X,i',l}\left(\bm{k}'\right).
\end{align}
The retarded electronic Green's function matrix is then
\begin{align}
\hat{G}^{\textrm{R}}_{\sigma,k,l}\left(\omega,\bm{k},\bm{k}'\right)=&\sum_{i,i'}\hat{U}_{f,k,i}\left(\bm{k}\right)\hat{U}_{X,k,i'}\left(\bm{k}'\right)\left[\frac{n_{\textrm{F}}\left(\xi_{\bm{k},i}\right)+n_{\textrm{B}}\left(\epsilon_{\bm{k},i'}\right)}{\omega+i0^+-\xi_{\bm{k},i}+\epsilon_{\bm{k}',i'}}+\frac{n_{\textrm{F}}\left(-\xi_{\bm{k},i}\right)+n_{\textrm{B}}\left(\epsilon_{\bm{k},i'}\right)}{\omega+i0^+-\xi_{\bm{k},i}-\epsilon_{\bm{k}',i'}}\right]\hat{U}^\ast_{f,i,l}\left(\bm{k}\right)\hat{U}^\ast_{X,i',l}\left(\bm{k}'\right).
\end{align}
The QSL electronic DOS can be calculated as
\begin{align}
\rho_\sigma\left(\omega,B,0\right)=&-\frac{1}{N^2\pi}\sum_{\bm{k},\bm{k}'}\textrm{tr}\textrm{Im}\hat{G}_\sigma^{\textrm{R}}\left(\omega,\bm{k},\bm{k}'\right).
\end{align}

\section{The Gauge Binding in the Quantum Spin Liquid}
In the low energy continuum limit, the QSL action in Eq. \ref{S_QSL2} changes the form to be
\begin{align}\nonumber
S_0=&\int_0^\beta d\tau\int d\bm{r}\sum_\sigma\left[f^\dagger_{\sigma,\bm{r}}\left(\partial_\tau+iea_{0,\bm{r}}-\mu_f\right)f_{\sigma,\bm{r}}+\frac{\hbar^2}{2m_f}\left(\partial_{\bm{r}}+i\frac{e}{\hbar}\bm{a}_{\bm{r}}\right)f^\dagger_{\sigma,\bm{r}}\cdot\left(\partial_{\bm{r}}-i\frac{e}{\hbar}\bm{a}_{\bm{r}}\right)f_{\sigma,\bm{r}}\right]\\\nonumber
&+\int_0^\beta d\tau\int d\bm{r}\left[a_{\bm{r}}\left(-\partial_\tau+iea_{0,\bm{r}}+\Delta\right)a^\dagger_{\bm{r}}+\frac{\hbar^2}{2m_X}\left(\partial_{\bm{r}}+i\frac{e}{\hbar}\bm{A}_{\bm{r}}-i\frac{e}{\hbar}\bm{a}_{\bm{r}}\right)a_{\bm{r}}\cdot\left(\partial_{\bm{r}}-i\frac{e}{\hbar}\bm{A}_{\bm{r}}+i\frac{e}{\hbar}\bm{a}_{\bm{r}}\right)a^\dagger_{\bm{r}}\right]\\
&+\int_0^\beta d\tau\int d\bm{r}\left[b^\dagger_{\bm{r}}\left(\partial_\tau-iea_{0,\bm{r}}+\Delta\right)b_{\bm{r}}+\frac{\hbar^2}{2m_X}\left(\partial_{\bm{r}}+i\frac{e}{\hbar}\bm{A}_{\bm{r}}-i\frac{e}{\hbar}\bm{a}_{\bm{r}}\right)b^\dagger_{\bm{r}}\cdot\left(\partial_{\bm{r}}+i\frac{e}{\hbar}\bm{A}_{\bm{r}}-i\frac{e}{\hbar}\bm{a}_{\bm{r}}\right)b_{\bm{r}}\right].
\end{align}
In the low energy effective action description, the high energy modes are integrated out and it generates a Maxwell term that controls the gauge field fluctuations~\cite{Lee02_Supp}:
\begin{align}
S_{\textrm{M}}=\int_0^\beta d\tau\int d\bm{r}\left[\frac{1}{2}\tilde{\epsilon}_0\left(\partial_{\bm{r}}a_{0,\bm{r}}+\frac{\partial_\tau\bm{a}_{\bm{r}}}{\hbar}\right)^2+\frac{1}{2\tilde{\mu}_0}\left(\partial_{\bm{r}}\times\bm{a}_{\bm{r}}\right)^2\right]
\end{align}
with $\tilde{c}=1/\sqrt{\tilde{\epsilon}_0\tilde{\mu}_0}$ being the speed of the gauge field propagation in the spin liquid. Now we get the low energy effective action: $S_{\textrm{eff}}=S_0+S_{\textrm{M}}$,  which includes both $S_0$ and $S_{\textrm{M}}$. Near the Hubbard band edge energies, the group velocity is small so the transverse component of gauge field fluctuations that arises from the current current correlation are negligible. The longitudinal component of gauge field fluctuations is dominant. By integrating out the scalar gauge field $a_{0,\bm{r}}$, one can get a Coulomb like interaction term:
\begin{align}
S_{\textrm{int}}=&\int_0^\beta d\tau\int d\bm{r}d\bm{r}'U_{\bm{r},\bm{r}'}\sum_\sigma f^\dagger_{\sigma,\bm{r}}f_{\sigma,\bm{r}}\left(a_{\bm{r}'}a^\dagger_{\bm{r}'}-b^\dagger_{\bm{r}'}b_{\bm{r}'}\right)
\end{align}
with $U_{\bm{r},\bm{r}'}=\frac{e^2}{4\pi^2\tilde{\epsilon}_0}\int\frac{e^{i\bm{q}\cdot\left(\bm{r}-\bm{r}'\right)}}{\bm{q}^2+\lambda^{-2}}d^2\bm{q}$ being the longitudinal gauge field fluctuations induced gauge binding interaction. Importantly, due to the spinon Fermi surface, the Coulomb like gauge binding interaction is screened and the screening parameter is $\lambda^{-1}=e\sqrt{\frac{\rho_f\left(0\right)}{\tilde{\epsilon}_0}}$. Here $\rho_f\left(0\right)$ is the spinon DOS at the Fermi level, so the gauge binding interaction strength is affected by the spinoin screening. In the long wave limit $\bm{q}\rightarrow0$, the onsite gauge binding is approximated as $U_{\textrm{b}}=U_{\bm{r}-\bm{r}'=\bm{0}}\approx\frac{e^2}{4\pi^2\tilde{\epsilon}_0}\int\frac{e^{i\bm{q}\cdot\left(\bm{r}-\bm{r}'\right)}}{\bm{q}^2+e^2\rho_f\left(0\right)/\tilde{\epsilon}_0}\delta\left(\bm{q}\right)d^2\bm{q}=1/\rho_f\left(0\right)$, which gives the value of the onsite gauge binding in the ideal case.

In the lattice model, the gauge binding introduces the interaction term
\begin{align}\label{H_interaction_term}
H_{\textrm{int}}=&U_{\textrm{n}}\sum_{\sigma,i}f^\dagger_{\sigma,i}f_{\sigma,i}\left(a_ia^\dagger_i-b^\dagger_ib_i\right)=\frac{U_{\textrm{b}}}{N}\sum_{\sigma,\bm{k},\bm{q},\bm{q}'}f^\dagger_{\sigma,\bm{k}-\bm{q}}f_{\sigma,\bm{k}-\bm{q}'}\left(a_{-\bm{q}}a^\dagger_{-\bm{q}'}-b^\dagger_{\bm{q}}b_{\bm{q}'}\right).
\end{align}
Now involving the gauge binding, the mean field Hamiltonian for the QSL is $H=H_0+H_{\textrm{int}}$, which takes the form
\begin{align}
H=\sum_{\bm{k}}\epsilon_{\bm{k}}\left(a_{-\bm{k}}a^\dagger_{-\bm{k}}+b^\dagger_{-\bm{k}}b_{\bm{k}}\right)+\sum_{\bm{k},\sigma}\xi_{\bm{k}}f^\dagger_{\sigma,\bm{k}}f_{\sigma,\bm{k}}+\frac{U_{\textrm{b}}}{N}\sum_{\sigma,\bm{k},\bm{q},\bm{q}'}f^\dagger_{\sigma,\bm{k}-\bm{q}}f_{\sigma,\bm{k}-\bm{q}'}\left(a_{-\bm{q}}a^\dagger_{-\bm{q}'}-b^\dagger_{\bm{q}}b_{\bm{q}'}\right).
\end{align}
In the case of zero $U_{\textrm{b}}$, it is known from Eq. \ref{GR0} that the QSL electronic retarded Green's function $G_\sigma^{\textrm{R}}\left(\omega,\bm{k},\bm{k}'\right)$ is composed of two terms
\begin{align}
G^{\textrm{R}}_{h,\sigma}\left(\omega,\bm{k},\bm{k}'\right)=&\frac{n_{\textrm{F}}\left(\xi_{\bm{k}}\right)+n_{\textrm{B}}\left(\epsilon_{\bm{k}'}\right)}{\omega+i0^+-\xi_{\bm{k}}+\epsilon_{\bm{k}'}},\quad\quad G^{\textrm{R}}_{d,\sigma}\left(\omega,\bm{k},\bm{k}'\right)=\frac{n_{\textrm{F}}\left(-\xi_{\bm{k}}\right)+n_{\textrm{B}}\left(\epsilon_{\bm{k}'}\right)}{\omega+i0^+-\xi_{\bm{k}}-\epsilon_{\bm{k}'}}.
\end{align}
The two terms correspond to the states in the LHB and UHB respectively. In the presence of a finite $U_{\textrm{b}}$, the retarded Green's function of a QSL electron with a quasi-momenta $\bm{k}$ can be obtained through the random phase approximation:
\begin{align}\label{G_tilde_retarded}
\tilde{G}^{\textrm{R}}_\sigma\left(\omega,\bm{k}\right)=\frac{\frac{1}{N}\sum_{\bm{q}}G_{h,\sigma}^{\textrm{R}}\left(\omega,\bm{k}-\bm{q},\bm{q}\right)}{1-\frac{U_{\textrm{b}}}{N}\sum_{\bm{q}}G^{\textrm{R}}_{h,\sigma}\left(\omega,\bm{k}-\bm{q},\bm{q}\right)}+\frac{\frac{1}{N}\sum_{\bm{q}}G_{d,\sigma}^{\textrm{R}}\left(\omega,\bm{k}-\bm{q},\bm{q}\right)}{1+\frac{U_{\textrm{b}}}{N}\sum_{\bm{q}}G^{\textrm{R}}_{d,\sigma}\left(\omega,\bm{k}-\bm{q},\bm{q}\right)},
\end{align}
where the summations over $\bm{q}$ in the numerators and denominators are to get the Green's functions of the LHB and UHB states with a quasi-momenta $\bm{k}$. The QSL electronic DOS in a finite gauge binding $U_{\textrm{b}}$ can then be derived from Eq. \ref{G_tilde_retarded} as
\begin{align}
\tilde{\rho}_\sigma\left(\omega\right)=-\frac{1}{N\pi}\sum_{\bm{k}}\textrm{Im}\tilde{G}_{\sigma}\left(\omega,\bm{k}\right)=-\frac{1}{N\pi}\sum_{\bm{k}}\textrm{Im}\frac{\frac{1}{N}\sum_{\bm{q}}G^{\textrm{R}}_{h,\sigma}\left(\omega,\bm{k}-\bm{q},\bm{q}\right)}{1-\frac{U_{\textrm{b}}}{N}\sum_{\bm{q}}G^{\textrm{R}}_{h,\sigma}\left(\omega,\bm{k}-\bm{q},\bm{q}\right)}-\frac{1}{N\pi}\sum_{\bm{k}}\textrm{Im}\frac{\frac{1}{N}\sum_{\bm{q}}G^{\textrm{R}}_{d,\sigma}\left(\omega,\bm{k}-\bm{q},\bm{q}\right)}{1+\frac{U_{\textrm{b}}}{N}\sum_{\bm{q}}G^{\textrm{R}}_{d,\sigma}\left(\omega,\bm{k}-\bm{q},\bm{q}\right)}.
\end{align}

For the lattice model of the QSL in an orbital magnetic field, given a rational magnetic flux ratio, the QSL Hamiltonian with zero $U_{\textrm{b}}$ has been done in Eq. \ref{hat_H0}. The gauge binding interaction term now takes the form
\begin{align}
\hat{H}_{\textrm{int}}=\frac{U_{\textrm{b}}}{N'}\sum_{\sigma,\bm{k},\bm{q},\bm{q}',i}\hat{f}^\dagger_{\sigma,\bm{k}-\bm{q},i}\hat{f}_{\sigma,\bm{k}-\bm{q}',i}\left(\hat{a}_{-\bm{q},i}\hat{a}^\dagger_{-\bm{q}',i}-\hat{b}^\dagger_{\bm{q},i}\hat{b}_{\bm{q}',i}\right),
\end{align}
so involving the gauge binding, the QSL mean field Hamiltonian $\hat{H}=\hat{H}_0+\hat{H}_{\textrm{int}}$ now becomes
\begin{align}
\hat{H}=&\sum_{\sigma,\bm{k}}\hat{f}^\dagger_{\sigma,\bm{k}}\hat{h}_f\left(\bm{k}\right)\hat{f}_{\sigma,\bm{k}}+\sum_{\bm{k}}\left[\hat{a}_{-\bm{k}}\hat{h}_X\left(\bm{k}\right)\hat{a}^\dagger_{-\bm{k}}+\hat{b}^\dagger_{\bm{k}}\hat{h}_X\left(\bm{k}\right)\hat{b}_{\bm{k}}\right]+\frac{U_{\textrm{b}}}{N'}\sum_{\sigma,\bm{k},\bm{q},\bm{q}',i}\hat{f}^\dagger_{\sigma,\bm{k}-\bm{q},i}\hat{f}_{\sigma,\bm{k}-\bm{q}',i}\left(\hat{a}_{-\bm{q},i}\hat{a}^\dagger_{-\bm{q}',i}-\hat{b}^\dagger_{\bm{q},i}\hat{b}_{\bm{q}',i}\right).
\end{align}
Here $N'=N/I$ is the number of magnetic unit cell and $I$ is the number of lattice sites in one magnetic unit cell. Similar to the case of zero magnetic flux, the retarded Green's function matrix of QSL states with a quasi-momenta $\bm{k}$ is obtained by the random phase approximation
\begin{align}\nonumber
\tilde{\hat{G}}^{\textrm{R}}_\sigma\left(\omega,\bm{k}\right)=&\left[\frac{1}{N'}\sum_{\bm{q}}\hat{G}^{\textrm{R}}_{h,\sigma}\left(\omega,\bm{k}-\bm{q},\bm{q}\right)\right]\left[1-\frac{U_{\textrm{b}}}{N'}\sum_{\bm{q}}\hat{G}^{\textrm{R}}_{h,\sigma}\left(\omega,\bm{k}-\bm{q},\bm{q}\right)\right]^{-1}\\
&+\left[\frac{1}{N'}\sum_{\bm{q}}\hat{G}^{\textrm{R}}_{d,\sigma}\left(\omega,\bm{k}-\bm{q},\bm{q}\right)\right]\left[1+\frac{U_{\textrm{b}}}{N'}\sum_{\bm{q}}\hat{G}^{\textrm{R}}_{d,\sigma}\left(\omega,\bm{k}-\bm{q},\bm{q}\right)\right]^{-1},
\end{align}
where $\hat{G}^{\textrm{R}}_{h,\sigma}\left(\omega,\bm{k},\bm{k}'\right)$ and $\hat{G}^{\textrm{R}}_{d,\sigma}\left(\omega,\bm{k},\bm{k}'\right)$ are
\begin{align}
\hat{G}^{\textrm{R}}_{h,\sigma}\left(\omega,\bm{k},\bm{k}'\right)=&\sum_{i,i'}\hat{U}_{f,k,i}\left(\bm{k}\right)\hat{U}_{X,k,i'}\left(\bm{k}'\right)\frac{n_{\textrm{F}}\left(\xi_{\bm{k},i}\right)+n_{\textrm{B}}\left(\epsilon_{\bm{k},i'}\right)}{i\omega_n-\xi_{\bm{k},i}+\epsilon_{\bm{k}',i'}}\hat{U}^\ast_{f,i,l}\left(\bm{k}\right)\hat{U}^\ast_{X,i',l}\left(\bm{k}'\right),\\
\hat{G}^{\textrm{R}}_{d,\sigma}\left(\omega,\bm{k},\bm{k}'\right)=&\sum_{i,i'}\hat{U}_{f,k,i}\left(\bm{k}\right)\hat{U}_{X,k,i'}\left(\bm{k}'\right)\frac{n_{\textrm{F}}\left(-\xi_{\bm{k},i}\right)+n_{\textrm{B}}\left(\epsilon_{\bm{k},i'}\right)}{i\omega_n-\xi_{\bm{k},i}-\epsilon_{\bm{k}',i'}}\hat{U}^\ast_{f,i,l}\left(\bm{k}\right)\hat{U}^\ast_{X,i',l}\left(\bm{k}'\right).
\end{align}
The QSL electronic DOS is then given as
\begin{align}\nonumber
\tilde{\rho}_\sigma\left(\omega,B,0\right)=&-\frac{1}{N'I^2\pi}\sum_{\bm{k}}\textrm{tr}\textrm{Im}\tilde{G}\left(\omega,\bm{k}\right)\\\nonumber
=&-\frac{1}{N\pi}\sum_{\bm{k}}\textrm{tr}\textrm{Im}\left\{\left[\frac{1}{N}\sum_{\bm{q}}\hat{G}^{\textrm{R}}_{h,\sigma}\left(\omega,\bm{k}-\bm{q},\bm{q}\right)\right]\left[1-\frac{U_{\textrm{b}}}{N'}\sum_{\bm{q}}\hat{G}^{\textrm{R}}_{h,\sigma}\left(\omega,\bm{k}-\bm{q},\bm{q}\right)\right]^{-1}\right\}\\
&-\frac{1}{N\pi}\sum_{\bm{k}}\textrm{tr}\textrm{Im}\left\{\left[\frac{1}{N}\sum_{\bm{q}}\hat{G}^{\textrm{R}}_{d,\sigma}\left(\omega,\bm{k}-\bm{q},\bm{q}\right)\right]\left[1+\frac{U_{\textrm{b}}}{N'}\sum_{\bm{q}}\hat{G}^{\textrm{R}}_{d,\sigma}\left(\omega,\bm{k}-\bm{q},\bm{q}\right)\right]^{-1}\right\}.
\end{align}
Here in the first line the number of lattice sites $I$ is considered because the QSL electronic DOS we concern is referred to the number of electronic states per unit cell at $B=0$T.

\begin{figure}
\centering
\includegraphics[width=5in]{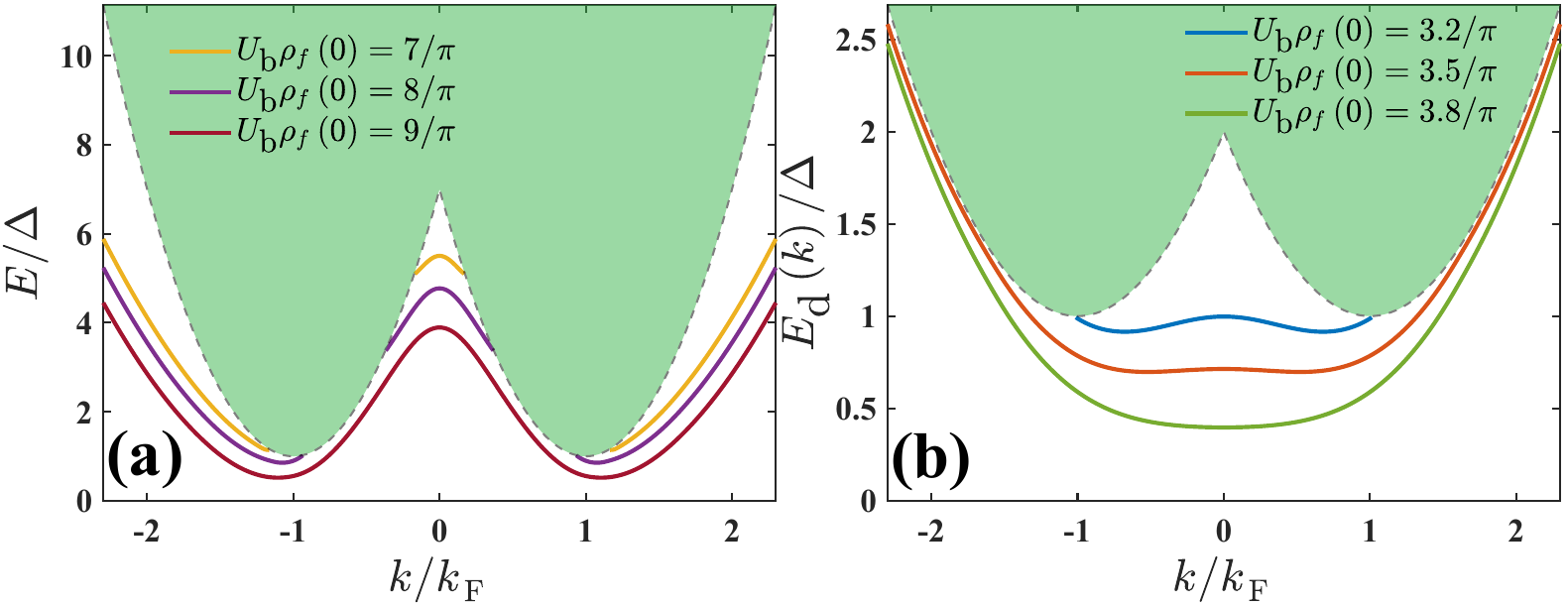}
\caption{(a) The evolution of the electronic bound state band dispersion as the gauge binding interaction increases. It is clear to see that the electronic bound state band emerges from the threshold energy band $E_{\textrm{th}}\left(\bm{k}\right)$, which is indicated by the black dashed line. The band parameters used in the calculation are $m_f/m_X=3$ and $\Delta/\mu_f=1/2$. (b) Given the band parameters $m_f/m_X=1/2$ and $\Delta/\mu_f=1/2$, the electronic bound state band evolves from a Mexican hat like dispersion at $U_{\textrm{b}}\rho_f\left(0\right)=3.2/\pi$ to a quadratic dispersion at $U_{\textrm{b}}\rho_f\left(0\right)=3.8/\pi$. In all the calculations, we take the momenta cut-off to be $k_{\textrm{c}}=2.6|\bm{k}_{\textrm{F}}|$ and the temperature is fixed to be $k_{\textrm{b}}T/\mu_f=0.05$.}\label{figureS2}
\end{figure}

\section{The Binding Equations of the In-gap Bound States}
\subsection{Threshold Energy}
In the absence of gauge binding, the many body ground state $\ket{\textrm{G}}$ for the QSL with SFS is the Fermi sea of occupied spinons: $\ket{\textrm{G}}=\prod_{\bm{k},\xi_{\bm{k}}\leqslant0}f^\dagger_{\uparrow,\bm{k}}f^\dagger_{\downarrow,\bm{k}}\ket{0}$. In the spin liquid, descroying a spinon (creating a spinon hole) and creating a holon is to excite a hole state: $\ket{h_\sigma,\bm{k}-\bm{q},\bm{q}}=f_{\sigma,\bm{k}-\bm{q}}a^\dagger_{-\bm{q}}\ket{\textrm{G}}$, while a correlated excitation of a spinon and a doublon is to excite a physical electronic state: $\ket{d_\sigma,\bm{k}-\bm{q},\bm{q}}=f^\dagger_{\sigma,\bm{k}-\bm{q}}b^\dagger_{\bm{q}}\ket{\textrm{G}}$. At zero temperature, the energy cost to excite a hole state and an electronic state are:
\begin{align}\label{Threshold001}
\bra{h_\sigma,\bm{k}-\bm{q},\bm{q}}H_0\ket{h_\sigma,\bm{k}-\bm{q},\bm{q}}&=-\xi_{\bm{k}-\bm{q}}+\epsilon_{\bm{q}},\quad\textrm{ with }\quad\xi_{\bm{k}-\bm{q}}\leqslant0,\\\label{Threshold002}
\bra{d_\sigma,\bm{k}-\bm{q},\bm{q}}H_0\ket{d_\sigma,\bm{k}-\bm{q},\bm{q}}&=\xi_{\bm{k}-\bm{q}}+\epsilon_{\bm{q}},\quad\textrm{with}\quad\xi_{\bm{k}-\bm{q}}\geqslant0.
\end{align}
Here the ground state energy $\bra{\textrm{G}}H_0\ket{\textrm{G}}$ has been subtracted in Eq. \ref{Threshold001} and \ref{Threshold002}. The energy cost takes the minimum value at $\bm{k}-\bm{q}=\bm{k}_{\textrm{F}}$. By definition, the minimum energy cost to add or remove an electron in the QSL is the threshold energy
\begin{align}
E_{\textrm{th}}\left(\bm{k}\right)=\epsilon_{\bm{k}-\bm{k}_{\textrm{F}}}=\frac{\hbar^2\left(|\bm{k}|-|\bm{k}_{\textrm{F}}|\right)^2}{2m_X}+\Delta.
\end{align}
Here in the last step we take a quadratic dispersion to approximate the chargon band.
\subsection{Binding Equations in Zero Magnetic Field}
Due to the finite gauge binding, there exist scatterings between $\ket{h_\sigma,\bm{k}-\bm{q},\bm{q}}$ and $\ket{h_\sigma,\bm{k}-\bm{q}',\bm{q}'}$, also scatterings between $\ket{d_\sigma,\bm{k}-\bm{q},\bm{q}}$ and $\ket{d_\sigma,\bm{k}-\bm{q}',\bm{q}'}$. As a result, in the presence of gauge binding, we need to consider a general anti-spinon holon pair state and a spinon doublon pair state:
\begin{align}
\ket{h_\sigma,\bm{k}}=\sum_{\bm{q}}C_{h,\bm{q}}\ket{h_\sigma,\bm{k}-\bm{q},\bm{q}}=\sum_{\bm{q}}C_{h,\bm{q}}f_{\sigma,\bm{k}-\bm{q}}a^\dagger_{-\bm{q}}\ket{\textrm{G}},\quad \ket{k_\sigma,\bm{k}}=\sum_{\bm{q}}C_{d,\bm{q}}\ket{d_\sigma,\bm{k}-\bm{q},\bm{q}}=\sum_{\bm{q}}C_{d,\bm{q}}f^\dagger_{\sigma,\bm{k}-\bm{q}}b^\dagger_{\bm{q}}\ket{\textrm{G}},
\end{align}
to construct the Hamiltonian for the correlated spinon chargon pair state. In the basis of $\ket{h_\sigma,\bm{k}}$ and $\ket{d_\sigma,\bm{k}}$, the eigen equations for the Hamiltonian are constructed to be
\begin{align}
\sum_{\bm{q}'}\bra{h_\sigma,\bm{k}-\bm{q},\bm{q}}H_0+H_{\textrm{int}}\ket{h_\sigma,\bm{k}-\bm{q}',\bm{q}'}=&-E_h\left(\bm{k}\right)C_{h,\bm{q}},\quad \sum_{\bm{q}'}\bra{d_\sigma,\bm{k}-\bm{q},\bm{q}}H_0+H_{\textrm{int}}\ket{d_\sigma,\bm{k}-\bm{q},\bm{q}}=E_d\left(\bm{k}\right)C_{d,\bm{q}}.
\end{align}
By calculating the matrix elements, we obtain the eigen equations of the anti-spinon holon pairs and the spinon doublon pairs as
\begin{align}\label{Int_eq1}
-\frac{A_{\textrm{c}}}{4\pi^2}\int \left[n_{\textrm{F}}\left(\xi_{\bm{k}-\bm{q}'}\right)+n_{\textrm{B}}\left(\epsilon_{\bm{q}'}\right)\right]U_{\textrm{b}}C_{h,\bm{q}'}d^2\bm{q}'=&\left[-E_h\left(\bm{k}\right)+\xi_{\bm{k}-\bm{q}}-\epsilon_{\bm{q}}\right]C_{h,\bm{q}},\\\label{Int_eq2}
-\frac{A_{\textrm{c}}}{4\pi^2}\int\left[n_{\textrm{F}}\left(-\xi_{\bm{k}-\bm{q}'}\right)+n_{\textrm{B}}\left(\epsilon_{\bm{q}'}\right)\right]U_{\textrm{b}}C_{d,\bm{q}'}d^2\bm{q}'=&\left[E_{d}\left(\bm{k}\right)-\xi_{\bm{k}-\bm{q}}-\epsilon_{\bm{q}}\right]C_{d,\bm{q}}.
\end{align}
To solve the integral equation, we need to introduce
\begin{align}
C_{h,\bm{q}}=\frac{A_{h,\bm{q}}}{-E_h\left(\bm{k}\right)+\xi_{\bm{k}-\bm{q}}-\epsilon_{\bm{q}}},\quad C_{d,\bm{q}}=\frac{A_{d,\bm{q}}}{E_d\left(\bm{k}\right)-\xi_{\bm{k}-\bm{q}}-\epsilon_{\bm{q}}}
\end{align}
and then the integral equations Eq. \ref{Int_eq1} and \ref{Int_eq2} become
\begin{align}
-\frac{U_{\textrm{b}}A_{\textrm{c}}}{4\pi^2}\int\frac{n_{\textrm{F}}\left(\xi_{\bm{k}-\bm{q}'}\right)+n_{\textrm{B}}\left(\epsilon_{\bm{q}'}\right)}{-E_h\left(\bm{k}\right)+\xi_{\bm{k}-\bm{q}'}-\epsilon_{\bm{q}'}}A_{h,\bm{q}'}d^2\bm{q}'=&A_{h,\bm{q}},\\
-\frac{U_{\textrm{b}}A_{\textrm{c}}}{4\pi^2}\int\frac{n_{\textrm{F}}\left(-\xi_{\bm{k}-\bm{q}'}\right)+n_{\textrm{B}}\left(\epsilon_{\bm{q}'}\right)}{E_d\left(\bm{k}\right)-\xi_{\bm{k}-\bm{q}'}-\epsilon_{\bm{q}'}}A_{d,\bm{q}'}d^2\bm{q}'=&A_{d,\bm{q}}.
\end{align}
The two integral equations can be factorized to give the self-consistent equations:
\begin{align}\label{Binding01}
\frac{1}{U_{\textrm{b}}}-\frac{A_{\textrm{c}}}{4\pi^2}\int\frac{n_{\textrm{F}}\left(\xi_{\bm{k}-\bm{q}}\right)+n_{\textrm{B}}\left(\epsilon_{\bm{q}}\right)}{E_h\left(\bm{k}\right)+i0^+-\xi_{\bm{k}-\bm{q}}+\epsilon_{\bm{q}}}d^2\bm{q}&=0,\\\label{Binding02}
\frac{1}{U_{\textrm{b}}}+\frac{A_{\textrm{c}}}{4\pi^2}\int\frac{n_{\textrm{F}}\left(-\xi_{\bm{k}-\bm{q}}\right)+n_{\textrm{B}}\left(\epsilon_{\bm{q}}\right)}{E_d\left(\bm{k}\right)+i0^+-\xi_{\bm{k}-\bm{q}}-\epsilon_{\bm{q}}}d^2\bm{q}&=0.
\end{align}
Eq. \ref{Binding01} and Eq. \ref{Binding02} are the binding equations of the hole bound state above the LHB and the electronic bound state below the UHB. The hole and electronic bound state band dispersions are given by $E_h\left(\bm{k}\right)$ and $E_d\left(\bm{k}\right)$ respectively. In the continuum limit, the spinon and chargon bands are approximated by quadratic dispersions: $\xi_{\bm{k}}=\frac{\hbar^2\bm{k}^2}{2m_f}-\mu_f$ and $\epsilon_{\bm{k}}=\frac{\hbar^2\bm{k}^2}{2m_X}+\Delta$. A momenta cut-off $|\bm{q}|=k_{\textrm{c}}$ is needed to introduce to the binding equations:
\begin{align}\label{hole_bound_band001}
\frac{1}{U_{\textrm{b}}}-\frac{A_{\textrm{c}}}{4\pi^2}\int_0^{|\bm{q}|=k_{\textrm{c}}}\frac{n_{\textrm{F}}\left(\xi_{\bm{k}-\bm{q}}\right)+n_{\textrm{B}}\left(\epsilon_{\bm{q}}\right)}{E_h\left(\bm{k}\right)+i0^+-\xi_{\bm{k}-\bm{q}}+\epsilon_{\bm{q}}}d^2\bm{q}&=0,\\\label{Electronic_bound_band001}
\frac{1}{U_{\textrm{b}}}+\frac{A_{\textrm{c}}}{4\pi^2}\int_0^{|\bm{q}|=k_{\textrm{c}}}\frac{n_{\textrm{F}}\left(-\xi_{\bm{k}-\bm{q}}\right)+n_{\textrm{B}}\left(\epsilon_{\bm{q}}\right)}{E_d\left(\bm{k}\right)+i0^+-\xi_{\bm{k}-\bm{q}}-\epsilon_{\bm{q}}}d^2\bm{q}&=0.
\end{align}
In Fig. \ref{figureS2} (a) and (b), a set of electronic bound state band dispersions solved from Eq. \ref{Electronic_bound_band001} are plotted. At small gauge binding, it can be seen that the electronic bound state band emerges from the threshold energy band $E_{\textrm{th}}\left(\bm{k}\right)$. The specific bound state band dispersion is affected by many factors like spinon band dispersion, chargon band dispersion, spinon chemical potential, gauge binding strength, temperature, etc. In Fig. \ref{figureS2} (a), the in-gap electronic bound state band has a Mexican hat like shape. Such Mexican hat like dispersion is one representative bound state band dispersion. In Fig. \ref{figureS2} (b), given different dispersions of the spinon band and the chargon band, the Mexican hat like bound state band evolves into a quadratic band as the gauge binding increases.

\section{Landau Levels Spectrum of the In-gap Bound States}
\subsection{Landau Quantization of the Mean Field Hamltonian}
An orbital magnetic field applied to the QSL can induce an emergent gauge magnetic field (EGMF) $b$ on the spinons, and the remaining magnetic field on the chargons is $B-b$. The EGMF $b$ on the spinons and the remaining magnetic field $B-b$ on the chargons generate the LLs spectrum. The Landau quantized mean field Hamiltonian has been written out in Eq. \ref{H_LLs}: $H_0\left(b,B-b\right)=\sum_{\sigma,n,m}\xi_nf^\dagger_{\sigma,n,m}f_{\sigma,n,m}+\sum_{n,m}\epsilon_n\left(a_{n,m}a^\dagger_{n,m}+b^\dagger_{n,m}b_{n,m}\right)$. The eigenstates of the the spinon and boson LLs are
\begin{align}
\psi_{f,n,m}\left(\bm{r}\right)=&\left\{\begin{matrix}
\frac{1}{\sqrt{2\pi l_b^2}}\sqrt{\frac{n!}{m!}}\left(\frac{iz}{\sqrt{2}l_b}\right)^{m-n}L_n^{m-n}\left(\frac{zz^\ast}{2l_b^2}\right)e^{-\frac{zz^\ast}{4l_b^2}}, & m\geqslant n \\ 
\frac{1}{\sqrt{2\pi l_b^2}}\sqrt{\frac{m!}{n!}}\left(\frac{iz^\ast}{\sqrt{2}l_b}\right)^{n-m}L_m^{n-m}\left(\frac{zz^\ast}{2l_b^2}\right)e^{-\frac{zz^\ast}{4l_b^2}} & m<n
\end{matrix}\right.,\\
\psi_{X,n,m}\left(\bm{r}\right)=&\left\{\begin{matrix}
\frac{1}{\sqrt{2\pi l^2_{B-b}}}\sqrt{\frac{n!}{m!}}\left(\frac{iz}{\sqrt{2}l_{B-b}}\right)^{m-n}L_n^{m-n}\left(\frac{zz^\ast}{2l^2_{B-b}}\right)e^{-\frac{zz^\ast}{4l^2_{B-b}}}, & m\geqslant n \\ 
\frac{1}{\sqrt{2\pi l^2_{B-b}}}\sqrt{\frac{m!}{n!}}\left(\frac{iz^\ast}{\sqrt{2}l^2_{B-b}}\right)^{n-m}L_m^{n-m}\left(\frac{zz^\ast}{2l^2_{B-b}}\right)e^{-\frac{zz^\ast}{4l^2_{B-b}}}, & m<n
\end{matrix}\right.,
\end{align}
with $l_b=\sqrt{\frac{\hbar}{eb}}$ and $l_{B-b}=\sqrt{\frac{\hbar}{e\left(B-b\right)}}$. The function $L_n^{m}\left(x\right)$ is the associated Laguerre function and $z$ is $z=x+iy$. In the LL basis, the gauge binding interaction term takes the form
\begin{align}
H_{\textrm{int}}\left(b,B-b\right)=&\sum_{\sigma,n_1,n_2,n_1',n_2',m_1,m_2,m_1',m_2'}\tilde{U}_{\textrm{b},n_1,n_2,n_1',n_2',m_1,m_2,m_1',m_2'}f^\dagger_{\sigma,n_1,m_1}f_{\sigma,n_2,m_2}\left(a_{n_1',m_1'}a^\dagger_{n_2',m_2'}-b^\dagger_{n_1',m_1'}b_{n_2',m_2'}\right)
\end{align}
where the gauge binding interaction matrix is
\begin{align}
\tilde{U}_{\textrm{b},n_1,n_2,n_1',n_2',m_1,m_2,m_1',m_2'}=\int d\bm{r}\int d\bm{r}'U_{\textrm{b}}\delta\left(\bm{r}-\bm{r}'\right)\psi^\ast_{f,n_1,m_1}\left(\bm{r}\right)\psi^\ast_{X,n_1',m_1'}\left(\bm{r}'\right)\psi_{f,n_2,m_2}\left(\bm{r}\right)\psi_{X,n_2',m_2'}\left(\bm{r}'\right).
\end{align}
Now the QSL mean field Hamiltonian that involves the gauge binding interaction is
\begin{align}\nonumber\label{H_b_B_b}
H\left(b,B-b\right)=&\sum_{\sigma,n,m}\xi_nf^\dagger_{\sigma,n,m}f_{\sigma,n,m}+\sum_{n,m}\epsilon_n\left(a_{n,m}a^\dagger_{n,m}+b^\dagger_{n,m}b_{n,m}\right)\\
&+\sum_{\sigma,n_1,n_2,n_1',n_2',m_1,m_2,m_1',m_2'}\tilde{U}_{\textrm{b},n_1,n_2,n_1',n_2',m_1,m_2,m_1',m_2'}f^\dagger_{\sigma,n_1,m_1}f_{\sigma,n_2,m_2}\left(a_{n_1',m_1'}a^\dagger_{n_2',m_2'}-b^\dagger_{n_1',m_1'}b_{n_2',m_2'}\right).
\end{align}
Given a sufficiently large gauge binding, diagonalizing the Hamiltonian $H\left(b,B-b\right)$ in Eq. \ref{H_b_B_b} gives the in-gap bound state eigen energy levels. However, the matrix dimension of $H\left(b,B-b\right)$ is extremely large: $n_1\times m_1\times n_1'\times m_1'\times n_2\times m_2\times n_2'\times m_2'$, so it goes beyond our computing power limit. In the below, we analyze the problem in the two limiting cases: $b\rightarrow B$ and $b\rightarrow0$T.

\begin{figure}
\centering
\includegraphics[width=5in]{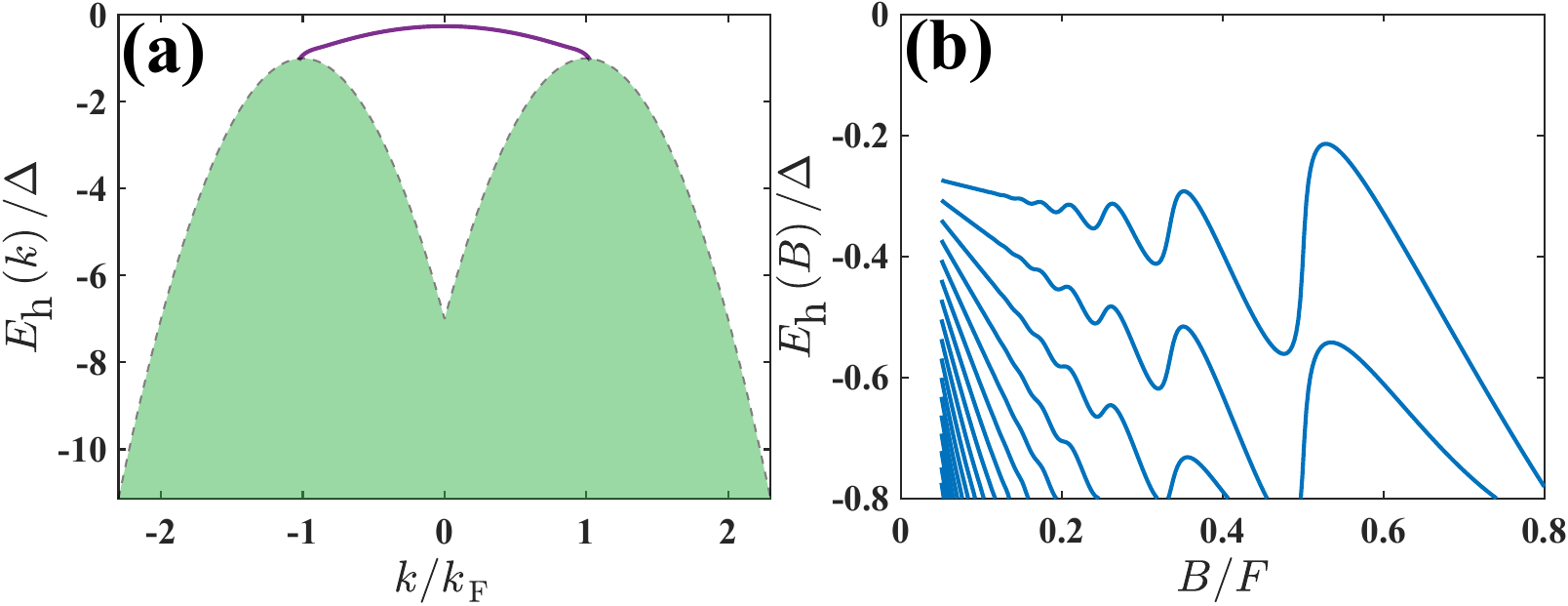}
\caption{(a) The hole bound state band dispersion solved from Eq. \ref{hole_bound_band001}. The bound state band shows a quadratic dispersion. (b) The hole bound state LLs specrum corresponding to the band dispersion in (a). The LLs spectrum is solved from Eq. \ref{Integral_Eq01} in the $b\rightarrow B$ limit. The LLs spectrum $E_{\textrm{h},n}\left(B\right)$ exhibits the linear increase with $B$ in the region of $B\rightarrow0$T. As $B$ increases, the envelop energy of the LLs spectrum $E_{\textrm{h},n}\left(B\right)$ turns from decreasing in $B$ to increasing in $B$. The band parameters used in the calculations are $m_f/m_X=3$, $\Delta/\mu_f=1/2$, $U_{\textrm{b}}\rho_f\left(0\right)=7/\pi$, $k_{\textrm{b}}T/\mu_f=0.03$. The momenta cut-off is set to be $k_{\textrm{c}}=2.6|\bm{k}_{\textrm{F}}|$.}\label{figureS3}
\end{figure}

\subsection{Binding Equations of the In-gap Bound States in the limit of $b\rightarrow B$}
In the limit of $b\rightarrow B$, all the external magnetic field is applied on the spinons, while the chargons feel no magnetic field. The QSL mean field Hamiltonain in the limit of $b\rightarrow B$ takes the form
\begin{align}\nonumber
H\left(B,0\right)=&H_0\left(B,0\right)+H_{\textrm{int}}\left(B,0\right)\\\nonumber
=&\sum_{\sigma,n,m}\xi_nf^\dagger_{\sigma,n,m}f_{\sigma,n,m}+\sum_{\bm{k}}\epsilon_{\bm{k}}\left(a_{-\bm{k}}a^\dagger_{-\bm{k}}+b^\dagger_{\bm{k}}b_{\bm{k}}\right)\\
&+\sum_{\sigma,n,m,n',m'}\int d\bm{r}\int d\bm{r}'U_{\textrm{b}}\delta\left(\bm{r}-\bm{r}'\right)f^\dagger_{\sigma,n,m}f_{\sigma,n',m'}\psi^\ast_{X,n,m}\left(\bm{r}\right)\psi_{X,n',m'}\left(\bm{r}\right)\left(a_{\bm{r}'}a^\dagger_{\bm{r}'}-b^\dagger_{\bm{r}'}b_{\bm{r}'}\right).
\end{align}
After Fourier transformation, the Hamiltonian $H\left(B,0\right)$ is simplified to be
\begin{align}\nonumber
H\left(B,0\right)=&\sum_{\sigma,n,m}\xi_nf^\dagger_{\sigma,n,m}f_{\sigma,n,m}+\sum_{\bm{k}}\epsilon_{\bm{k}}\left(a_{-\bm{k}}a^\dagger_{-\bm{k}}+b^\dagger_{\bm{k}}b_{\bm{k}}\right)\\
&+\frac{U_{\textrm{b}}}{N}\sum_{\bm{q},\bm{q}',\sigma}\sum_{n,m,n',m'}\sum_{l_1,l_2}D_{n,l_1}\left(-q\right)D_{m,l_2}\left(-q^\ast\right)D_{l_1,n'}\left(q'\right)D_{l_2,m'}\left(q'^\ast\right)e^{-\frac{l_B^2\left(\bm{q}^2+\bm{q}'^2\right)}{2}}f^\dagger_{\sigma,n,m}f_{\sigma,n',m'}\left(a_{-\bm{q}}a^\dagger_{-\bm{q}'}-b^\dagger_{\bm{q}}b_{\bm{q}'}\right).
\end{align}
Here the function $D_{n,m}\left(q\right)$ is
\begin{align}
D_{n,m}\left(q\right)=\left\{\begin{matrix}
\sqrt{\frac{m!}{n!}}\left(-\frac{l_Bq}{\sqrt{2}}\right)^{n-m}L_m^{n-m}\left(\frac{l_B^2qq^\ast}{2}\right), & n\geqslant m \\ 
\sqrt{\frac{n!}{m!}}\left(-\frac{l_Bq^\ast}{\sqrt{2}}\right)^{m-n}L_n^{m-n}\left(\frac{l_B^2qq^\ast}{2}\right), & n<m
\end{matrix}\right.,\quad\textrm{with}\quad q=q_x+iq_y.
\end{align}
Following the spirit of the $B=0$T case, we introduce the general anti-spinon holon pair state and spinon doublon pair state:
\begin{align}
\ket{h_\sigma}=&\sum_{n,m,\bm{q}}C_{h,n,m,\bm{q}}\ket{h_\sigma,n,m,\bm{q}}=\sum_{n,m,\bm{q}}C_{h,n,m,\bm{q}}f_{\sigma,n,m}a^\dagger_{-\bm{q}}\ket{\textrm{G}},\\
\ket{d_\sigma}=&\sum_{n,m,\bm{q}}C_{d,n,m,\bm{q}}\ket{d_\sigma,n,m,\bm{q}}=\sum_{n,m,\bm{q}}C_{d,n,m,\bm{q}}f^\dagger_{\sigma,n,m}b^\dagger_{\bm{q}}\ket{\textrm{G}}.
\end{align}
By calculating the matrix elements $\bra{h_\sigma,n,m,\bm{q}}H\left(B,0\right)\ket{h_\sigma,n',m',\bm{q}'}$ and $\bra{d_\sigma,n,m,\bm{q}}H\left(B,0\right)\ket{d_\sigma,n',m',\bm{q}'}$, one can get eigen equations of the anti-spinon holon pairs and the spinon doublon pairs as
\begin{align}\label{Integral_matrix01}
-\frac{U_{\textrm{b}}}{N}\sum_{\bm{q}',l_1,l_2,n',m'}D_{n,l_1}\left(-q\right)D_{m,l_2}\left(-q^\ast\right)e^{-\frac{l_B^2\bm{q}^2}{2}}D_{l_1,n'}\left(q'\right)D_{l_2,m'}\left(q'^\ast\right)e^{-\frac{l_B^2q'q'^\ast}{2}}\frac{n_{\textrm{F}}\left(\xi_{n'}\right)+n_{\textrm{B}}\left(\epsilon_{\bm{q}'}\right)}{-E_{h,n}\left(B\right)+\xi_{n'}-\epsilon_{\bm{q}'}}A_{h,n',m',\bm{q}'}&=A_{h,n,m,\bm{q}},\\\label{Integral_matrix02}
-\frac{U_{\textrm{b}}}{N}\sum_{\bm{q}',l_1,l_2,n',m'}D_{n,l_1}\left(-q\right)D_{m,l_2}\left(-q^\ast\right)e^{-\frac{l_B^2\bm{q}^2}{2}}D_{l_1,n'}\left(q'\right)D_{l_2,m'}\left(q'^\ast\right)e^{-\frac{l_B^2\bm{q}'^2}{2}}\frac{n_{\textrm{F}}\left(-\xi_{n'}\right)+n_{\textrm{B}}\left(\epsilon_{\bm{q}}\right)}{E_{d,n}\left(B\right)-\xi_{n'}-\epsilon_{\bm{q}'}}A_{d,n',m',\bm{q}'}&=A_{d,n,m,\bm{q}},
\end{align}
with
\begin{align}
A_{h,n',m',\bm{q}'}=\left[-E_{h,n}\left(B\right)+\xi_{n'}-\epsilon_{\bm{q}'}\right]C_{h,n',m',\bm{q}'},\quad\textrm{and}\quad A_{d,n',m',\bm{q}'}=\left[E_{d,n}\left(B\right)-\xi_{n'}-\epsilon_{\bm{q}'}\right]C_{d,n',m',\bm{q}'}.
\end{align}
The matrix integral equations in Eq. \ref{Integral_matrix01} and \ref{Integral_matrix02} are of the type: $\int\hat{K}\left(\bm{q},\bm{q}'\right)\hat{A}\left(\bm{q}'\right)d^2\bm{q}'=\hat{A}\left(\bm{q}\right)$ with a separable kernal $\hat{K}\left(\bm{q},\bm{q}'\right)=\hat{d}\left(\bm{q}\right)\hat{g}\left(\bm{q}'\right)$, so the solution of the integral equation is given byu the self-consistent equation: $\int\hat{g}\left(\bm{q}\right)\hat{d}\left(\bm{q}\right)d^2\bm{q}=1$. In this way, the binding equations for the in-gap bound states in the limit $b\rightarrow B$ are obtained to be
\begin{align}\label{Integral_Eq01}
\frac{1}{U_{\textrm{b}}}-\frac{A_{\textrm{c}}}{4\pi^2}\sum_{n'}\int_0^{|\bm{q}|=k_{\textrm{c}}}\frac{n_{\textrm{F}}\left(\xi_{n'}\right)+n_{\textrm{B}}\left(\epsilon_{\bm{q}}\right)}{E_{h,n}\left(B\right)+i0^+-\xi_{n'}+\epsilon_{\bm{q}}}D_{n,n'}\left(q\right)D_{n',n}\left(-q\right)\exp\left(-l_B^2\bm{q}^2/2\right)d^2\bm{q}=&0,\\\label{Integral_Eq02}
\frac{1}{U_{\textrm{b}}}+\frac{A_{\textrm{c}}}{4\pi^2}\sum_{n'}\int_0^{|\bm{q}|=k_{\textrm{c}}}\frac{n_{\textrm{F}}\left(-\xi_{n'}\right)+n_{\textrm{B}}\left(\epsilon_{\bm{q}}\right)}{E_{d,n}\left(B\right)+i0^+-\xi_{n'}-\epsilon_{\bm{q}}}D_{n,n'}\left(q\right)D_{n',n}\left(-q\right)\exp\left(-l_B^2\bm{q}^2/2\right)d^2\bm{q}=&0.
\end{align}
By numerically solving the integral equations in Eq. \ref{Integral_Eq01} and \ref{Integral_Eq02}, one can get the LLs spectrum $E_{h,n}\left(B\right)$ and $E_{d,n}\left(B\right)$ respectively. The energy levels of $E_{h,n}\left(B\right)$ correspond to the LLs spectrum of the hole bound state above the LHB, and the levels of $E_{d,n}\left(B\right)$ correspond to the LLs spectrum of the electronic bound state below the UHB. In Fig. \ref{figureS3}, the LLs spectrum of a hole bound state is shown. Given a quadratic hole bound state band in Fig. \ref{figureS3} (a), the LLs spectrum solved from Eq. \ref{Integral_Eq01} is plotted in Fig. \ref{figureS3} (b). The edge of the LLs spectrum $E_{h,n}\left(B\right)$ oscillates as the spinon chemical potential oscillates with $B$. In the $B\rightarrow0$T region, the LLs in $E_{h,n}\left(B\right)$ decrease linearly with $B$, which means that the excitation energy of a hole state from the LLs spectrum $E_{h,n}\left(B\right)$ increases linearly with $B$. As the magnetic field increases, one can see that the envelop of the oscillation in $E_{h,n}\left(B\right)$ turns from decreasing in $B$ to increasing in $B$. Accordingly, it means that the envelop energy to have a hole excitation turns from increasing with $B$ to decreasing with $B$. It indicates that the orbital magnretic field induced energy saving in the binding plays the dominant role when the magnetic field $B$ is sufficiently large. It matches the intiution that forming an in-gap bound state is energetically more favorable as $B$ increases, which has been analyzed in the maintext.

\begin{figure}
\centering
\includegraphics[width=6.9in]{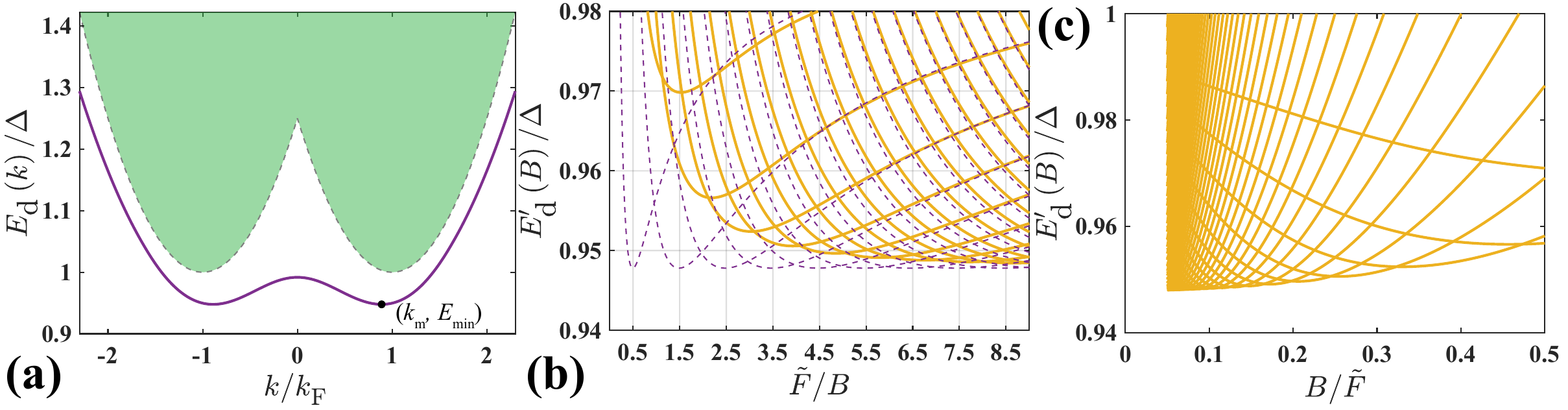}
\caption{(a) The electronic bound state band dispersion solved from Eq. \ref{Electronic_bound_band001}. It has a Mexican hat like shape. For an ordinary charged particle that has the band dispersion in (a), the corresponding Landau levels spectrum $\tilde{E}_n\left(B\right)$ is plotted as the dashed purple lines in (b). Specifically, the energy level $\tilde{E}_n\left(B\right)$ takes the minimum value at $\tilde{F}/B=n+\frac{1}{2}$ with $n=0,1,2,3,\dots$. The Landau levels spectrum $E'_{\textrm{d},n}\left(B\right)$ solved from the binding equation in Eq. \ref{Bound_Eq_electron_b_B} is plotted as the orange lines in (b). In the small magnetic field region $B\rightarrow0$, the energy levels spectrum $E_{\textrm{d},n}\left(B\right)\rightarrow\tilde{E}_n\left(B\right)$. (c) The electronic bound state LLs spectrum in (b) plotted as a function of $B/\tilde{F}$. In the calculation, the band parameters are taken to be $m_f/m_X=1$, $\Delta/\mu_f=4$, $U_{\textrm{b}}\rho_f\left(0\right)=5/\pi$, and $k_{\textrm{b}}T/\mu_f=0.05$. The momenta cut-off is set to be $k_{\textrm{c}}=2.6|\bm{k}_{\textrm{F}}|$.}\label{figureS4}
\end{figure}

\subsection{Binding Equations of the In-gap Bound States in the limit of $b\rightarrow0$T}
In the limit of $b\rightarrow0$T, the spinons do not feel the external orbital magnetic field and the orbital magnetic field is fully acted on the chargons. The QSL mean field Hamiltonian in the limit of $b\rightarrow0$T takes the form
\begin{align}\nonumber
H\left(0,B\right)=&H_0\left(0,B\right)+H_{\textrm{int}}\left(0,B\right)\\
=&\sum_{\sigma,\bm{k}}\xi_{\bm{k}}f^\dagger_{\sigma,\bm{k}}f_{\sigma,\bm{k}}+\sum_{n,m}\epsilon_n\left(a_{n,m}a^\dagger_{n,m}+b^\dagger_{n,m}b_{n,m}\right)\\
&+\sum_{\sigma,n,m,n',m'}\int d\bm{r}\int d\bm{r}'U_{\textrm{b}}\delta\left(\bm{r}-\bm{r}'\right)f^\dagger_{\sigma,\bm{r}}f_{\sigma,\bm{r}}\left(a_{n,m}a^\dagger_{n',m'}-b^\dagger_{n,m}b_{n',m'}\right)\psi_{n',m'}\left(\bm{r}'\right)\psi^\ast_{n,m}\left(\bm{r}'\right).
\end{align}
After Fourier transformation, the Hamiltonian $H\left(0,B\right)$ is simplified to be
\begin{align}\nonumber
H\left(0,B\right)=&\sum_{\sigma,\bm{k}}\xi_{\bm{k}}f^\dagger_{\sigma,\bm{k}}f_{\sigma,\bm{k}}+\sum_{n,m}\epsilon_n\left(a_{n,m}a^\dagger_{n,m}+b^\dagger_{n,m}b_{n,m}\right)\\
&+\frac{U_{\textrm{b}}}{N}\sum_{\bm{q},\bm{q}',\sigma}\sum_{n,m,n',m'}\sum_{l_1,l_2}D_{n,l_1}\left(q\right)D_{m,l_2}\left(q^\ast\right)D_{l_1,n'}\left(-q'\right)D_{l_2,m'}\left(-q'^\ast\right)e^{-\frac{l_B^2\left(\bm{q}^2+\bm{q}'^2\right)}{2}}f^\dagger_{\sigma,\bm{q}}f_{\sigma,\bm{q}'}\left(a_{n,m}a^\dagger_{n',m'}-b^\dagger_{n,m}b_{n',m'}\right).
\end{align}
Similar to the case of $b\rightarrow B$, we introduce the general anti-spinon holon pair state and spinon doublon pair state:
\begin{align}
\ket{h'_\sigma}=&\sum_{\bm{q},n,m}C_{h,\bm{q},n,m}\ket{h_\sigma,\bm{q},n,m}=\sum_{\bm{q},n,m}C_{h,\bm{q},n,m}f_{\sigma,\bm{q}}a^\dagger_{n,m}\ket{\textrm{G}},\\
\ket{d'_\sigma}=&\sum_{\bm{q},n,m}C_{d,\bm{q},n,m}\ket{d_\sigma,\bm{q},n,m}=\sum_{\bm{q},n,m}C_{d,\bm{q},n,m}f^\dagger_{\sigma,\bm{q}}b^\dagger_{n,m}\ket{\textrm{G}}.
\end{align}
By calculating the matrix elements $\bra{h_\sigma,\bm{q},n,m}H\left(0,B\right)\ket{h_\sigma,\bm{q},n,m}$ and $\bra{d_\sigma,\bm{q},n,m}H\left(0,B\right)\ket{d_\sigma,\bm{q},n,m}$, one can get the eigen equations of the anti-spinon holon pairs and the spinon doublon pairs in the limit of $b\rightarrow0$T as
\begin{align}\label{Matrix_integral_eq03}
-\frac{U_{\textrm{b}}}{N}\sum_{\bm{q},l_1,l_2,n',m'}D_{n,l_1}\left(q\right)D_{m,l_2}\left(q^\ast\right)e^{-\frac{l_B^2\bm{q}^2}{2}}D_{l_1,n'}\left(-q'\right)D_{l_2,m'}\left(-q'^\ast\right)e^{-\frac{l_B^2\bm{q}'^2}{2}}\frac{n_{\textrm{F}}\left(\xi_{\bm{q}}\right)+n_{\textrm{B}}\left(\epsilon_{n'}\right)}{-E'_{h,n}\left(B\right)+\xi_{\bm{q}'}-\epsilon_{n'}}A_{h,\bm{q}',n',m'}=&A_{h,\bm{q},n,m},\\\label{Matrix_integral_eq04}
-\frac{U_{\textrm{b}}}{N}\sum_{\bm{q},l_1,l_2,n',m'}D_{n,l_1}\left(q\right)D_{m,l_2}\left(q^\ast\right)e^{-\frac{l_B^2\bm{q}^2}{2}}D_{l_1,n'}\left(-q'\right)D_{l_2,m'}\left(-q'^\ast\right)e^{-\frac{l_B^2\bm{q}'^2}{2}}\frac{n_{\textrm{F}}\left(-\xi_{\bm{q}}\right)+n_{\textrm{B}}\left(\epsilon_{n'}\right)}{E'_{d,n}\left(B\right)-\xi_{\bm{q}'}-\epsilon_{n'}}A_{d,\bm{q}',n',m'}=&A_{d,\bm{q},n,m},
\end{align}
with
\begin{align}
A_{h,\bm{q}',n',m'}=\left[-E'_{h,n}\left(B\right)+\xi_{\bm{q}'}-\epsilon_{n'}\right]C_{h,\bm{q}',n',m',},\quad\textrm{and}\quad A_{d,\bm{q}',n',m'}=\left[E'_{d,n}\left(B\right)-\xi_{\bm{q}'}-\epsilon_{n'}\right]C_{d,\bm{q}',n',m'}.
\end{align}
The matrix integral equations in Eq. \ref{Matrix_integral_eq03} and \ref{Matrix_integral_eq04} are also the type $\hat{K}\left(\bm{q},\bm{q}'\right)\hat{C}\left(\bm{q}'\right)d^2\bm{q}'=\hat{C}\left(\bm{q}\right)$, so the self-consistent equations that give the bound state LLs spectrum derived to be
\begin{align}
\frac{1}{U_{\textrm{b}}}-\frac{A_{\textrm{c}}}{4\pi^2}\sum_{n'}\int_0^{|\bm{q}|=k_{\textrm{c}}}\frac{n_{\textrm{F}}\left(\xi_{\bm{q}}\right)+n_{\textrm{B}}\left(\epsilon_{n'}\right)}{E'_{h,n}\left(B\right)+i0^+-\xi_{\bm{q}}+\epsilon_{n'}}D_{n,n'}\left(-q\right)D_{n',n}\left(q\right)\exp\left(-l_B^2\bm{q}^2/2\right)\frac{d^2\bm{q}}{4\pi^2}=&0,\\\label{Bound_Eq_electron_b_B}
\frac{1}{U_{\textrm{b}}}+\frac{A_{\textrm{c}}}{4\pi^2}\sum_{n'}\int_0^{|\bm{q}|=k_{\textrm{c}}}\frac{n_{\textrm{F}}\left(-\xi_{\bm{q}}\right)+n_{\textrm{B}}\left(\epsilon_{n'}\right)}{E'_{d,n}\left(B\right)+i0^+-\xi_{\bm{q}}-\epsilon_{n'}}D_{n,n'}\left(-q\right)D_{n',n}\left(q\right)\exp\left(-l_B^2\bm{q}^2/2\right)\frac{d^2\bm{q}}{4\pi^2}=&0.
\end{align}
The energy levels of $E'_{h,n}\left(B\right)$ represent the LLs spectrum of the hole bound state above the LHB in the limit of $b\rightarrow0$T, and those of $E'_{d,n}\left(B\right)$ give the LLs spectrum of the electronic bound state below the UHB in the limit of $b\rightarrow0$T. Given an electronic bound state band dispersion in Fig. \ref{figureS4} (a), the electronic bound state LLs spectrum solved in Eq. \ref{Bound_Eq_electron_b_B} in the $b\rightarrow0$T limit is plotted in Fig. \ref{figureS4} (b) and (c). As $B\rightarrow0$T, the envelop of the LLs spectrum $E'_{\textrm{d},n}\left(B\right)$ approaches to the electronic bound state band mininum $E_{\textrm{min}}$ as expected. Interestingly, the envelop of $E'_{\textrm{d},n}\left(B\right)$ is found to increase quadratically with $B$ as is seen in Fig. \ref{figureS4} (c). In the $b\rightarrow0$T limit, all the magnetic field is acted on the chargons. As the magnetic field increases, all the chargon LL increases linearly. Although the magnetic field localizes the chargons and makes the energy saved in the binding increase with $B$, the energy saved in the binding cannot compete with the energy increase in the chargon LLs. Since multiple chargon LLs are involved in the binding, the total energy increase of the resulting bound states is then integrated to be the $B^2$ type. Such quadratic increase of the bound state LLs envelop energy in the $b\rightarrow0$T limit is in sharp contrast to the quadratic decrease in the $b\rightarrow B$ limit.

\section{Landau Quantization of an Isotropic Electronic Band}
For an isotropic electronic band that has the form $E\left(\bm{k}\right)=f\left(\bm{k}^2\right)$, it can always be expanded as
\begin{align}\label{Ek}
E\left(\bm{k}\right)=f\left(0\right)+f'\left(0\right)\bm{k}^2+\frac{1}{2!}f''\left(0\right)\bm{k}^4+\frac{1}{3!}f'''\left(0\right)\bm{k}^6+\cdots.
\end{align}
In the presence of a magnetic field along $z$ direction, the momentum is replaced by the canonical momentum: $\hbar\bm{k}\rightarrow\bm{\pi}$, and the canonical momentum $\bm{\pi}$ respects the commutation relation $\left[\pi_x, \pi_y\right]=ie\hbar B$. The ladder operators are defined as
\begin{align}
a=\sqrt{\frac{1}{2\hbar eB}}\left(\pi_x+i\pi_y\right)\rightarrow\frac{l_B}{\sqrt{2}}\left(k_x+ik_y\right),\quad\quad\quad a^\dagger=\sqrt{\frac{1}{2\hbar eB}}\left(\pi_x-i\pi_y\right)\rightarrow\frac{l_B}{\sqrt{2}}\left(k_x-ik_y\right),
\end{align}
which respects the bosonic commutation relation: $\left[a, a^\dagger\right]=1$. Applying the substitution $k_x\rightarrow\frac{a+a^\dagger}{\sqrt{2}l_B}$ and $k_y\rightarrow-i\frac{a-a^\dagger}{\sqrt{2}l_B}$ back to the $E\left(\bm{k}\right)$ in Eq. \ref{Ek}, we can obtain
\begin{align}
E\left[\frac{a+a^\dagger}{\sqrt{2}l_B},-i\frac{a-a^\dagger}{\sqrt{2}l_B}\right]=f\left(0\right)+f'\left(0\right)2l_B^{-2}\left(a^\dagger a+\frac{1}{2}\right)+\frac{1}{2!}f''\left(0\right)\left[2l_B^{-2}\left(a^\dagger a+\frac{1}{2}\right)\right]^2+\frac{1}{3!}f'''\left(0\right)\left[2l_B^{-2}\left(a^\dagger a+\frac{1}{2}\right)\right]^3+\cdots.
\end{align}
We know that the eigenvalue of $a^\dagger a$ is $n$, which is the Landau level index. As a result, after Landau quantization of an ordinary electronic band $E\left(\bm{k}\right)$, its $n$th Landau level is
\begin{align}\label{En_B}
\tilde{E}_n\left(B\right)=E\left[\sqrt{2l_B^{-2}\left(n+\frac{1}{2}\right)}\right].
\end{align}
This way to obtain the Landau levels spectrum applies to the electron that has never been fractionalized. Given an electron that has the same band dispersion as the Mexican hat like electronic bound state band shown in Fig. \ref{figureS4} (a), the LLs spectrum $\tilde{E}_n\left(B\right)$ is plotted as the dashed purple lines in Fig. \ref{figureS4} (b). Importantly, as the Mexican hat like band has the band mininum at $|\bm{k}|=k_m$, the resulting LLs spectrum $\tilde{E}_n\left(B\right)$ takes the same minimum value at $1/B=\frac{2\pi e}{\hbar\pi k^2_m}\left(n+\frac{1}{2}\right)$. Therefore, the edge of the spectrum $\tilde{E}_n\left(B\right)$ oscillates in $1/B$ as can be seen in Fig. \ref{figureS4} (b). The oscillation frequency is $\tilde{F}=\frac{\pi\hbar k^2_m}{2\pi e}$, where the wave vector at the band minimum plays the role of the Fermi wave vector in a metal.

\end{document}